\documentclass[a4paper,11pt]{article}
\pdfoutput=1 

\usepackage{jheppub} 

\usepackage[T1]{fontenc} 

\usepackage{amsfonts,slashed,tikz,bm,psfrag,color,dsfont,soul, multicol,multirow,amsthm,xcolor,float,esint,mathrsfs,mathtools,mathdots,dsfont,listings,subfigure,makecell}
\usepackage[normalem]{ulem}
\usepackage{placeins}

\allowdisplaybreaks
\renewcommand{\arraystretch}{1.2}

\title{\boldmath SCET sum rules for $B\to D_1(2420)$ and $B\to D_1'(2430)$ form factors at next-to-leading order  }

\author[a]{Jun-Wei Zhang}
\author[a]{and Yong-Kang Huang}

\affiliation[a]{School of Physics, Nankai University, Tianjin 300071, China}

\emailAdd{zhjw@mail.nankai.edu.cn}

\emailAdd{huangyongkang@mail.nankai.edu.cn}

\abstract{
We present the first calculation of the $B \to D_1(2420)$ and $B \to D'_1(2430)$
transition form factors at ${\cal O}(\alpha_s)$ using light-cone sum rules within
the framework of soft-collinear effective theory (SCET). We first match the QCD transition currents onto ${\rm SCET_I}$ and
then factorize the corresponding vacuum-to-$B$-meson correlation
functions in ${\rm SCET_{II}}$. The resulting factorization formulae
are used to construct the leading-power sum rules for the effective
SCET form factors $\xi^R_{\parallel/\perp}$ and
$\Xi^R_{\parallel/\perp}$ ($R=D_1,D'_1$). In particular, we calculate the additional longitudinal form factor
$\xi^R_{\parallel,m_c}$ induced by the finite charm-quark mass, whose
contribution depends only on the $B$-meson light-cone distribution
amplitude $\phi_B^+(\omega,\mu)$. To disentangle the mixed $D_1$ and $D'_1$ states, we
introduce dedicated combinations of interpolating currents, with their decay
constants determined via the equations of motion. To isolate the orbitally excited states from ground-state contamination, we subtract the ground-state contribution from the total sum rules and examine the stability of the resulting sum rules. Furthermore, the $q^2$-dependence of the physical form factors
is extrapolated over the full kinematic region using the
Bourrely--Caprini--Lellouch parameterization. Finally, we provide
phenomenological predictions for the branching fractions, differential decay
widths, and lepton flavor universality ratios. Numerically, we obtain
$R(D_1) = 0.070^{+0.028}_{-0.018}$ and
$R(D'_1) = 0.159^{+0.032}_{-0.025}$, which can be confronted with the future measurements
at Belle~II and LHCb.}

\begin{document} 
\maketitle
\flushbottom

\section{Introduction}
\label{sec:intro}

Precision tests of the Standard Model (SM) and searches for new physics heavily
rely on the rich phenomenology of semileptonic $B$-meson decays. These channels,
particularly the $b \to c \ell \bar{\nu}_\ell$ transitions, offer a clean theoretical
and experimental environment for extracting the Cabibbo-Kobayashi-Maskawa (CKM)
matrix element $|V_{cb}|$ and conducting stringent tests of lepton flavor universality (LFU), as the
strong interaction effects can be factorized into a limited number of
form factors. For decades, theoretical and experimental efforts have been predominantly
directed towards the decays into the ground-state charm mesons, e.g., $B \to D$ and
$B \to D^*$ (see review \cite{Gambino:2020jvv}). Accordingly, the theoretical
determinations of these ground-state form factors have reached a high level of
precision. On the lattice QCD frontier, first-principles evaluations have evolved
from the static zero-recoil limits to comprehensive calculations at nonzero recoil
across the full kinematic phase space. State-of-the-art unquenched lattice simulations,
utilizing improved fermion actions (such as the recent developments by the Fermilab,
MILC, and HPQCD collaborations \cite{MILC:1503.07237, MILC:2105.14019,
HPQCD:2304.03137}; see also the comprehensive review by FLAG
\cite{FLAG:2411.04268}), now provide tight constraints on both the normalization
and the dependence on $q^2$ of the form factors governing these ground-state transitions.
On the analytical side, building upon the foundational evaluations based on light-cone sum rules (LCSRs) \cite{Faller:0809.0222}, the theoretical framework has been
systematically advanced through the calculation of next-to-leading order (NLO)
perturbative corrections \cite{Wang:1701.06810} and the detailed exploration of
subleading-power effects \cite{Gao:2112.12674}.

\begin{table}[ht!]
\centering
\begin{tabular}{l c c c c}
\hline\hline
State & $j_q$ & $J^P$ & Mass [MeV] & Width [MeV] \\ [0.5ex] 
\hline
$D_0^*(2300)$ & $1/2$ & $0^+$ & $2343 \pm 10$       & $229 \pm 16$ \\
$D_1'(2430)$  & $1/2$ & $1^+$ & $2412 \pm 9$       & $314 \pm 29$ \\
$D_1(2420)$   & $3/2$ & $1^+$ & $2422.1 \pm 0.6$    & $31.3 \pm 1.9$ \\
$D_2^*(2460)$ & $3/2$ & $2^+$ & $2461.1 \pm 0.8$    & $47.3 \pm 0.8$ \\
\hline\hline
\end{tabular}
\caption{Properties of the four orbitally excited $P$-wave charm mesons ($D^{**}$ states). The mass and width values are quoted from the Particle Data Group (PDG) world averages \cite{PDG:2024}.}
\label{tab:D**_properties}
\end{table}

Moving beyond the ground-state mesons, the transitions to orbitally excited P-wave states---collectively
denoted as $D^{**}$ mesons---present a much more intricate dynamical picture. The $D^{**}$ system
comprises four physical states, among which the axial-vector $D_1$ and $D'_1$ mesons constitute a
``$J^P=1^+$ doublet''. In the heavy-quark symmetry (HQS) limit ($m_c \to \infty$), they belong to distinct
spin-symmetry doublets characterized by the total angular momentum of the light degrees of freedom: $j_q=3/2$ and $j_q=1/2$.
These two basis states exhibit quite different decay behaviors dictated by angular momentum conservation \cite{Neubert:1993mb}.
Since the $D^*\pi$ final state consists of a vector ($J^P=1^-$) and a pseudoscalar ($0^-$),
the conservation of parity and total angular momentum restricts the allowed partial waves for
a $1^+$ decay to $L=0$ (S-wave) or $L=2$ (D-wave). Furthermore, HQS dictates that the light-quark
angular momentum $j_q$ must be conserved in the decay to the ground-state $D^*$ ($j_q=1/2$).
Consequently, the $j_q=3/2$ state is kinematically restricted to decay strictly via a D-wave ($1/2 \otimes 2 \to 3/2$), where
the centrifugal barrier significantly suppresses its decay width, rendering it a narrow resonance.
Conversely, the $j_q=1/2$ state can decay predominantly via an S-wave ($1/2 \otimes 0 \to 1/2$),
unhindered by such a barrier, resulting in a remarkably broad width \cite{Isgur:1991wq,Lu:1991px}. At a finite charm-quark mass, these basis states undergo physical mixing. It is instructive to compare this with the light-quark sector, where mixing is typically driven by $SU(3)$ flavor symmetry breaking. A well-known example is the physical $K_1(1270)$ and $K_1(1400)$ doublet \cite{Suzuki:1993yc,Cheng:1110.2249}, which are substantial mixtures of the $^1P_1$ ($K_{1B}$) and $^3P_1$ ($K_{1A}$) basis states. In contrast, the mixing in the heavy-light $D^{**}$ system is driven by HQS breaking and is parametrically suppressed by $\mathcal{O}(1/m_c)$. Under this suppression, the physical mixing angle is expected to be small, and it is a common convention to directly label the physical $D_1$ and $D_1'$ states as the $j_q = 3/2$ and $1/2$ pure states, respectively. Indeed, as summarized in Table~\ref{tab:D**_properties}, experimental measurements of these two physical states corroborate this HQS prediction, confirming the existence of one relatively narrow state ($\Gamma_{D_1} \sim 30$ MeV) and one extremely broad state ($\Gamma_{D'_1} \sim 300$ MeV).

Obtaining high-precision theoretical predictions for the $B \to D^{**}$ transition
form factors has become an important task in heavy-flavor physics for several reasons.
Firstly, a quantitative knowledge of these orbitally excited channels is essential for
saturating the inclusive $B \to X_c \ell \bar{\nu}_\ell$ decay width. While the
ground-state transitions $B \to D^{(*)} \ell \bar{\nu}_\ell$ account for approximately
$70\%-75\%$ of the total inclusive rate \cite{PDG:2024}, the remaining gap must be
predominantly filled by decays into these $D^{**}$ states. Secondly, the
$B \to D^{**} \ell \nu$ processes constitute a complex ``feed-down'' background in
the measurement of $R(D^{(*)})$. The substantial systematic uncertainties associated
with this background render a precise theoretical determination of the $R(D^{**})$
ratios essential for a reliable extraction of $R(D^{(*)})$
\cite{BaBar:1205.5442, Belle:1904.08794}. Finally, in lattice QCD, while variational
methods \cite{Michael:1985ne, Luscher:1990ck, Blossier:0902.1265} are employed to
extract excited states, controlling excited-state contaminations in three-point
functions remains challenging \cite{Barca:2508.09006}. In particular, for the
$J^P=1^+$ doublet studied in this work, the broad and unstable $D'_1$ state decays
predominantly via S-wave into $D^*\pi$, so that extracting its form factors on the
lattice requires a coupled-channel finite-volume analysis of the relevant scattering
amplitudes \cite{Briceno:1706.06223}. The near-degenerate masses of $D_1$ and $D'_1$
further complicate their clean separation, making the lattice evaluation of this
coupled system a significant challenge.

To systematically separate the widely disparate energy scales at large hadronic
recoil, soft-collinear effective theory (SCET) factorization \cite{Beneke:0311335,Bauer:0109045} has gradually become
the standard procedure, particularly in the context of heavy-to-light transitions.
In recent years, the NLO radiative corrections and subleading-power effects have been
systematically computed for these processes
\cite{Gao:2019lta,huang:2212.11624,cui:2301.12391}. In the kinematic region where
the momentum transfer $q^2$ is small, the charm meson in the final state is
highly boosted with an energy $E \sim m_B / 2$. Within the SCET framework, this
energetic charm quark can be effectively treated as a hard-collinear particle, since
the finite charm-quark mass $m_c$ is essentially invisible at the hard scale
$\mu_h \sim m_b$. Its dynamical effects only emerge at the hard-collinear scale
$\mu_{hc} \sim \sqrt{m_b \Lambda_{\rm QCD}}$, where the charm quark behaves as a
massive ``light-flavor'' particle in the boosted regime, but with a different analytic
structure of the jet functions compared to the massless case. Given this physical
picture, a systematic evaluation of the $B \to D_1^{(\prime)}$ form factors within
the SCET factorization framework is both feasible and essential.

In this work, we present the first calculation of the $B \to D_1$ and $B \to D'_1$
form factors at $\mathcal{O}(\alpha_s)$ accuracy. We apply and extend the standard
SCET factorization approach for heavy-to-light transitions to a scenario where the
final states are excited and involve physical mixing. To disentangle the $D_1$ and
$D'_1$ states, we employ specific linear combinations of interpolating currents, a
strategy introduced in~\cite{Gubernari:2203.08493} at the tree level. Notably, the
SCET framework necessitates a separate treatment of longitudinal and transverse
polarization contributions, which requires the construction of a more extensive set
of current structures compared to the tree-level analysis. Furthermore, constructing the vacuum-to-$B$-meson correlation functions
using these interpolating currents inevitably incorporates unwanted
contributions from the ground states. To address this, we construct
the full sum rules that encompass both the ground-state and $P$-wave
contributions, and subtract the corresponding ground-state sum rules
to obtain dedicated sum rules for the orbitally excited $D_1^{(\prime)}$
states. By systematically incorporating NLO radiative corrections and performing
multi-scale resummation, we achieve a consistent perturbative description of the
$B \to D_1^{(\prime)}$ form factors at large hadronic recoil.

The outline of this paper is as follows. In Sec.~\ref{sec2}, we present
the ${\rm SCET_I}$ factorization formulae for the $B \to D_1^{(\prime)}$
form factors. In Sec.~\ref{sec3}, we construct the disentangling
interpolating currents and perform the ${\rm SCET_{II}}$ matching to
obtain the final LCSRs for the ``effective'' form factors at
$\mathcal{O}(\alpha_s)$. In Sec.~\ref{sec4}, we present numerical
results for the physical form factors and phenomenological predictions
for branching fractions, LFU ratios, and differential decay widths.
Finally, we conclude in Sec.~\ref{sec5}.

\section{SCET factorization for $B\rightarrow D^{(\prime)}_1$ form factors}\label{sec2}
In this section, we derive the SCET factorization formulae for the
$B\to D_1^{(\prime)}$ form factors at large hadronic recoil. We first define the $B\rightarrow D^{(\prime)}_1$ form factors through
the standard Lorentz decomposition of the QCD matrix elements of the
generic heavy-to-light currents $\bar\psi\Gamma_i Q$
\cite{Beneke:0008255}:
\begin{align}
    \left\langle R\left(p, \epsilon^*\right)\left|\bar{q} \gamma_\mu  b\right| \bar{B}(p+q)\right\rangle ={}&\frac{2 m_{R} \epsilon^* \cdot q}{q^2} q_\mu V^R_0\left(q^2\right) 
\nonumber\\
    & +\left(m_B+m_{R}\right)\left[\epsilon_\mu^*-\frac{\epsilon^* \cdot q}{q^2} q_\mu\right] V^R_1\left(q^2\right) 
\nonumber\\
    & -\frac{\epsilon^* \cdot q}{m_B+m_{R}}\left[(2 p+q)_\mu-\frac{m_B^2-m_{R}^2}{q^2} q_\mu\right] V^R_2\left(q^2\right), 
\nonumber\\
    \left\langle R\left(p, \epsilon^*\right)\left|\bar{q}  \gamma_\mu \gamma_5 b\right| \bar{B}(p+q)\right\rangle ={}&-\frac{2 i A^R\left(q^2\right)}{m_B+m_{R}} \epsilon_{\mu \nu \rho \sigma} \epsilon^{* \nu} p^\rho q^\sigma, 
\nonumber\\
    \left\langle R\left(p, \epsilon^*\right)\left|\bar{q} i \sigma_{\mu \nu}  q^\nu b\right| \bar{B}(p+q)\right\rangle  ={}&T^R_2\left(q^2\right)\left[\left(m_B^2-m_{R}^2\right) \epsilon_\mu^*-\left(\epsilon^* \cdot q\right)(2 p+q)_\mu\right] 
\nonumber\\
    & +T^R_3\left(q^2\right)\left(\epsilon^* \cdot q\right)\left[q_\mu-\frac{q^2}{m_B^2-m_{R}^2}(2 p+q)_\mu\right],
    \nonumber\\
    \left\langle R\left(p, \epsilon^*\right)\left|\bar{q} i \sigma_{\mu \nu}\gamma_5 q^\nu b\right| \bar{B}(p+q)\right\rangle  ={}&2 i T^R_1\left(q^2\right) \epsilon_{\mu \nu \rho \sigma} \epsilon^{* \nu} p^\rho q^\sigma \,,\qquad(R=D_1,D'_1)
\label{qcdff}
\end{align}
where $p$ is the momentum of the charmed meson and $q$ is the momentum
transfer carried by the weak current and we use $F^R\equiv F^{B\to R}$ throughout for brevity. We use the convention
$\epsilon_{0123}=-1$. At maximal hadronic recoil, the form-factor
definitions imply the kinematic constraints
\begin{align}
\frac{m_B+m_R}{2 m_{R}} V^R_1(0)-\frac{m_B-m_R}{2 m_{R}} V^R_2(0)=V^R_0(0), \quad T^R_1(0)=T^R_2(0) .
\label{eq:kinematic-constraints}
\end{align}
We use two light-like vectors $n$ and $\bar n$, satisfying
$n^2=\bar n^2=0$ and $n\cdot\bar n=2$, and choose
$v=(n+\bar n)/2$, such that $n\cdot p$ is the large light-cone
component of the final-state momentum. Choosing $p_\perp=0$, the remaining light-cone component satisfies
$\bar n\cdot p=m_R^2/(n\cdot p)$. The momentum transfer is therefore
related to the light-cone components through
\begin{align}
q^2=m_B^2+m_R^2-m_B
\left(n\cdot p+\bar n\cdot p\right).
\end{align}

For a massless quark in the final state, after the multipole expansion,
the ${\rm SCET_I}$ representation of the generic color-singlet
heavy-to-light current $\bar\psi\Gamma_i Q$ reads
\cite{Beneke:0508250}
\begin{align}
    \left(\bar{\psi} \Gamma_i Q\right)(0)= & \int d \hat{s} \sum_j \tilde{C}_{i j}^{(\mathrm{A} 0)}(\hat{s}) O_j^{(\mathrm{A} 0)}(s ; 0)+\int d \hat{s} \sum_j \tilde{C}_{i j \mu}^{(\mathrm{A} 1)}(\hat{s}) O_j^{(\mathrm{A} 1) \mu}(s ; 0) \nonumber\\
& +\int d \hat{s}_1 \int d \hat{s}_2 \sum_j \tilde{C}_{i j \mu}^{(\mathrm{B} 1)}\left(\hat{s}_1, \hat{s}_2\right) O_j^{(\mathrm{B} 1) \mu}\left(s_1, s_2 ; 0\right)+\ldots,\label{obasisexpansion}
\end{align}
where the ellipses indicate the further suppressed power effects. The explicit expressions for the gauge-invariant operator basis are given by
\begin{align}
    O_j^{(\mathrm{A} 0)}(s ; 0) \equiv{}& (\bar \xi W_c)(sn)\Gamma'_jh_v(0)\,,\nonumber\\
    O_j^{(\mathrm{A} 1) \mu}(s ; 0)\equiv{}&(\bar \xi i \overleftarrow{D}_{\perp c \mu}(i n\cdot\overleftarrow{D}_c)^{-1}W_c)(s n)
\Gamma'_jh_v(0)\,,\nonumber\\
    O_j^{(\mathrm{B} 1) \mu} (s_1,s_2 ; 0)\equiv{}&{1 \over m_b}(\bar \xi W_c)(s_1 n)(W_c^\dagger iD_{\perp c \mu}W_c)(s_2 n)\Gamma'_j h_v(0)\,, 
\end{align}
where $\hat s_{(i)}=m_b s_{(i)}$. Momentum conservation in the hard-collinear sector permits setting
$s=0$ for the A-type operators and $s_1=0$ for the B-type operators. The light-cone Wilson line is introduced to maintain collinear gauge invariance:
\begin{align}
     W_c(x)={\rm P \,\,exp} \left[ ig_s\int^0_{-\infty }ds \, n \cdot A_c(x+s n) \right].\label{collineargaugelink}
\end{align}
The A1-type operator contains a transverse covariant derivative,
whereas the B1-type operator contains an additional transverse
collinear building block. When the quark in the final state is
massless, the A1 contribution can be removed from the factorization
formulae at leading power by a redefinition of the operator basis
\cite{Becher:0408344,Hill:0404217}. Retaining a finite charm-quark mass gives rise to additional A1
operators \cite{cui:2301.12391}, whose longitudinal and transverse matrix elements define
$\xi^R_{\parallel,m_c}$ and $\xi^R_{\perp,m_c}$:
\begin{align}
\left\langle R(p,\epsilon^*)\left|
O_{\parallel}^{(\mathrm{A1},m_c)}
\right|\bar B_v\right\rangle
={}&-n\cdot p\,(\epsilon^*\cdot v)\,
\xi^R_{\parallel,m_c}(n\cdot p),
\nonumber\\
\left\langle R(p,\epsilon^*)\left|
O_{\perp,\mu}^{(\mathrm{A1},m_c)}
\right|\bar B_v\right\rangle
={}&-n\cdot p\,
\left(\epsilon_\mu^*-\epsilon^*\cdot v\,\bar n_\mu\right)
\xi^R_{\perp,m_c}(n\cdot p),
\end{align}
where
\begin{align}
O_{\parallel}^{(\mathrm{A1},m_c)}
\equiv{}&
(\bar\xi W_c)\frac{\not\! n}{2}
\frac{m_c}{-i n\cdot\overleftarrow D_c}h_v,
\nonumber\\
O_{\perp,\mu}^{(\mathrm{A1},m_c)}
\equiv{}&
(\bar\xi W_c)\frac{\not\! n}{2}
\frac{m_c}{-i n\cdot\overleftarrow D_c}
\gamma_{\mu\perp}h_v.\label{Omc}
\end{align}
These contributions remain at leading power for
$m_c\sim\sqrt{m_b\Lambda_{\rm QCD}}$. We notice that $\xi^R_{\perp,m_c}$ vanishes at $\mathcal{O}(\alpha_s)$ in four
dimensions.

The effective form factors $\xi^{R}_a$ and $\Xi^R_a$
$(a=\|,\perp)$ are defined through the matrix elements of the A0- and B1-type
operators:
\begin{align}
    & \left\langle R\left(p, \epsilon^*\right)\left|\left(\bar{\xi} W_c\right)  h_v\right| \bar{B}_v\right\rangle=-n \cdot p\left(\epsilon^* \cdot v\right) \xi^R_{\|}(n \cdot p), \nonumber\\
& \left\langle R\left(p, \epsilon^*\right)\left|\left(\bar{\xi} W_c\right)  \gamma_{\mu \perp} h_v\right| \bar{B}_v\right\rangle=-n \cdot p\left(\epsilon_\mu^*-\epsilon^* \cdot v \bar{n}_\mu\right) \xi^R_{\perp}(n \cdot p), \nonumber\\
& \left\langle R\left(p, \epsilon^*\right)\left|\left(\bar{\xi} W_c\right) \left(W_c^{\dagger} i \not\!\! D_{c \perp} W_c\right)(r n) h_v\right| \bar{B}_v\right\rangle=-n \cdot p m_b \epsilon^* \cdot v \int_0^1 d \tau e^{i \tau n \cdot p r} \Xi^R_{\|}(\tau, n \cdot p), \nonumber\\
& \left\langle R\left(p, \epsilon^*\right)\left|\left(\bar{\xi} W_c\right)  \gamma_{\mu \perp}\left(W_c^{\dagger} i \not\!\! D_{c \perp} W_c\right)(r n) h_v\right| \bar{B}_v\right\rangle \nonumber\\
& =-n \cdot p m_b\left(\epsilon_\mu^*-\epsilon^* \cdot v \bar{n}_\mu\right) \int_0^1 d \tau e^{i \tau n \cdot p r} \Xi^R_{\perp}(\tau, n \cdot p).\label{SCETmatrixelements}
\end{align}
Matching the QCD matrix elements onto the SCET matrix elements gives
the factorization formulae at leading power:
\begin{align}
F_i^R(n\cdot p)
={}&C_i^{(\mathrm{A0})}(n\cdot p)\xi_a^R(n\cdot p)
+C_i^{(\mathrm{A1},m_c)}(n\cdot p)
 \xi_{a,m_c}^R(n\cdot p)
\nonumber\\
&+\int_0^1d\tau\,
 C_i^{(\mathrm{B1})}(\tau,n\cdot p)
 \Xi_a^R(\tau,n\cdot p),
\qquad (a=\parallel,\perp) 
\label{facformulae}
\end{align}
where the hard coefficients are independent of the final hadronic
state. With $\bar\tau=1-\tau$, the explicit factorization formulae
read
\begin{align}
&\begin{aligned}
{m_B \over m_B + m_R} \, A^R(n \cdot p) ={}&
C_{A}^{(\rm A0)} \, \left ({n \cdot p \over m_b}, \mu \right )
\xi^{R}_{\perp}(n \cdot p) \\
&+ \int_0^1 d\tau \, C_{A}^{(\rm B1)}
\left ({n \cdot p \, \bar\tau \over m_b},
{n \cdot p \, \tau \over m_b},\mu \right )
\Xi^R_{\perp}(\tau, n \cdot p) \,,
\end{aligned}
\nonumber\\
&\begin{aligned}
{2m_R \over n \cdot p} \, V^R_0(n \cdot p) ={}&
C_{f_0}^{(\rm A0)} \, \left ({n \cdot p \over m_b}, \mu \right )
\xi^R_{\parallel}(n \cdot p)+ C_{f_0}^{(\rm A1,m_c)} \,
\left ({n \cdot p \over m_b}, \mu \right )
\xi^R_{\parallel,m_c}(n \cdot p) \\
&+ \int_0^1 d\tau \, C_{f_0}^{(\rm B1)}
\left ({n \cdot p \, \bar\tau \over m_b},
{n \cdot p \, \tau \over m_b},\mu \right )
\Xi^R_{\parallel}(\tau, n \cdot p) \,,
\end{aligned}
\nonumber\\
&\begin{aligned}
{m_B + m_R \over n \cdot p} \, V^R_1(n \cdot p) ={}&
C_{A}^{(\rm A0)} \, \left ({n \cdot p \over m_b}, \mu \right )
\xi^R_{\perp}(n \cdot p) \\
&+ \int_0^1 d\tau \, C_{A}^{(\rm B1)}
\left ({n \cdot p \, \bar\tau \over m_b},
{n \cdot p \, \tau \over m_b},\mu \right )
\Xi^R_{\perp}(\tau, n \cdot p) \,,
\end{aligned}
\nonumber\\
&\begin{aligned}
&{m_B + m_R \over n \cdot p} \, V^R_1(n \cdot p)
- {m_B - m_R \over m_B} \, V^R_2(n \cdot p)\\ &={}
C_{f_+}^{(\rm A0)} \, \left ({n \cdot p \over m_b}, \mu \right )
\xi^R_{\parallel}(n \cdot p)+ C_{f_+}^{(\rm A1,m_c)} \,
\left ({n \cdot p \over m_b}, \mu \right )
\xi^R_{\parallel,m_c}(n \cdot p) \\
&\quad+ \int_0^1 d\tau \, C_{f_+}^{(\rm B1)}
\left ({n \cdot p \, \bar\tau \over m_b},
{n \cdot p \, \tau \over m_b},\mu \right )
\Xi^R_{\parallel}(\tau, n \cdot p) \,,
\end{aligned}
\nonumber\\
&\begin{aligned}
T^R_1(n \cdot p) ={}&
C_{T_1}^{(\rm A0)} \, \left ({n \cdot p \over m_b}, \mu \right )
\xi^R_{\perp}(n \cdot p) \\
&+ \int_0^1 d\tau \, C_{T_1}^{(\rm B1)}
\left ({n \cdot p \, \bar\tau \over m_b},
{n \cdot p \, \tau \over m_b},\mu \right )
\Xi^R_{\perp}(\tau, n \cdot p) \,,
\end{aligned}
\nonumber\\
&\begin{aligned}
{m_B \over n \cdot p} \, T^R_2(n \cdot p) ={}&
C_{T_1}^{(\rm A0)} \, \left ({n \cdot p \over m_b}, \mu \right )
\xi^R_{\perp}(n \cdot p) \\
&+ \int_0^1 d\tau \, C_{T_1}^{(\rm B1)}
\left ({n \cdot p \, \bar\tau \over m_b},
{n \cdot p \, \tau \over m_b},\mu \right )
\Xi^R_{\perp}(\tau, n \cdot p) \,,
\end{aligned}
\nonumber\\
&\begin{aligned}
&{m_B \over n \cdot p} \, T^R_2(n \cdot p)-T^R_3(n \cdot p) \\
&={}
C_{f_T}^{(\rm A0)} \, \left ({n \cdot p \over m_b}, \mu \right )
\xi^R_{\parallel}(n \cdot p)+ C_{f_T}^{(\rm A1,m_c)} \,
\left ({n \cdot p \over m_b}, \mu \right )
\xi^R_{\parallel,m_c}(n \cdot p)  \\
&\quad+ \int_0^1 d\tau \, C_{f_T}^{(\rm B1)}
\left ({n \cdot p \, \bar\tau \over m_b},
{n \cdot p \, \tau \over m_b},\mu \right )
\Xi^R_{\parallel}(\tau, n \cdot p) \,,
\end{aligned}
\label{SCET-I-factorization-formulae}
\end{align}
where we neglect the $\xi^R_{\perp,m_c}$ terms since they vanish at the considered accuracy.
The hard coefficient functions in these expressions are given in
momentum space \cite{cui:2301.12391,Beneke:0311335}. It is straightforward to see from these factorization formulae that
 \begin{align}
{m_B \over m_B + m_R} \, A^R(n \cdot p)  = {m_B + m_R \over n \cdot p} \, V^R_1(n \cdot p) \,,
\qquad  T^R_1(n \cdot p) =  {m_B \over n \cdot p} \,T^R_2(n \cdot p) \,.
\label{large-recoil-relations}
\end{align}
These relations hold to all orders in $\alpha_s$ at leading power for
the A0- and B1-type contributions
\cite{Beneke:0008255,Burdman:2000ku}. 
 
\section{LCSRs for $B\rightarrow D^{(\prime)}_1$ form factors}\label{sec3}

In this section, we construct the SCET sum rules for the effective
form factors introduced in Sec.~\ref{sec2}. The $D_1$ and $D'_1$
mesons have the same quantum numbers $J^P=1^+$, while their light
degrees of freedom carry different angular momenta in the heavy-quark
limit. Since each interpolating current generally couples to both physical
states, a single current cannot isolate either state in the
correlation functions. We therefore construct separate linear
combinations for the longitudinal and transverse polarizations by
requiring each combination to have a vanishing matrix element with
the other state.
\subsection{Interpolating currents, decay constants and correlation functions}

We start with four linearly independent currents 
\begin{align}
\label{cJ1}
    &j^1_\mu  =\left(m_c+m_q\right) \bar{q} \gamma_\mu \gamma_5 c 
    \,,
    \\* 
    &j_{\mu\nu}^{2}  =i \bar{q} \gamma_\mu \overleftrightarrow{D_\nu}\gamma_5  c
    \,,\label{cJ2}\\
    &j^3_{\mu\rho}  =\left(m_c+m_q\right) \bar{q} \gamma_\mu \gamma_{\rho_{\perp}}\gamma_5 c 
    \,,\label{cJ3}
    \\* 
    &j_{\mu\nu\rho}^{4}  =i \bar{q} \gamma_\mu\gamma_{\rho_{\perp}}\gamma_5 \overleftrightarrow{D_\nu}  c
    \,.\label{cJ4}
\end{align}
Here $q$ denotes the light quark in charm mesons, and
\begin{align}
\overleftrightarrow D_\mu
=\overrightarrow D_\mu-\overleftarrow D_\mu,
\qquad
g_{\perp}^{\mu\nu}
=g^{\mu\nu}
-\frac{n^\mu\bar n^\nu+\bar n^\mu n^\nu}{2}.
\end{align}
We further define
$\gamma_\perp^\mu=g_\perp^{\mu\nu}\gamma_\nu$ and
$\epsilon_\perp^\mu=g_\perp^{\mu\nu}\epsilon_\nu$.
The currents $j_\mu^1$ and $j_{\mu\nu}^2$ enter the longitudinal
projections, while $j_{\mu\rho}^3$ and $j_{\mu\nu\rho}^4$ enter the
transverse projections. Each current can couple to both $D_1$ and
$D'_1$. We parametrize the corresponding matrix elements as
\begin{align}
        &\langle 0|n^{\mu} j^{1}_\mu|D_1(p,\epsilon)\rangle =\langle 0|\left(m_c+m_q\right) \bar{q} \not \! n \gamma_5 c |D_1(p,\epsilon)\rangle=f_1\, m^2_{D_1}(n\cdot\epsilon^{D_1})\,,\nonumber\\
        &\langle 0|n^{\mu} j^{1}_\mu|D^{\,\prime}_1(p,\epsilon)\rangle =\langle 0|\left(m_c+m_q\right) \bar{q} \not \! n \gamma_5 c |D^{\prime}_1(p,\epsilon)\rangle=f_2\,m^2_{D_1^{\prime}}(n\cdot \epsilon^{D_1^{\prime}})\,,\nonumber\\
        &\langle 0|n^\mu n^\nu j^{2}_{\mu\nu}|D_1(p,\epsilon)\rangle =\langle 0| \bar{q}\not \! n \gamma_5 (i n\cdot\overleftrightarrow{D})  c|D_1(p,\epsilon)\rangle=g_1\, m_{D_1}(n\cdot p)(n\cdot\epsilon^{D_1})\,,\nonumber\\
        &\langle 0|n^\mu n^\nu j^{2}_{\mu\nu}|D^{\prime}_1(p,\epsilon)\rangle= \langle 0| \bar{q}\not \! n \gamma_5 (i n\cdot\overleftrightarrow{D})  c|D^{\prime}_1(p,\epsilon)\rangle=g_2\, m_{D^{\prime}_1}(n\cdot p)(n\cdot\epsilon^{D^{\prime}_1})\,,\nonumber\\
         &\langle 0|n^\mu j^{3}_{\mu\rho}|D_1(p,\epsilon)\rangle =\langle 0|\left(m_c+m_q\right) \bar{q} \not \! n  \gamma_{\rho_\perp}\gamma_5 c |D_1(p,\epsilon)\rangle =h_1 \,m_{D_1}(n\cdot p) \,\epsilon^{D_1}_{\rho_\perp} \,,\nonumber\\
       &\langle 0|n^\mu j^{3}_{\mu\rho}|D^\prime_1(p,\epsilon)\rangle=\langle 0|\left(m_c+m_q\right) \bar{q} \not \! n  \gamma_{\rho_\perp}\gamma_5 c |D^\prime_1(p,\epsilon)\rangle  =h_2 \,m_{D^\prime_1}(n\cdot p) \,\epsilon^{D^\prime_1}_{\rho_\perp}\,,\nonumber\\
        &\langle 0|n^\mu n^\nu j^{4}_{\mu\nu\rho}|D_1(p,\epsilon)\rangle =\langle 0| \bar{q} \not \! n  \gamma_{\rho_\perp} \gamma_5 (i n\cdot\overleftrightarrow{D})  c|D_1(p,\epsilon)\rangle =r_1\,(n\cdot p)^2 \epsilon_{\rho_\perp}^{D_1}\,,\nonumber\\
        &\langle 0|n^\mu n^\nu j^{4}_{\mu\nu\rho}|D^{\prime}_1(p,\epsilon)\rangle =\langle 0| \bar{q} \not \! n  \gamma_{\rho_\perp} \gamma_5 (i n\cdot\overleftrightarrow{D})  c|D^{\prime}_1(p,\epsilon)\rangle =r_2\,(n\cdot p)^2 \epsilon_{\rho_\perp}^{D^\prime_1}\,.\label{primary_hadronic_inputs}
   \end{align}
The parameters $f_{1,2}$, $g_{1,2}$, $h_{1,2}$, and $r_{1,2}$ are
decay constants with the same mass dimension. Following the strategy adopted in \cite{Gubernari:2203.08493}, we construct separate linear
combinations for the two states and the two polarizations:
\begin{align}
  & j^{D_1}_{\|}=n^\mu j^1_{\mu}+a\,n^\mu n^\nu\,j^2_{\mu\nu}\,,\\
   &  j^{D'_1}_{\|}=n^\mu j^1_{\mu}+b\,n^\mu n^\nu\,j^2_{\mu\nu}\,,\\
   & j^{D_1}_{\perp,\rho}=n^\mu\, j^3_{\mu\rho}+c \,n^\mu n^\nu\,j^4_{\mu\nu\rho}\,,\\
 & j^{D'_1}_{\perp,\rho}=n^\mu\, j^3_{\mu\rho}+d \,n^\mu n^\nu\,j^4_{\mu\nu\rho}\,.
   \end{align}
The coefficients $a$, $b$, $c$, and $d$ are fixed by requiring the
current associated with either state to have a vanishing matrix
element with the other state:
\begin{align}
 \langle 0|j^{D_1}_{\|}|D'_1(p,\epsilon)\rangle =0 \,,\nonumber\\
 \langle 0|j^{D'_1}_{\|}|D_1(p,\epsilon)\rangle =0 \,,\nonumber\\
   \langle 0|j^{D_1}_{\perp,\rho}|D'_1(p,\epsilon)\rangle =0 \,,\nonumber\\
    \langle 0|j^{D'_1}_{\perp,\rho}|D_1(p,\epsilon)\rangle =0 \,.
   \end{align}
Using the matrix elements defined above, we obtain
\begin{align}
     j^{D_1}_{\|}={}&n^\mu j^1_{\mu}-{f_2 \,m_{D'_1} \over g_2(n\cdot p)}\,n^\mu n^\nu\,j^2_{\mu\nu}\,,\\
     j^{D'_1}_{\|}={}&n^\mu j^1_{\mu}-{f_1 \,m_{D_1}\over g_1(n\cdot p)}\,n^\mu n^\nu\,j^2_{\mu\nu}\,,\\
     j^{D_1}_{\perp,\rho}={}&n^\mu\, j^3_{\mu\rho}- {h_2 \,m_{D'_1}\over r_2(n\cdot p)}\,n^\mu n^\nu\,j^4_{\mu\nu\rho}\,,\\
      j^{D'_1}_{\perp,\rho}={}&n^\mu\, j^3_{\mu\rho}- {h_1 \,m_{D_1}\over r_1(n\cdot p)}\,n^\mu n^\nu\,j^4_{\mu\nu\rho}\,.
\end{align}
We then define the decay constants associated with the resulting
interpolating currents by
\begin{align}
\langle 0|j^{D_1}_{\parallel}|D_1(p,\epsilon)\rangle
&=m_{D_1}^2f^{D_1}_{\parallel}
(n\cdot\epsilon^{D_1}),
\nonumber\\
\langle 0|j^{D'_1}_{\parallel}|D'_1(p,\epsilon)\rangle
&=m_{D'_1}^2f^{D'_1}_{\parallel}
(n\cdot\epsilon^{D'_1}),
\nonumber\\
\langle 0|j^{D_1}_{\perp,\rho}|D_1(p,\epsilon)\rangle
&=m_{D_1}(n\cdot p)f^{D_1}_{\perp}
\epsilon_{\rho\perp}^{D_1},
\nonumber\\
\langle 0|j^{D'_1}_{\perp,\rho}|D'_1(p,\epsilon)\rangle
&=m_{D'_1}(n\cdot p)f^{D'_1}_{\perp}
\epsilon_{\rho\perp}^{D'_1},
\end{align}
and thereby obtain the algebraic relations
\begin{align}
&f^{D_1}_{\|}=f_1-{f_2 g_1\over g_2}{m_{D_1'}\over m_{D_1}}\,,\nonumber\\
&f^{D^{\prime}_1}_{\|}=f_2-{f_1 g_2\over g_1}{m_{D_1}\over m_{D'_1}}\,,\nonumber\\
&f^{D_1}_{\perp}=h_1-{h_2 r_1\over r_2}{m_{D_1'}\over m_{D_1}}\,,\nonumber\\
&f^{D^{\prime}_1}_{\perp}=h_2-{h_1 r_2\over r_1}{m_{D_1}\over m_{D'_1}}\,.
   \end{align}

The construction of the linear combinations follows
\cite{Gubernari:2203.08493}, but we use a different second
longitudinal current. The current employed there contains the
derivative structure
$\gamma_5\overleftrightarrow D_\nu$, whereas our current contains the
tensor structure
$\gamma_\mu\overleftrightarrow D_\nu\gamma_5$.
The $n^\mu n^\nu$ projection of our current gives the leading-power
collinear structure
$\bar\xi\slashed n\gamma_5
(i n\cdot\overleftrightarrow D)\xi$.
The derivative current used in \cite{Gubernari:2203.08493}
contributes only at subleading power in the present SCET expansion and
is therefore not suitable for the leading-power sum rules considered
here. We then introduce the vacuum-to-$B$-meson correlation functions
\begin{align}
    \Pi^{R}_{ \|}(p, q)  ={}& i\int d^4 x \, e^{i p \cdot x} \,
 \langle 0 | {\rm T}  \left \{ j_{\|}^{R}(x),  \,\, \left (\bar \xi \, W_c \right)(0) \, h_v(0) \,    \right \}   | \bar B_v \rangle ,\label{correlationfunction1}  
 \\
 \Pi^{R}_{\rho\sigma, \perp}(p, q) ={}& i\int d^4 x \, e^{i p \cdot x} \,
 \langle 0 | {\rm T}  \left \{ j_{\perp,\rho}^{R}(x),  \,\, \left (\bar \xi \, W_c \right)(0) \gamma_{\sigma \perp} \, h_v(0) \,    \right \}   | \bar B_v \rangle ,\label{correlationfunction2} 
 \\
\tilde{\Pi}^{R}_{\|}(p,q,\tau)={}&i\, {n \cdot p \over 2\pi}\, \int d^4 x \, e^{i p \cdot x} \, \int {d r }\, e^{- i \, n \cdot p \, \tau \, r} 
	\nonumber	\\& \langle 0 | {\rm T}  \left \{ j_{\|}^{R}(x),  \,\left (\bar \xi \, W_c \right)(0) \,  \left ( W_c^\dagger i \not\!\! D_{c\perp} W_c \right ) (r\, n) \, h_v(0) \,    \right \}   | \bar B_v \rangle , \label{correlationfunction3}\\
\tilde{\Pi}^{R}_{\rho\sigma,\perp}(p,q,\tau)={}& i\,{n \cdot p \over 2\pi}\, \int d^4 x \, e^{i p \cdot x} \, \int {d r }\, e^{- i \, n \cdot p \, \tau \, r} 
		\, \nonumber	\\&\langle 0 | {\rm T}  \left \{ j_{\perp,\rho}^{R}(x),  \,\left (\bar \xi \, W_c \right)(0) \,  \gamma_{\sigma\perp}\left ( W_c^\dagger i \not\!\! D_{c\perp} W_c \right ) (r\, n) \, h_v(0) \,    \right \}   | \bar B_v \rangle .\label{correlationfunction4}
\end{align}
These correlation functions inherit the same linear combinations as
the corresponding interpolating currents. The A0-type form factors
$\xi^R_{\parallel}$ and $\xi^R_{\perp}$ are extracted from
\eqref{correlationfunction1} and \eqref{correlationfunction2},
respectively, while the B1-type form factors $\Xi^R_{\parallel}$ and $\Xi^R_{\perp}$ are
obtained from \eqref{correlationfunction3} and
\eqref{correlationfunction4}. The additional correlation function
required for $\xi^R_{\parallel,m_c}$ will be introduced separately
below.

Now we are in a position to match these ${\rm SCET_I}$ correlation functions onto
${\rm SCET_{II}}$ at one-loop accuracy to determine the jet functions
and their spectral representations. The sum rules are then obtained
by matching the partonic and hadronic dispersion relations, followed
by continuum subtraction and the Borel transformation. We begin with
the longitudinal A0-type form factor $\xi^R_{\parallel}$.

\subsection{The $B$-meson LCSRs for $\xi^{R}_{\|}$}
Now we turn to determining the effective form factors $\xi^{D_1}_{\|}$ and $\xi^{D'_1}_{\|}$ which enter the QCD vector form factors of $D_1$ and $D'_1$ respectively. According to their definition in the SCET matrix elements \eqref{SCETmatrixelements}, they can be extracted from the correlation functions \eqref{correlationfunction1} which admit the decomposition
\begin{align}
     \Pi^{D_1}_{ \|}(p, q)  =&n^\mu \Pi_{\mu,\|}(p,q)-{f_2 \,m_{D'_1} \over g_2(n\cdot p)}\,n^\mu n^\nu\,\Pi^*_{\mu\nu,\|}(p,q)\,,\\
      \Pi^{D'_1}_{\|}(p, q)  =&n^\mu \Pi_{\mu,\|}(p,q)-{f_1 \,m_{D_1} \over g_1(n\cdot p)}\,n^\mu n^\nu\,\Pi^*_{\mu\nu,\|}(p,q)\,,
\end{align}
where the two building blocks on the right-hand side are constructed
with $j_\mu^1$ and $j_{\mu\nu}^2$, respectively:
\begin{align}
     \, \Pi^{}_{\mu, \|}(p, q)  =i&\int d^4 x \, e^{i p \cdot x} \,
 \langle 0 | {\rm T}  \left \{ j_{\mu}^{1}(x),  \,\, \left (\bar \xi \, W_c \right)(0) \, h_v(0) \,    \right \}   | \bar B_v \rangle \,,\label{correlationofJ1}\\
     \, \Pi^{*}_{\mu\nu, \|}(p, q)  =i&\int d^4 x \, e^{i p \cdot x} \,
 \langle 0 | {\rm T}  \left \{ j_{\mu\nu}^{2}(x),  \,\, \left (\bar \xi \, W_c \right)(0) \, h_v(0) \,    \right \}   | \bar B_v \rangle\,.\label{correlationofJ2}
\end{align}

We first consider $\Pi_{\mu,\parallel}$ constructed with the
non-derivative axial-vector current $j_\mu^1$. Its perturbative
calculation parallels that of the vector-current correlation function
in \cite{cui:2301.12391}. The
${\rm SCET_{II}}$ factorization formula for \eqref{correlationofJ1} is
 \begin{align}
&\Pi_{\mu, \|}^{i}(p, q) =i\, {\tilde{f}_B(\mu) \, m_B \over2}\,
\sum_{m = \pm } \, \int_0^{+\infty} \, d \omega \, J_{\|, m}^{i}
\left ({\mu^2 \over n \cdot p \, \omega}, {\omega \over \bar n \cdot p} \right )  \,
\phi_B^{m}(\omega, \mu) \,\, \bar n_{\mu},  \,\, \, (i=A, B, C) \,
\label{SCET factorization formula L}
\end{align}
where $J^i_{\parallel,m}$ and $\phi_B^m$ describe the hard-collinear and
soft contributions, respectively. The corresponding spectral
representation is
\begin{align}
    &\Pi_{\mu, \|}^{}(p, q) =i\, {\tilde{f}_B(\mu) \, m_B \,m_c \over 2}\,
 \int_0^{+\infty} \, {d\omega^{\prime}\over \omega^{\prime}- \bar n \cdot p-i0}\,\left[\phi^{-}_{\parallel,\rm eff}\left(\omega^{\prime},\mu\right) +\phi^{+}_{\parallel,\rm eff}\left(\omega^{\prime},\mu\right)  \right]\,\, \bar n_{\mu}. \,\label{specj1}
\end{align}
Here $\phi^\pm_{\parallel,\rm eff}$ combine the jet functions with the
standard $B$-meson light-cone distribution amplitudes (LCDAs). Their explicit spectral representations,
together with those required for the other correlation functions, are
collected in the appendix \ref{secA1}.

Now we concentrate on the correlation function \eqref{correlationofJ2}. The SCET representation of the QCD current \eqref{cJ2} is as follows
\begin{align}
    j^2_{\mu\nu}=j_{\xi \xi, \| \mu\nu}^{2(0)}+j_{\xi q_s, \| \mu\nu}^{2(2)}+...,
\end{align}
with the explicit expressions of the power-expanded ``effective'' currents
\begin{align}
    &j_{\xi \xi, \|\mu\nu}^{2(0)}=\bar{\xi} {\not \! n\over 4} \gamma_5 (i n\cdot\overleftrightarrow{D}) \xi {\bar{n}_\mu}{\bar{n}_\nu}, \label{j20}\\
    &j_{\xi q_s, \| \mu\nu}^{2(2)}=\left[\bar{q}_s Y_s{\not \! n\over4}\gamma_5 W_c^{\dagger}(i n\cdot D)\xi-\bar{\xi}(i n\cdot \overleftarrow{D}) W_c {\not \! n\over4}  \gamma_5 Y_s^{\dagger} q_s \right]{\bar{n}_\mu}{\bar{n}_\nu}.\label{j22}
\end{align}
Here we neglect the terms which are transversely polarized and power suppressed. The soft gauge link
\begin{align}
    Y_s(x)={\rm P \,\,exp} \left[ i g_s\int^0_{-\infty }ds \,\,\bar n \cdot A_s(x+s\bar n) \right],\label{softgaugelink}
\end{align}
together with the previously mentioned collinear Wilson line \eqref{collineargaugelink}, is introduced for SCET currents to maintain the general gauge. Following the prescription in \cite{Gao:2019lta}, the $\rm SCET_I$ representation of correlation function \eqref{correlationofJ2} at leading power can therefore be constructed as
\begin{align}
\Pi^{*}_{\mu\nu, \|}(p, q)  ={}&i \int d^4 x \, e^{i p \cdot x} \,
 \langle 0 | {\rm T}  \left \{ j_{\xi q_s, \parallel \, \mu\nu}^{2(2)}(x),  \,\, \left (\bar \xi \, W_c \right)(0) \,
 \, h_v(0) \,    \right \}   | \bar B_v \rangle \nonumber \\
& + i\int d^4 x \, e^{i p \cdot x} \, \int d^4 y \,
 \langle 0 | {\rm T}  \left \{ j_{\xi \xi, \|\mu\nu}^{2(0)}(x),  \,\, i \, {\cal L}_{\xi q_s}^{(2)}(y),  \,\,
\left (\bar \xi \, W_c \right)(0) \,  \, h_v(0) \,    \right \}   | \bar B_v \rangle  \nonumber \\
& +i \int d^4 x \, e^{i p \cdot x} \, \int d^4 y \, \int d^4 z \, \nonumber \\
& \langle 0 | {\rm T}  \left \{ j_{\xi \xi, \|\mu\nu}^{2(0)}(x),  \,\, i \, {\cal L}_{\xi q_s}^{(1)}(y),  \,\,
i \, {\cal L}_{\xi m_c}^{(0)}(z),  \,\,
\left (\bar \xi \, W_c \right)(0) \,  \, h_v(0) \,    \right \}   | \bar B_v \rangle  \nonumber \\
& \equiv \Pi_{\mu\nu, \|}^{*A}(p, q)  + \Pi_{\mu\nu, \|}^{*B}(p, q) + \Pi_{\mu\nu, \|}^{*C}(p, q) \,,\label{SCETIcorrelatorPimunu}
\end{align}
where the multipole-expanded Lagrangians up to $\mathcal{O}\left(\lambda^2\right)$ accuracy are given by \cite{Beneke:0211358,Boos:0504005,Leibovich:0303099}
\begin{align}
{\cal L}_{\xi}^{(0)} =& \bar \xi \, \left ( i \, \bar n \cdot D
+ (i \not \! \! D_{\perp c} -m_c)\,\, {1 \over i \, n \cdot D_c} \, (i \not \! \! D_{\perp c} +m_c) \right ) \,
{\not \! n \over 2 }  \,\, \xi  \,,  \nonumber \\
{\cal L}_{\xi m_c}^{(0)} =&
m_c\,\bar\xi\left[
i\not\!\!D_{\perp c},{1\over i n\cdot D_c}
\right]{\not\!n\over2}\xi
-m_c^2\,\bar\xi\,{1\over i n\cdot D_c}
{\not\!n\over2}\xi\,,
\nonumber\\
{\cal L}_{\xi m_c}^{(1)} =& m_c \,\, \bar \xi \, \left [ g_s \not \! \! A_{\perp s},  \, {1 \over i \, n \cdot D_c}   \right ] \,{\not \! n \over 2 }  \,\, \xi  \,, \nonumber \\
{\cal L}_{\xi m_c}^{(2)} =& m_c^2 \,\, \bar \xi \,  {1 \over i \, n \cdot D_c}\, g_s\, n\cdot A_s\, {1 \over i \, n \cdot D_c} \,
{\not \! n \over 2 }  \,\, \xi  \nonumber   \,, \\
{\cal L}_{\xi q_s}^{(1)} =& \bar q_s \,  W_c^{\dagger} \,\,  i \not \! \! D_{\perp c}\,\, \xi
- \bar \xi \,\,   i \not \! \! \overleftarrow{D}_{\perp c} \,\,  W_c \, q_s,   \nonumber   \\
{\cal L}_{\xi q_s}^{(2)} =& \bar q_s \,  W_c^{\dagger} \,\,   \left ( i \, \bar n \cdot D
+ i \not \! \! D_{\perp c} \,\, {1 \over i \, n \cdot D_c} \, i \not \! \! D_{\perp c} \right ) \,\,
{\not \! n \over 2 }  \,\, \xi   \nonumber \\
& -  \, \bar \xi \, {\not \! n \over 2 } \,\,  \left ( i \, \bar n \cdot \overleftarrow{D}
+ \, i \not \! \! \overleftarrow{D}_{\perp c} \,\, {1 \over i \, n \cdot \overleftarrow{D}_c} \,
i \not \! \! \overleftarrow{D}_{\perp c} \right ) \,\,
W_c \,\, q_s   \nonumber \\
& + \,  \bar q_s  \, \overleftarrow{D}_s^{\mu} \, x_{\perp \mu} \,  W_c^{\dagger} \,\,  i \not \! \! D_{\perp c}\,\, \xi
- \bar \xi  \,  i \not \! \! \! \overleftarrow{ D}_{\perp c} \,\,  W_c \, x_{\perp \mu}  \, D_s^{\mu} \,   q_s \,,
\end{align}
where \({\cal L}_{\xi m_c}^{(0)}\) denotes the part proportional to $m_c$ of \({\cal L}_{\xi}^{(0)}\).

Now we are in a position to perform the perturbative matching of the $\rm SCET_{I}$ correlation function \eqref{SCETIcorrelatorPimunu} onto $\rm SCET_{II}$ at tree level. Evidently, the entire tree-level contribution to $\Pi_{\mu\nu, \|}^{*}$ arises solely from $ \Pi_{\mu\nu, \|}^{*A}$, whereas $ \Pi_{\mu\nu, \|}^{*B}$ and $ \Pi_{\mu\nu, \|}^{*C}$ first contribute at $\mathcal{O}(\alpha_s)$. Therefore we consider the following partonic matrix element 

\begin{align}
    F^{*}_{ \|} & = \int d^4 x \, e^{i p \cdot x} \,
 \left\langle 0 | {\rm T}  \left \{ \bar{q}_s(x)Y_s{\not \! n\over4}\gamma_5 W_c^{\dagger}(i n\cdot D)\xi(x),  \,\, \left (\bar \xi \, W_c \right)(0) \,
 \, h_v(0) \,    \right \}   | \bar q_s(k) h_v \right\rangle .
\end{align}
Evaluating the tree-level contribution results in
\begin{align}
     F^{*}_{ \|,\rm LO}=   {i\,n\cdot p\over \bar n\cdot p -\bar n\cdot k-{m^2_c\over n\cdot p} +i0} \bar q_s (k){\not\! n\over 4}\gamma_5\,h_v,
\end{align}
which, in principle, can be written schematically as a convolution integral denoted by an asterisk over the soft momentum
\begin{align}
     F^{*}_{ \|,\rm LO}={i\,n\cdot p\over \bar n\cdot p -\omega'-\omega_c+i0} * \langle O\left( \omega,\omega' \right)\rangle,
\end{align}
where the soft momentum component $\bar n\cdot k'$ in the hard-collinear
propagator is identified with the convolution variable $\omega'$, while
$\bar n\cdot k=\omega$ denotes the external soft momentum component. We note that here we use the external partonic state $| \bar q_s(k) h_v \rangle $ instead of the external hadronic state $|\bar B_v \rangle$  in order to perturbatively extract the jet function which is free of the external soft state.
The soft function is then obtained by using the external hadronic state $|\bar B_v \rangle$:
\begin{align}
    \langle 0| \bar q_s (\bar nz){\not\!  n\over 4}\gamma_5\,h_v(0)|\bar B_v\rangle={i\over4} \tilde f_B(\mu) m_B{\tilde \phi}_B^-(\bar n z,\mu).
\end{align}
To express this, we employ the standard definition of the two-particle $B$-meson LCDAs \cite{Grozin:9607366,Beneke:0008255}
\begin{align}
  &\langle0|\bar q\left(\bar nz\right) \Gamma \left[ \bar nz,0\right] h_v\left( 0\right)|\bar B_v\rangle
  \nonumber\\
  &=-{i \over 2} \tilde f_B(\mu) m_B {\rm Tr}\left\{ \gamma_5\Gamma \left({1+\not \!v\over2}\right) \left[{\tilde \phi}_B^+\left(\bar n z\right)-{\not \!\bar n\over 2}\left({\tilde \phi}_B^+\left(\bar n z\right)-{\tilde \phi}_B^- \left(\bar n z\right)  \right)  \right]\right\},
\end{align}
where the Fourier expansions of the ${\tilde \phi}_B^{\pm}$ in coordinate space read
\begin{align}
    {\tilde \phi}_B^{\pm}\left(\bar nz\right)=\int^\infty_0 e^{-i\omega \bar nz}\,\phi_B^{\pm}\left(\omega\right).
\end{align}
The resulting $\rm SCET_{II}$ factorization formula for \eqref{SCETIcorrelatorPimunu} is then represented as
\begin{align}
    \Pi^{*}_{\mu\nu,\|}(p,q)=i\,{\tilde f_B(\mu)m_B \over 2} \sum_{m=\pm} \int^\infty_0 d\omega \, J_{\|,m}^{*(0)}\left({\mu^2\over n\cdot p\,\omega},{\omega\over \bar n\cdot p}\right)\,\phi_B^m\left(\omega,\mu\right)\bar n_\mu \bar n_\nu\,+\mathcal{O}(\alpha_s),
\end{align}
with corresponding jet functions 
\begin{align}
    &J_{\|,+}^{*(0)}=0,\nonumber\\ &J_{\|,-}^{*(0)}=-{n\cdot p\over 2(\bar n\cdot p - \omega-\omega_c+i0)}.
\end{align}

\subsubsection{The NLO corrections to $ \Pi^{*A}_{\mu\nu,\|}(p,q)$}
\begin{figure}[htbp]
    \centering
    \includegraphics[width=0.3\linewidth]{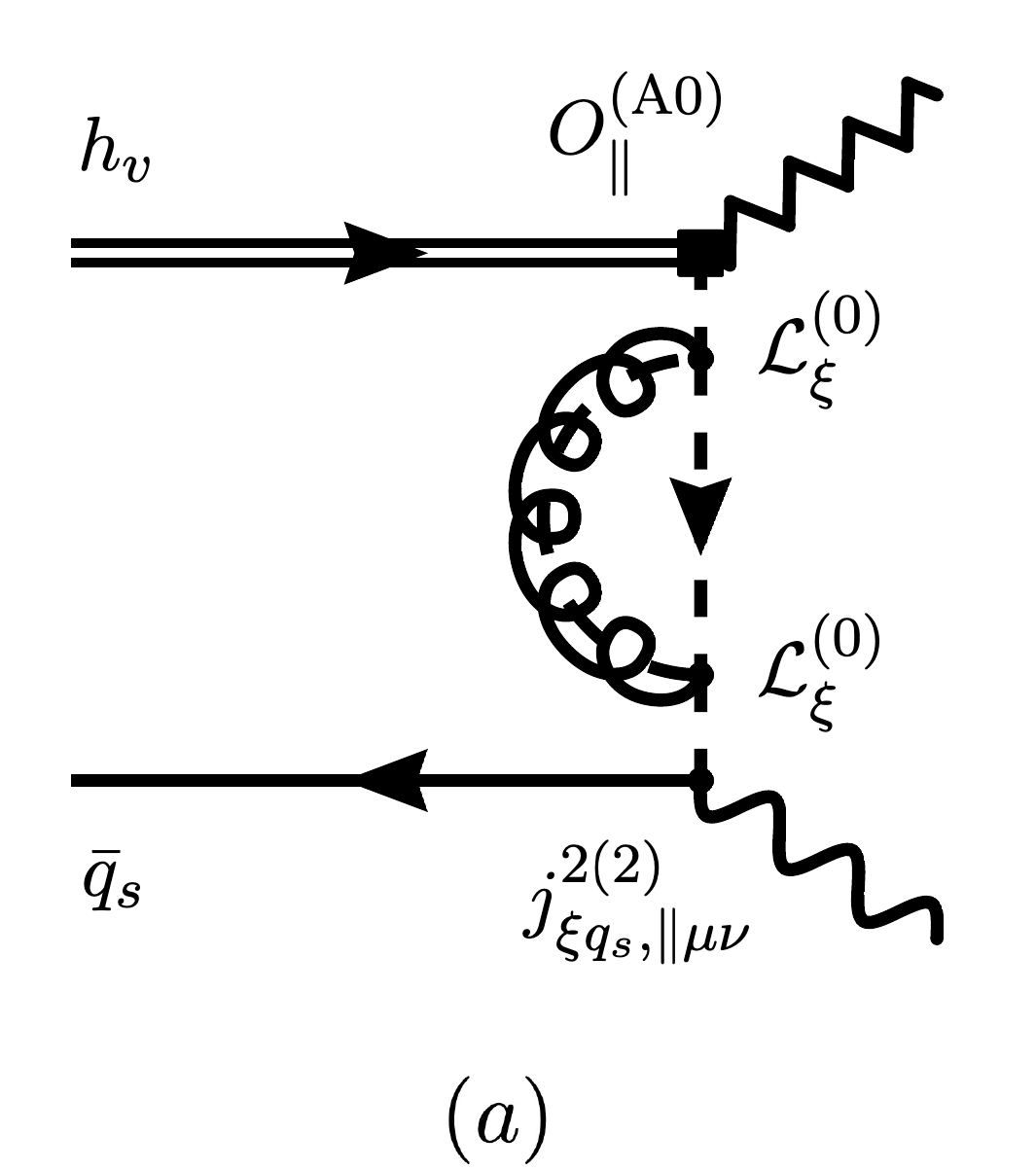}
    \hfill
    \includegraphics[width=0.3\linewidth]{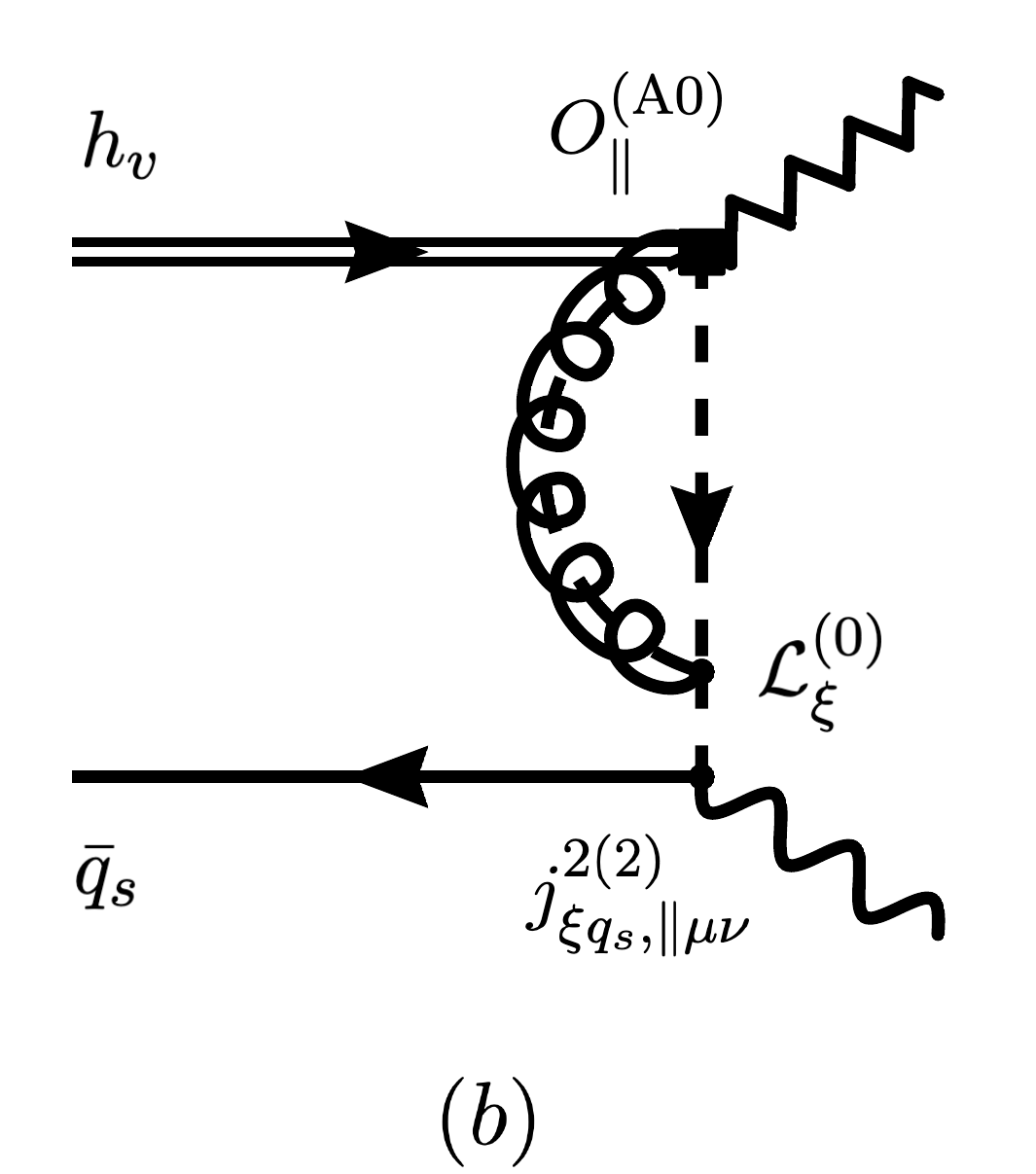}
    \hfill
    \includegraphics[width=0.3\linewidth]{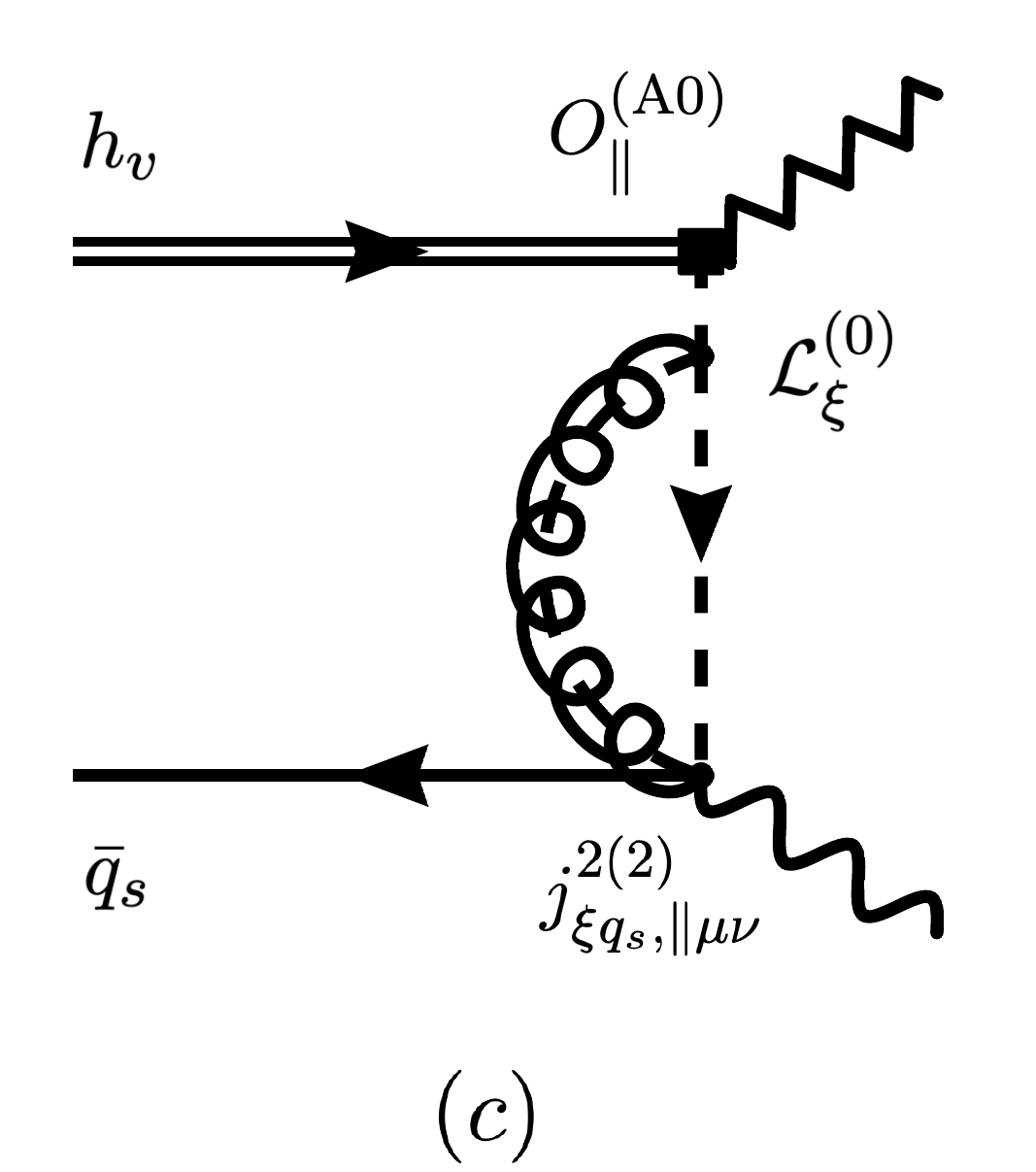}
    \caption{One-loop diagrams contributing to
    $\Pi_{\mu\nu,\|}^{*A}(p,q)$.}
    \label{fig:my_feynman_simple}
\end{figure}

The one-loop diagrams contributing to $\Pi_{\mu\nu,\|}^{*A}$ are
shown in Fig.~\ref{fig:my_feynman_simple}.

Using the subleading-power SCET Feynman rules collected in
\cite{Beneke:1808.04742}, the amplitude of diagram (a) is
\begin{align}
    F_{ \|,\rm NLO}^{*A(a)}(p, q) = &{g_s^2 C_F \mu^{2\epsilon} \over \bar n \cdot p - \omega - \omega_c} \, \bar q_s {\not\! n \over 4} \gamma_5 \, h_v \nonumber\\&\times\int {d^D l \over (2 \pi)^D} {m_c^2 \left[(D+2) n \cdot p + (2 - D) n \cdot l\right] + (2 - D)\bar n \cdot (p -k+ l) (n \cdot p)^2 \over [(p - k + l)^2 - m_c^2 + i0] [l^2 + i 0] \, [(p - k)^2 - m_c^2] } \nonumber\\
    =&{\tilde{f}_B(\mu) \, m_B \over 2}\left({\alpha_s C_F\over 4\pi}\right)\,
 \int_0^{+\infty} \, {d\omega} \, J^{*A(a)}_{\|,-}\left ({\mu^2 \over n \cdot p \, \omega}, {\omega \over \bar n \cdot p} \right )  \phi^{-}_{B}\left( \omega,\mu \right)  \label{Faa},
\end{align}
where
\begin{align}
 \omega_c\equiv {m_c^2\over n\cdot p}\,,\qquad
 \eta\equiv-{\omega\over\bar n\cdot p}\,,\qquad
 \eta_c\equiv-{m_c^2\over p^2}
 =-{\omega_c\over\bar n\cdot p}\,.
\end{align}
In the second line the
partonic matrix element is matched onto the $B$-meson LCDA. The
hard-collinear contribution from diagram (b) is
\begin{align}
    F_{ \|,\rm NLO}^{*A(b)}(p, q) =&  { g_s^2 C_F \mu^{2\epsilon}  \over \bar n \cdot p - \omega - \omega_c}\bar q_s \, {\not \! n \over 4} \gamma_5 \, h_v \int {d^D l \over (2 \pi)^D} { \, n \cdot (p + l) n\cdot p \over [(p - k + l)^2 - m_c^2][l^2 + i0]} \,  \left [ {2 \over n \cdot l}\right ] \nonumber\\
		  =&{\tilde{f}_B(\mu) \, m_B \over 2}\left({\alpha_s C_F\over 4\pi}\right)\,
 \int_0^{+\infty} \, {d\omega} \, J^{*(A)(b)}_{\|,-}\left ({\mu^2 \over n \cdot p \, \omega}, {\omega \over \bar n \cdot p} \right )  \phi^{-}_{B}\left( \omega,\mu \right)  \label{Fab}.
\end{align}
Diagram (c) contains the contribution of diagram (b) and an extra
term generated by the gluon field in the covariant derivative. The
additional term is
\begin{align}
    F_{ \|,\rm NLO}^{*A(\rm ex)}(p, q) =&  -{ g_s^2 C_F \mu^{2\epsilon}  \over \bar n \cdot p - \omega - \omega_c}\bar q_s \, {\not \! n \over 4} \gamma_5 \, h_v \int {d^D l \over (2 \pi)^D} { 2\ n \cdot (p + l) n\cdot p \over [(p - k + l)^2 - m_c^2][l^2 + i0]} \,  \nonumber \\
		  =&{\tilde{f}_B(\mu) \, m_B \over 2}\left({\alpha_s C_F\over 4\pi}\right)\,
 \int_0^{+\infty} \, {d\omega} \, J^{*(A)(\rm ex)}_{\|,-}\left ({\mu^2 \over n \cdot p \, \omega}, {\omega \over \bar n \cdot p} \right )  \phi^{-}_{B}\left( \omega,\mu \right)  .
\label{Fex}\end{align}
Combining the three diagrams gives
\begin{align}
    J^{*A}_{\|,+}=&0\nonumber\\
     J^{*A}_{\|,-}=& J^{*A(a)}_{\|,-}+2J^{*A(b)}_{\|,-}+J^{*A(\rm ex)}_{\|,-}\nonumber \\
 =&- \frac{1}{\epsilon} \left[ \frac{1}{2} + \frac{3\eta_c}{1+\eta+\eta_c} + \ln(1+\eta+\eta_c)-\ln\left({\mu^2\over -p^2}\right) \right] + \frac{1}{\epsilon^2}-{1\over2}+{\pi^2\over12}+ \frac{\eta_c}{2(1+\eta)} \nonumber\\&- \frac{4\eta_c}{1+\eta+\eta_c}   
+ \left[ \frac{1}{2} + \frac{2\eta_c}{1+\eta} - \frac{\eta_c^2}{2(1+\eta)^2} \right] \ln\left( \frac{1+\eta+\eta_c}{\eta_c} \right) \nonumber \\&+ \frac{1}{2} \ln^2(1+\eta+\eta_c) 
  +\left[ -\frac{1}{2} - \frac{3\eta_c}{1+\eta+\eta_c} - \ln(1+\eta+\eta_c) \right]\ln\left({\mu^2\over -p^2}\right)\nonumber\\&+{1\over 2} \ln^2\left({\mu^2\over -p^2}\right)+ \left( \frac{1}{2} + \frac{3\eta_c}{1+\eta+\eta_c} \right) \ln(\eta_c)-  \operatorname{Li}_2 \left( \frac{1+\eta}{1+\eta+\eta_c} \right)  
\end{align}

\subsubsection{The one-loop calculation of $ \Pi^{*B}_{\mu\nu,\|}(p,q)$}
\begin{figure}[htbp]
    \centering
    \includegraphics[width=0.4\linewidth]{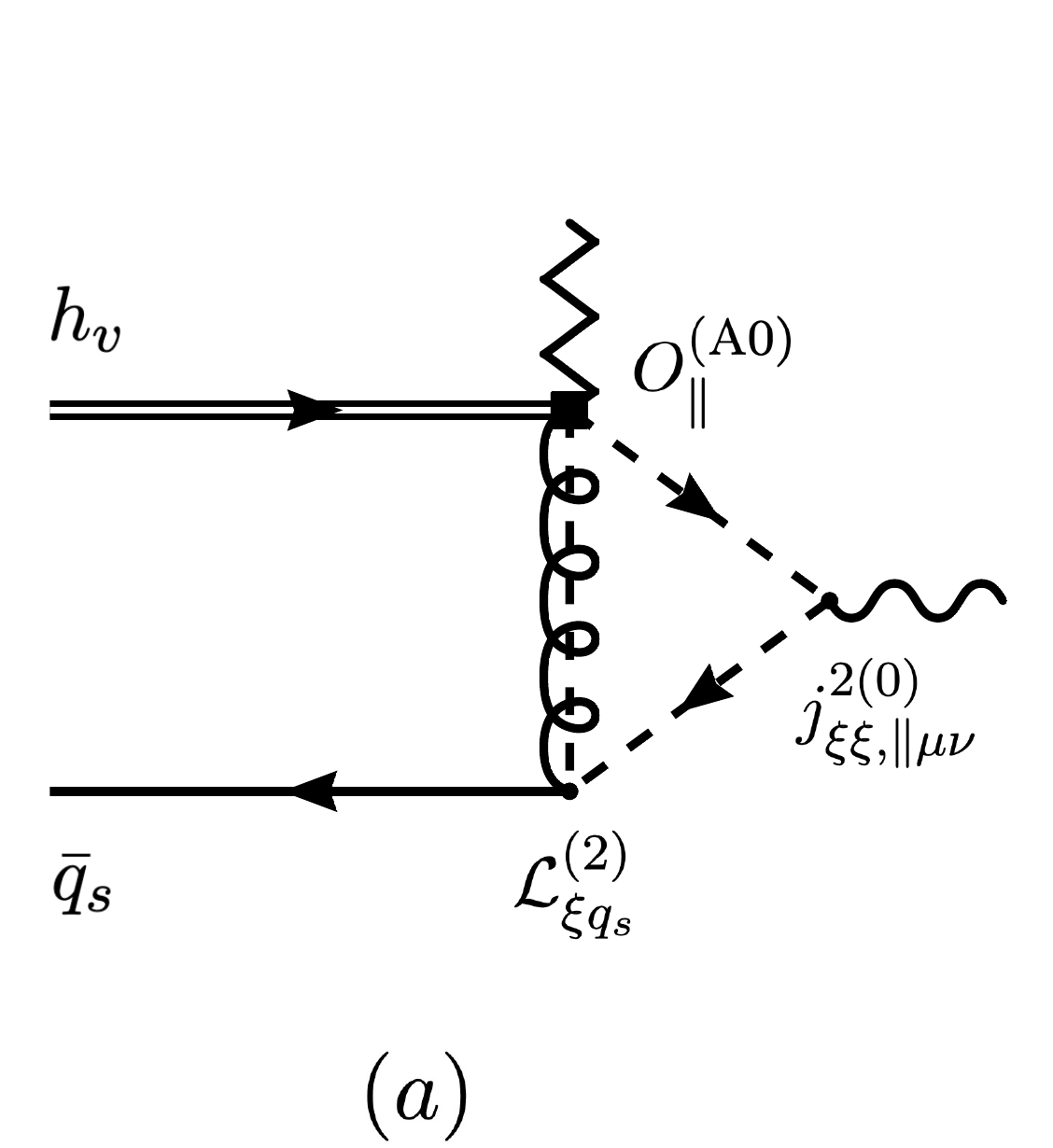}
    \hspace{1.3cm}
    \includegraphics[width=0.4\linewidth]{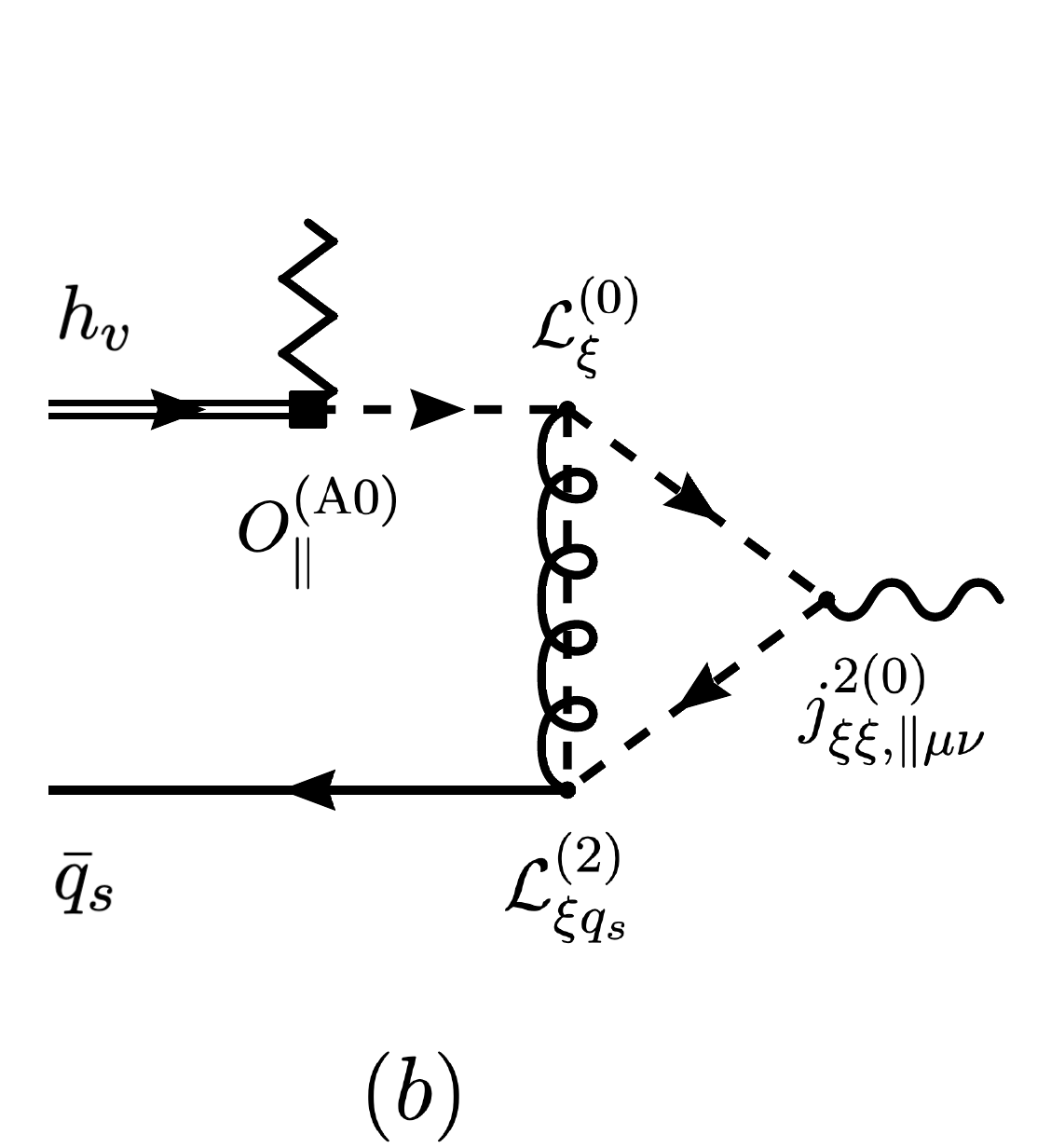}
    \caption{One-loop diagrams contributing to
    $\Pi_{\mu\nu,\|}^{*B}(p,q)$.}
    \label{fig:pi-star-b}
\end{figure}
The jet function $J^{*B}_{\|,\pm}$ is extracted from the partonic
matrix element
\begin{align}
    F^{*B}_{ \|} & = \int d^4 x \,\int d^4 y\, e^{i p \cdot x} \,
 \langle 0 | {\rm T}  \left \{ \bar{\xi}(x){\not \! n\over4}\gamma_5 (i n\cdot D)\xi(x), i{\cal L}_{\xi q_s}^{(2)}(y), \, \left (\bar \xi \, W_c \right)(0) \,
 \, h_v(0) \,    \right \}   | \bar q_s(k) h_v \rangle .
\end{align}

The first nonvanishing contribution to
$\Pi^{*B}_{\mu\nu,\|}$ occurs at one loop. Diagram (a) in
Fig.~\ref{fig:pi-star-b} gives
\begin{align}
    F^{*B(a)}_{ \|} =& g_s^2\,C_F \,\mu^{2\epsilon}\bar q_s \, {\not\! n \over 4} \gamma_5 \, h_v \, \int {d^D l \over (2 \pi)^D} \, {2 n \cdot (p + l) \,n\cdot (p+2l)\over (l - k)^2 \, [(p - k + l)^2 - m_c^2] (l^2 + i0)} \, \nonumber\\
    =&{\tilde{f}_B(\mu) \, m_B \over 2}\left({\alpha_s C_F\over 4\pi}\right)\,
 \int_0^{+\infty} \, {d\omega} \, J^{*B(a)}_{\|,-}\left ({\mu^2 \over n \cdot p \, \omega}, {\omega \over \bar n \cdot p} \right )  \phi^{-}_{B}\left( \omega,\mu \right)  \label{Fba}.
\end{align}
The amplitude of diagram (b) is
\begin{align}
F^{*B(b)}_{\|} ={}&
g_s^2 C_F \mu^{2\epsilon}
\int {d^D l\over(2\pi)^D}
\Bigg\{
{n\cdot(p+2l)\over {\cal D}(p,l,k)}
\left[
2m_c^2 n\cdot l
+(D-2)l_\perp^2 n\cdot p
\right]
\bar q_s{\slashed n\over4}\gamma_5h_v
\nonumber\\
&
-k^\mu{\partial\over\partial k^\mu}\,
\bar q_s\,{n\cdot(p+2l)\over{\cal D}(p,l,k)}
\nonumber\\
&\times
\left\{
(2-D)n\cdot l\,n\cdot(p+l)\slashed{k}_\perp
+\left[(2-D)n\cdot l-2n\cdot p\right]\slashed{l}_\perp
\right\}
{\slashed{\bar n}\over4}\gamma_5h_v
\Bigg\}
\nonumber\\
={}&
{\tilde f_B(\mu)m_B\over2}
\left({\alpha_sC_F\over4\pi}\right)
\int_0^\infty d\omega\,
J^{*B(b)}_{\|,-}
\left(
{\mu^2\over n\cdot p\,\omega},
{\omega\over\bar n\cdot p}
\right)
\phi_B^-(\omega,\mu)\,,
\label{Fbb}
\end{align}
where
\begin{align}
{\cal D}(p,l,k)
=(l^2+i0)
\left[(p+l)^2-m_c^2+i0\right]
\left[(l+k)^2+i0\right]
\left[(p-k)^2-m_c^2\right].
\end{align}
The sum of the two diagrams yields
\begin{align}
     J^{*B}_{\|,+}=&0\nonumber\\
     J^{*B}_{\|,-}=& J^{*B(a)}_{\|,-}+J^{*B(b)}_{\|,-}\nonumber\\
                  =& \frac{1}{\epsilon} \left( -\frac{5}{6} + \ln(1+\eta_c) - \ln(1+\eta+\eta_c) \right) - \frac{\eta_c}{1+\eta} - \frac{26}{9} + \frac{7\eta_c}{6} + \frac{2\eta_c^2}{3} \nonumber \\
& + \frac{1}{6} \left( 6\eta_c - 9\eta_c^2 - 4\eta_c^3 \right) \ln\left( \frac{1+\eta_c}{\eta_c} \right) + \left( \frac{\eta_c}{1+\eta} - \frac{\eta_c^2}{(1+\eta)^2} \right) \ln\left( \frac{\eta_c}{1+\eta+\eta_c} \right) \nonumber \\
& + \frac{1-\eta+\eta_c}{\eta} \ln\left( \frac{1+\eta+\eta_c}{1+\eta_c} \right) + \frac{5}{6} \ln(1+\eta_c) - \frac{1}{2} \ln^2(1+\eta_c)  \nonumber \\
&  + \frac{1}{2} \ln^2(1+\eta+\eta_c)+ \left( -\frac{5}{6} + \ln(1+\eta_c) - \ln(1+\eta+\eta_c) \right) \ln\left( \frac{\mu^2}{-p^2} \right) 
\nonumber\\&+ \operatorname{Li}_2\left( \frac{1}{1+\eta_c} \right) - \operatorname{Li}_2\left( \frac{1+\eta}{1+\eta+\eta_c} \right)   
\end{align}

\subsubsection{The one-loop calculation of $ \Pi^{*C}_{\mu\nu,\|}(p,q)$}
\begin{figure}[H]
  \centering
  \includegraphics[width=6cm]{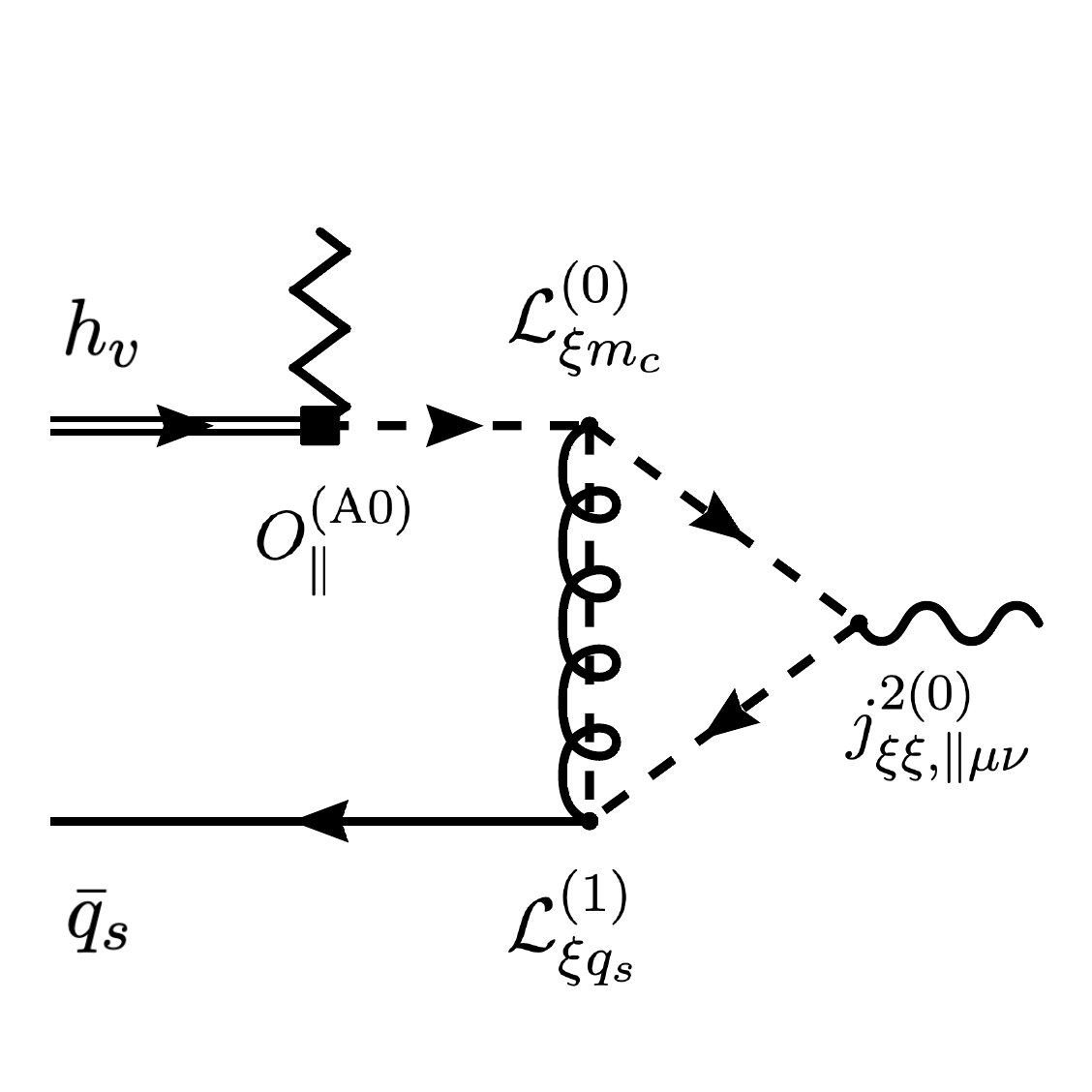}
  \caption{One-loop diagram contributing to
  $\Pi_{\mu\nu,\|}^{*C}(p,q)$.}\label{JC}
\end{figure}
The jet function $J^{*C}_{\|,\pm}$ is extracted from
\begin{align}
 F^{*C}_{ \|}  =& \int d^4 x \, e^{i p \cdot x} \, \int d^4 y \, \int d^4 z \, \nonumber \\
& \langle 0 | {\rm T}  \left \{ \bar{\xi}(x){\not \! n\over4}\gamma_5 (i n\cdot D)\xi(x),  \,\, i \, {\cal L}_{\xi q_s}^{(1)}(y),  \,
i \, {\cal L}_{\xi m_c}^{(0)}(z),  \,\,
\left (\bar \xi \, W_c \right)(0) \,  \, h_v(0) \,    \right \}   | \bar q_s(k)h_v \rangle  , 
\end{align}
Evaluating the diagram in Fig.~\ref{JC} gives
\begin{align}
  F^{*C}_{ \|}   =&   g_s^2\,C_F \,\mu^{2\epsilon} \, \bar q_s \,{\not \!{\bar n} \over 4} \gamma_5 \, h_v\nonumber\\
  &\times\int {d^D l \over (2 \pi)^D} \,  {m_c \, (2 - D) (n \cdot l)^2\,n\cdot(p+2l)\over (l^2 + i 0) [(p + l )^2 - m_c^2 + i0] ((l + k)^2 + i0) \, [(p - k)^2 - m_c^2]} \nonumber \\
		=& {\tilde{f}_B(\mu) \, m_B \over 2}\left({\alpha_s C_F\over 4\pi}\right)\,
 \int_0^{+\infty} \, {d\omega} \, J^{*(C)}_{\|,+}\left ({\mu^2 \over n \cdot p \, \omega}, {\omega \over \bar n \cdot p} \right )  \phi^{+}_{B}\left( \omega,\mu \right) .
\label{Fc}\end{align}
The resulting jet functions are
\begin{align}
    J^{*C}_{\|,+}=&{2\eta_c^2\over3(1+\eta)^2}+{7\eta_c+4\eta_c^2\over6(1+\eta)}-\left[ {(1+\eta+\eta_c)^2(1+\eta+4\eta_c)\over 6\eta (1+\eta)^3}\right]\ln\left[{\eta_c\over 1+\eta+\eta_c}\right]\nonumber\\&+\left({1+6\eta_c+9\eta_c^2+4\eta_c^3\over6\eta}\right)\ln\left[{\eta_c\over 1+\eta_c}\right]  \,,\nonumber\\
     J^{*C}_{\|,-}=& 0\,.
\end{align}
The nonzero $J^{*C}_{\|,+}$ originates from the finite charm-quark
mass. The complete A0-type jet function is then
\begin{align}
    J_{\parallel,-}^{*(\mathrm{A0})}
=&J_{\parallel,-}^{*(0)}
+\frac{\alpha_sC_F}{4\pi}
\left(J_{\parallel,-}^{*A}
+J_{\parallel,-}^{*B}
+J_{\parallel,-}^{*C}\right),  \nonumber \\
   J_{\parallel,+}^{*(\mathrm{A0})}=&\frac{\alpha_s C_F}{4\pi} J_{\parallel,+}^{*C}.
\end{align}
Now we can obtain the $\rm SCET_{II}$ factorization formula for $\Pi^{*}_{\mu\nu, \|}$ 
\begin{align}
    \Pi^{*}_{\mu\nu, \|}(p, q)={}&i\, {\tilde{f}_B(\mu) \, m_B \,\bar n_\mu \, \bar n_\nu\over 2}\,
 \int_0^{+\infty} \, {d\omega}\,\bigg[J^{*(\rm A0)}_{\|,-}\left ({\mu^2 \over n \cdot p \, \omega}, {\omega \over \bar n \cdot p} \right )\phi^{-}_{B}\left( \omega,\mu \right)\nonumber\\&+ J^{*(\rm A0)}_{\|,+}\left ({\mu^2 \over n \cdot p \, \omega}, {\omega \over \bar n \cdot p} \right )  \phi^{+}_{B}\left( \omega,\mu \right)\bigg]\nonumber\\
  ={}&i\,{\tilde{f}_B(\mu) \, m_B \,n\cdot p\over 2}\,
 \int_0^{+\infty} \, {d\omega^{\prime}\over \omega^{\prime}- \bar n \cdot p-i0}\,\left[\phi^{*-}_{\parallel,\rm eff}\left(\omega^{\prime},\mu\right) +\phi^{*+}_{\parallel,\rm eff}\left(\omega^{\prime},\mu\right)  \right]\,\, \bar n_{\mu}\bar n_{\nu},  \label{specj2}
\end{align}
together with \eqref{specj1}, constructing the partonic dispersion relations
\begin{align}
       \Pi_{ \|}^{D_1}(p, q) ={}&i\, {\tilde{f}_B(\mu) \, m_B \,m_c\over 2}\,
 \int_0^{+\infty} \, {d\omega^{\prime}\over \omega^{\prime}- \bar n \cdot p-i0}\,\nonumber\\&\left[2\phi^{-}_{\parallel,\rm eff}\left(\omega^{\prime},\mu\right) +2\phi^{+}_{\parallel,\rm eff}\left(\omega^{\prime},\mu\right) -{4f_2m_{D'_1}\over g_2\, m_c}\left(\phi^{*-}_{\parallel,\rm eff}\left(\omega^{\prime},\mu\right) +\phi^{*+}_{\parallel,\rm eff}\left(\omega^{\prime},\mu\right)\right) \right]\, , \\
  \Pi_{\|}^{D'_1}(p, q) ={}&i\, {\tilde{f}_B(\mu) \, m_B \, m_c \over 2}\,
 \int_0^{+\infty} \, {d\omega^{\prime}\over \omega^{\prime}- \bar n \cdot p-i0}\,\nonumber\\&\left[2\phi^{-}_{\parallel,\rm eff}\left(\omega^{\prime},\mu\right) +2\phi^{+}_{\parallel,\rm eff}\left(\omega^{\prime},\mu\right) -{4f_1m_{D_1}\over g_1\, m_c}\left(\phi^{*-}_{\parallel,\rm eff}\left(\omega^{\prime},\mu\right) +\phi^{*+}_{\parallel,\rm eff}\left(\omega^{\prime},\mu\right)\right) \right]\, . 
\end{align}

The currents constructed above separate the $D_1$ and $D'_1$ states,
but they also interpolate the ground-state $D^{(*)}$ mesons. The
hadronic representation of each correlation function therefore
contains both a ground-state pole and the selected $P$-wave pole. To
remove the ground-state contribution, we use two sum rules constructed
from the same correlation function. For an invariant amplitude
$\Pi_a^R$, we denote the ground-state and $P$-wave pole residues by
${\cal R}_{a,G}^R$ and ${\cal R}_{a,P}^R$, respectively, and its OPE
spectral density by $\rho_a^{R,\rm OPE}$. After the Borel
transformation, the sum rule containing both poles reads
\begin{align}
 {\cal R}_{a,G}^R\,
 e^{-m_G^2/(n\cdot p\,\omega_M)}
 +{\cal R}_{a,P}^R\,
 e^{-m_R^2/(n\cdot p\,\omega_M)}
 =
 \int_0^{\omega_{s,1}}d\omega'\,
 e^{-\omega'/\omega_M}\,
 \rho_a^{R,\rm OPE}(\omega',n\cdot p)\,,
 \label{eq:full-sum-rule}
\end{align}
whereas the corresponding ground-state sum rule is
\begin{align}
 {\cal R}_{a,G}^R\,
 e^{-m_G^2/(n\cdot p\,\omega_M)}
 =
 \int_0^{\omega_{s,0}}d\omega'\,
 e^{-\omega'/\omega_M}\,
 \rho_a^{R,\rm OPE}(\omega',n\cdot p)\,.
 \label{eq:ground-state-sum-rule}
\end{align}
where
\begin{align}
 \omega_{s,i}={s_i\over n\cdot p}\,,\qquad
 \omega_M={M^2\over n\cdot p}\,.\qquad (i=0,1)
\end{align}
Here $G=D$ or $D^*$, depending on the polarization. Specifically, the
longitudinal currents couple to the pseudoscalar $D$ meson, whereas the
transverse currents couple to the vector $D^*$ meson. Subtracting \eqref{eq:ground-state-sum-rule} from
\eqref{eq:full-sum-rule} gives
\begin{align}
 {\cal R}_{a,P}^R\,
 e^{-m_R^2/(n\cdot p\,\omega_M)}
 =
 \int_{\omega_{s,0}}^{\omega_{s,1}}d\omega'\,
 e^{-\omega'/\omega_M}\,
 \rho_a^{R,\rm OPE}(\omega',n\cdot p)\,.
 \label{eq:p-wave-subtracted-sum-rule}
\end{align}
We use this subtraction for all the effective form factors below.
Applying \eqref{eq:p-wave-subtracted-sum-rule} to the longitudinal
A0-type amplitudes gives
\begin{align}
    \xi^{D_1}_{\|} =& { -2 \tilde f_B(\mu)  \over f^{D_1}_{\|} } \, { m_B\, m_c \over (n \cdot p)^2} \,\int_{\omega_{s,0}}^{\omega_{s,1}} d \omega' \, {\rm exp} \left [ {m_{D_1}^2 - n \cdot p \, \omega' \over n \cdot p \, \omega_M} \right ] \nonumber\\& \left [ \phi_{\|,\,\rm{eff}}^-  + \phi_{\|,\,\rm{eff}}^+ \,  - {2f_2 \,m_{D'_1}\over g_2\, m_c }(\phi_{\|,\,\rm{eff}}^{*-} +\phi_{\|,\,\rm{eff}}^{*+})\right ](\omega', \mu) \,,
    \\
     \xi^{D'_1}_{\|} =& { -2 \tilde f_B(\mu)  \over f^{D'_1}_{\|} } \, { m_B \, m_c\over (n \cdot p)^2} \,\int_{\omega_{s,0}}^{\omega_{s,1}} d \omega' \, {\rm exp} \left [ {m_{D'_1}^2 - n \cdot p \, \omega' \over n \cdot p \, \omega_M} \right ] \nonumber\\& \left [ \phi_{\|,\,\rm{eff}}^-  + \phi_{\|,\,\rm{eff}}^+ \,  - {2f_1 \,m_{D_1}\over g_1\, m_c }(\phi_{\|,\,\rm{eff}}^{*-} +\phi_{\|,\,\rm{eff}}^{*+})\right ](\omega', \mu)\,.
\end{align}

In contrast to the massless light-quark case, the charm-quark mass effect induces an additional contribution, which manifests as an extra operator \eqref{Omc} when matching the QCD heavy-to-light currents onto ${\rm SCET_I}$ currents.
\begin{figure}[H]
  \centering
  \includegraphics[width=6cm]{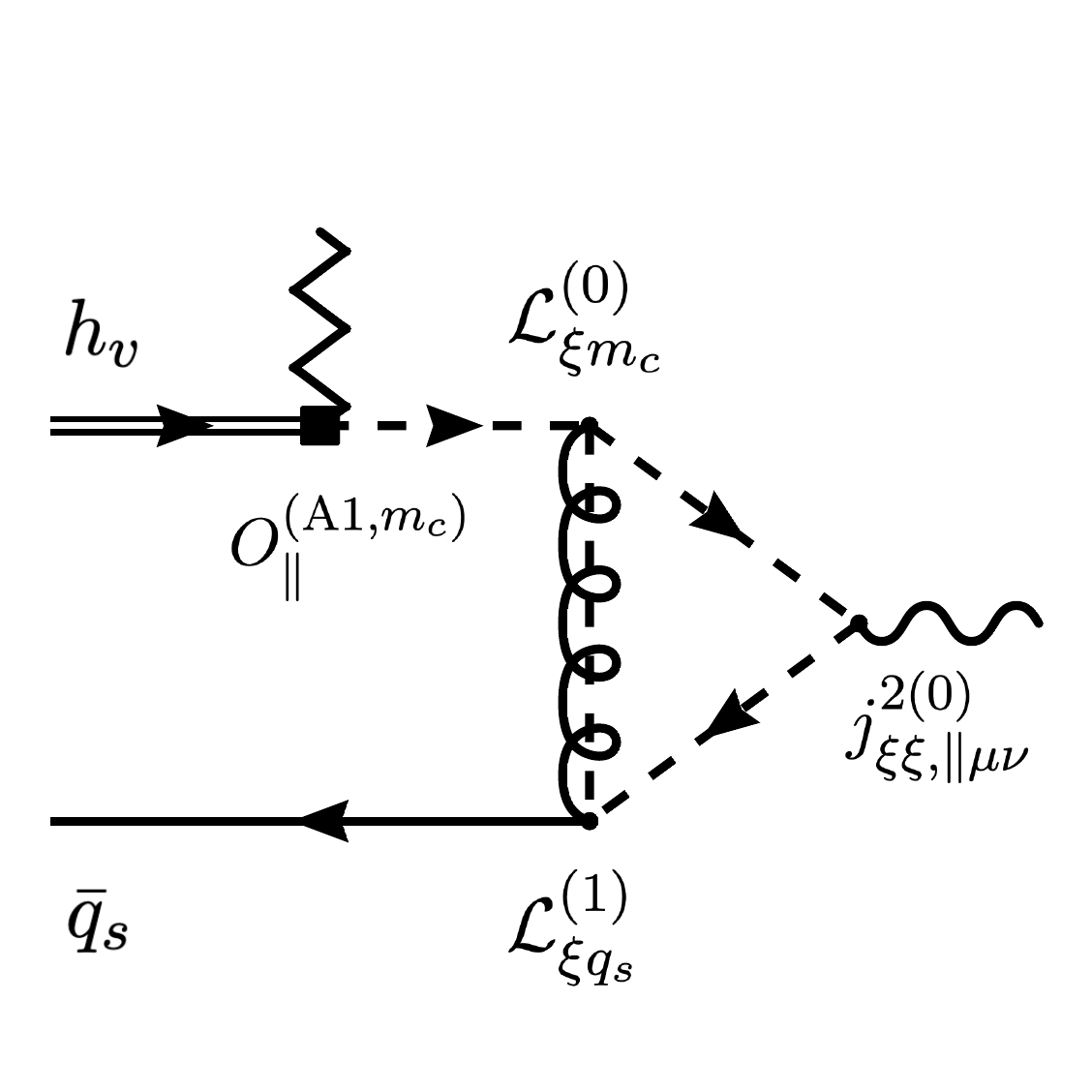}
  \caption{Diagram contributing to
  $\Pi^{*({\rm A1},m_c)}_{\mu\nu,\|}(p,q)$ at
  $\mathcal{O}(\alpha_s)$.}\label{fig:a1-mass}
\end{figure}
Observing that the corresponding correlation function
\begin{align}
    \Pi^{*({\rm A1},\,m_c)}_{\mu\nu\,,\|}={}&i\int d^4 x \, e^{i p \cdot x} \, \int d^4 y \, \int d^4 z \, 
\nonumber\\
&\langle 0 | {\rm T}  \left \{ j_{\xi \xi, \|\mu\nu}^{2(0)}(x),  \,\, i \, {\cal L}_{\xi q_s}^{(1)}(y),  \,\,
i \, {\cal L}_{\xi m_c}^{(0)}(z),  \,\,
O_{\|}^{(\mathrm{A} 1,m_c)}(0)\,    \right \}   | \bar B_v \rangle  ,
\end{align}
is proportional to $\Pi_{\|}^{*C}$, and the correlation function $\Pi_{\parallel}^{({\rm A1},m_c)}$ is likewise
proportional to $\Pi_{\parallel}^{C}$, we can therefore immediately
write down the sum rules for $\xi_{\parallel,m_c}^{R}$
\begin{align}
    \xi^{D_1}_{\|,m_c} =& { -2 \tilde f_B(\mu)  \over f^{D_1}_{\|} } \, { m_B\, m_c \over (n \cdot p)^2} \,\int_{\omega_{s,0}}^{\omega_{s,1}} d \omega' \, {\rm exp} \left [ {m_{D_1}^2 - n \cdot p \, \omega' \over n \cdot p \, \omega_M} \right ] \nonumber\\& \left( -{m_c\over n \cdot p} \right) \left [  \phi_{\|,\,\rm{eff}}^+ \,  - {2f_2 \,m_{D'_1}\over g_2\, m_c }\phi_{\|,\,\rm{eff}}^{*+}\right ](\omega', \mu) \,,
    \\
     \xi^{D'_1}_{\|,m_c} =& { -2 \tilde f_B(\mu)  \over f^{D'_1}_{\|} } \, { m_B \, m_c\over (n \cdot p)^2} \,\int_{\omega_{s,0}}^{\omega_{s,1}} d \omega' \, {\rm exp} \left [ {m_{D'_1}^2 - n \cdot p \, \omega' \over n \cdot p \, \omega_M} \right ] \nonumber\\& \left( -{m_c\over n \cdot p} \right) \left [  \phi_{\|,\,\rm{eff}}^+ \,  - {2f_1 \,m_{D_1}\over g_1\, m_c }\phi_{\|,\,\rm{eff}}^{*+}\right ](\omega', \mu)\,.
\end{align}

\subsection{The $B$-meson LCSRs for $\Xi^{R}_{\|}$ }
\begin{figure}[H]
  \centering
  \includegraphics[width=6cm]{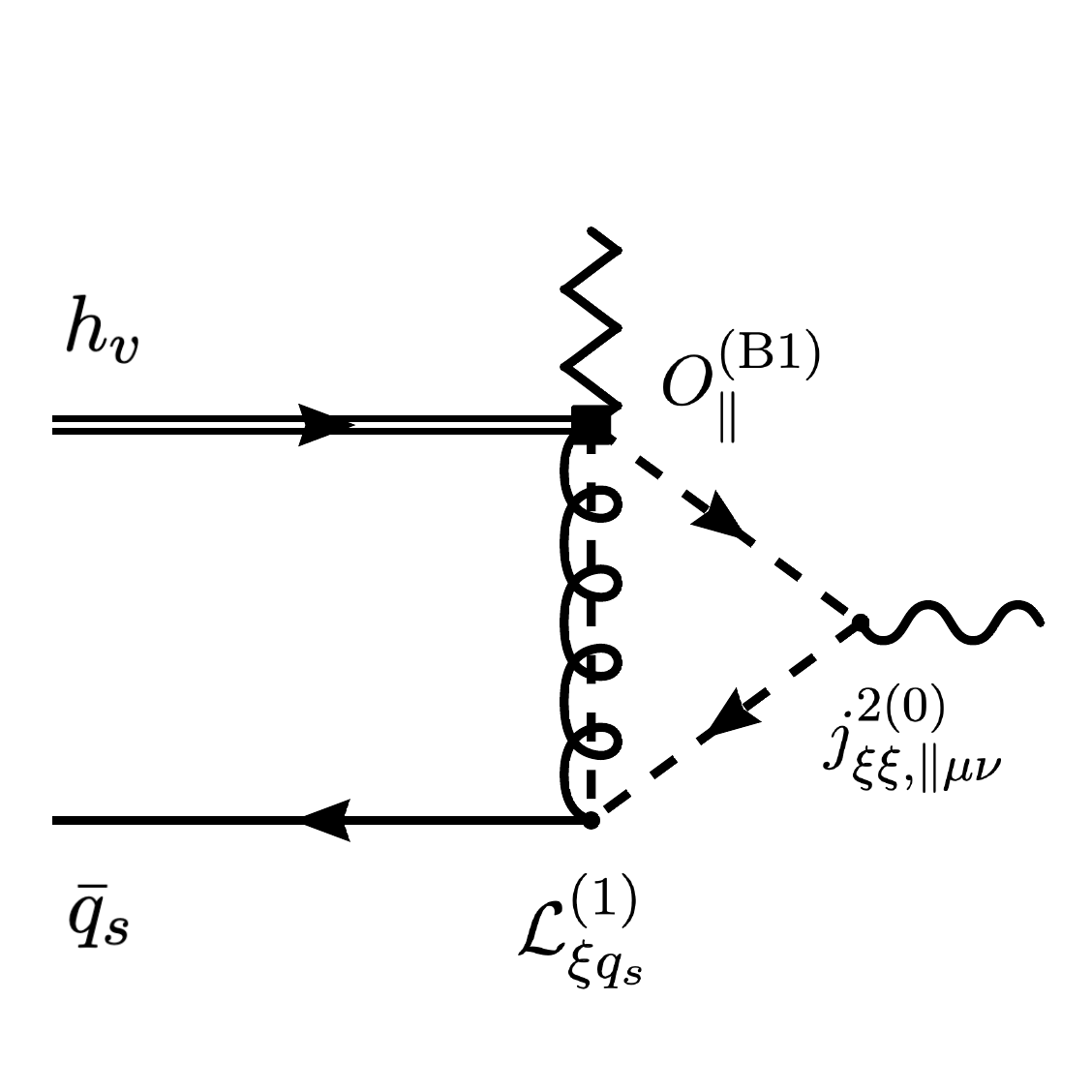}
  \caption{Diagram contributing to the B1-type correlation function
  $\tilde\Pi^{*}_{\mu\nu,\|}(p,q,\tau)$.}
  \label{fig:b1-longitudinal}
\end{figure}
We continue to calculate the B1-type effective form factors $\Xi^{D_1}_{\|}$ and $\Xi^{D'_1}_{\|}$, which enter the correlation functions \eqref{correlationfunction3}. Still we can decompose them as 
\begin{align}
     \tilde{\Pi}^{D_1}_{ \|}(p,q,\tau)  =&n^\mu \tilde{\Pi}_{\mu,\|}(p,q,\tau)-{f_2 \,m_{D'_1} \over g_2(n\cdot p)}\,n^\mu n^\nu\,\tilde{\Pi}^*_{\mu\nu,\|}(p,q,\tau)\,,\\
     \tilde{\Pi}^{D'_1}_{\|}(p,q,\tau)  =&n^\mu \tilde{\Pi}_{\mu,\|}(p,q,\tau)-{f_1 \,m_{D_1} \over g_1(n\cdot p)}\,n^\mu n^\nu\,\tilde{\Pi}^*_{\mu\nu,\|}(p,q,\tau)\,,
\end{align}
where the two building blocks on the right-hand side are constructed with longitudinal currents $j_{\mu}^{1}(x)$, $j_{\mu\nu}^{2}(x)$ and B1-type SCET operator $\left( \bar \xi \, W_c \right)(0) \,  \left ( W_c^\dagger i \not\!\! D_{c\perp} W_c \right ) (r\, n) \, h_v(0) $:
\begin{align}
    \tilde{\Pi}_{\mu,\|}(p,q,\tau)={}&i\, {n \cdot p \over 2\pi}\, \int d^4 x \, e^{i p \cdot x} \, \int {d r }\, e^{- i \, n \cdot p \, \tau \, r} 
	\nonumber	\\&\langle 0 | {\rm T}  \left \{ j_{\mu}^{1}(x),  \,\left (\bar \xi \, W_c \right)(0) \,  \left ( W_c^\dagger i \not\!\! D_{c\perp} W_c \right ) (r\, n) \, h_v(0) \,    \right \}   | \bar B_v \rangle ,\label{BtypecorrelationofJ1} \\
       \tilde{\Pi}^{*}_{\mu\nu,\|}(p,q,\tau)={}&i\, {n \cdot p \over 2\pi}\, \int d^4 x \, e^{i p \cdot x} \, \int {d r }\, e^{- i \, n \cdot p \, \tau \, r} 
	\nonumber	\\&\langle 0 | {\rm T}  \left \{ j_{\mu\nu}^{2}(x),  \,\left (\bar \xi \, W_c \right)(0) \,  \left ( W_c^\dagger i \not\!\! D_{c\perp} W_c \right ) (r\, n) \, h_v(0) \,    \right \}   | \bar B_v \rangle \,.\label{BtypecorrelationofJ2}
\end{align}
Due to the aforementioned reason, we can readily write down the final spectral representation of \eqref{BtypecorrelationofJ1}
\begin{align}
   & \tilde{\Pi}_{\mu, \|}(p, q,\tau) \nonumber\\
   &= 
   i\,{\tilde f_B(\mu) \, m_B \,m_c\,(n \cdot p)\,\over 2}\left({\alpha_s C_F\over 4\pi}\right) \, \int_{\omega_c / \bar \tau}^\infty {d \omega' \over \omega' - \bar n \cdot p-i0} \, \int_{\omega' - \omega_c / \bar \tau}^\infty d \omega \, {\phi_B^+ \over \omega } \,  2\,\theta(\tau) \theta(\bar\tau) \,  \bar\tau \,\bar n_{\mu},
   \label{specj1Btype}
\end{align}
where $\bar \tau\equiv 1-\tau$. We further write down the leading power contribution of the tensor correlation function \eqref{BtypecorrelationofJ2}
\begin{align}
     \tilde{\Pi}^{*}_{\mu\nu,\|}(p,q,\tau)={}&i\, {n \cdot p \over 2\pi}\, \int d^4 x \, e^{i p \cdot x} \,\int d^4 y\, \int {d r }\, e^{- i \, n \cdot p \, \tau \, r} 
	\nonumber	\\&\langle 0 | {\rm T}  \left \{ j_{\xi \xi, \|\mu\nu}^{2(0)}(x),  \,\, i \, {\cal L}_{\xi q_s}^{(1)}(y),  \,\left (\bar \xi \, W_c \right)(0) \,  \left ( W_c^\dagger i \not\!\! D_{c\perp} W_c \right ) (r\, n) \, h_v(0) \,    \right \}   | \bar B_v \rangle ,
\end{align}
and its $\rm SCET_I$ factorization formula takes the form
\begin{align}
    \tilde{\Pi}^{*}_{\mu\nu,\|}(p,q,\tau)=i\,{\tilde{f}_B(\mu) \, m_B \over 2}\,
 \sum_{m=\pm} \int_0^{+\infty} \, {d\omega} \, \tilde{J}^{*}_{\|,m}\left ({\mu^2 \over n \cdot p \, \omega}, {\omega \over \bar n \cdot p},\tau \right )  \phi^{m}_{B}\left( \omega,\mu \right)\bar n_{\mu} \bar n_{\nu}.\label{SCETfacformBtype}
\end{align}
Along the same vein, we can extract the jet functions $\tilde{J}^{*}_{\|,m}$ containing hard-collinear dynamics by using an external partonic state
\begin{align}
     \tilde{F}^{*}_{\|}(p,q,\tau)=& {n \cdot p \over 2\pi}\, \int d^4 x \, e^{i p \cdot x} \,\int d^4 y\, \int {d r }\, e^{- i \, n \cdot p \, \tau \, r} 
	\bigg\langle 0 \bigg| {\rm T}  \bigg\{ \bar{\xi}(x){\not \! n\over4}\gamma_5 (i n\cdot D)\xi(x),  \,\, \nonumber	\\&i \, {\cal L}_{\xi q_s}^{(1)}(y),  \,\bigg(\bar \xi \, W_c \bigg)(0) \, \bigg ( W_c^\dagger i \not\!\! D_{c\perp} W_c \bigg) (r\, n) \, h_v(0) \,    \bigg \}   \bigg| \bar q_s(k) h_v \bigg\rangle ,
\end{align}
Evaluating the diagram in Fig.~\ref{fig:b1-longitudinal} gives
\begin{align}
    \tilde{F}^{*}_{\|}(p,q,\tau)=& g_s^2 C_F \mu^{2\epsilon} \bar q_s \, {\not \! \bar n \over 4} \gamma_5 \, h_v \int {d^D l \over (2 \pi)^D} (D-2){  n\cdot l \,n \cdot (p + l) \,n\cdot (p+2l) \delta(\tau+{n\cdot l\over n \cdot p})\over [l-k]^2\,[(p-k+l)^2-m^2_c]\,[l^2 + i0]} \,   \nonumber\\
		  =&{\tilde{f}_B(\mu) \, m_B \over 2}\,
 \int_0^{+\infty} \, {d\omega} \, \tilde J^{*}_{\|,+}\left ({\mu^2 \over n \cdot p \, \omega}, {\omega \over \bar n \cdot p},\tau \right )  \phi^{+}_{B}\left( \omega,\mu \right) .
\end{align}
Evaluating the contour integral yields the $\mathcal{O}(\alpha_s)$ results for the jet functions $\tilde{J}^{*}_{\|,m}$
\begin{align}
    &\tilde{J}^{*}_{\|,+}=\left({\alpha_s C_F\over 4\pi}\right){(n\cdot p)^2\over {\bar n}\cdot k}\left[(\bar \tau)(\bar\tau-\tau)\theta(\tau)\theta(\bar \tau)\right]\ln\left[1+{(\bar \tau)\eta \over(\bar \tau)+\eta_c}\right]\,,
    \nonumber\\
   &\tilde{J}^{*}_{\|,-}=0. 
\end{align}
We further derive the spectral representation of the correlation function \eqref{SCETfacformBtype} at partonic level
\begin{align}
    \tilde{\Pi}_{\mu\nu, \|}^{*}(p, q,\tau) ={}&i\,{\tilde f_B(\mu) \, m_B \,(n \cdot p)^2\,\over 2} \, \left({\alpha_s C_F\over 4\pi}\right)\,\int_{\omega_c / \bar \tau}^\infty {d \omega' \over \omega' - \bar n \cdot p-i0} \, \nonumber\\&\int_{\omega' - \omega_c / \bar \tau}^\infty d \omega \, {\phi_B^+\left( \omega,\mu \right) \over \omega } \,  \theta(\tau) \theta(\bar\tau) \,  \bar\tau(\bar\tau-\tau) \,\bar n_{\mu}\bar n_{\nu}, 
\end{align}
together with \eqref{specj1Btype}, giving the total partonic dispersion relations of $\tilde{\Pi}^{D_1}_{ \|}$ and $\tilde{\Pi}^{D'_1}_{\|}$
\begin{align}
    \tilde{\Pi}^{D_1}_{ \|}(p, q,\tau) ={}&i\,
   {\tilde f_B(\mu) \, m_B \,m_c\,(n \cdot p)\,\over 2}\left({\alpha_s C_F\over 4\pi}\right) \, \int_{\omega_c / \bar \tau}^\infty {d \omega' \over \omega' - \bar n \cdot p-i0} \, \int_{\omega' - \omega_c / \bar \tau}^\infty d \omega 
   \nonumber\\
   & \, {\phi_B^+(\omega,\mu) \over \omega } \,  2\,\theta(\tau) \theta(\bar\tau) \,  \bar\tau \left[ 2- {2f_2 \,m_{D'_1} \over g_2\,m_c }(\bar\tau-\tau)  \right] ,
   \\
    \tilde{\Pi}^{D'_1}_{\|}(p, q,\tau) ={}&i\,
   {\tilde f_B(\mu) \, m_B \,m_c\,(n \cdot p)\,\over 2}\left({\alpha_s C_F\over 4\pi}\right) \, \int_{\omega_c / \bar \tau}^\infty {d \omega' \over \omega' - \bar n \cdot p-i0} \, \int_{\omega' - \omega_c / \bar \tau}^\infty d \omega 
   \nonumber\\
   & \, {\phi_B^+(\omega,\mu) \over \omega } \,  2\,\theta(\tau) \theta(\bar\tau) \,  \bar\tau \left[ 2- {2f_1 \,m_{D_1} \over g_1\,m_c }(\bar\tau-\tau)  \right] .
\end{align}
Applying the subtraction in
\eqref{eq:p-wave-subtracted-sum-rule} to these B1-type amplitudes, we
obtain
\begin{align}
   \Xi^{D_1}_{\|} = 	& - {\tilde f_B(\mu)  \over f^{D_1}_{ \|} } \,
{ m_B \,m_c \over n \cdot p \, m_b} {\alpha_s C_F \over \pi}\,
\theta(\tau)\theta(\bar\tau)\,\bar \tau
\left[ 1-{f_2\,m_{D'_1}\over g_2\,m_c}(\bar \tau-\tau) \right] \,\nonumber\\
&\times\int_{\omega_{s,0}}^{\omega_{s,1}} d \omega'\,
\theta\left(\omega'-{\omega_c\over\bar\tau}\right)
{\rm exp} \left [ {m_{D_1}^2 - n \cdot p \, \omega' \over n \cdot p \, \omega_M} \right ]
\int_{\omega' - \omega_c / \bar \tau}^\infty d \omega \,
{\phi_B^+ (\omega, \mu) \over \omega } \,,
\\
  \Xi^{D'_1}_{\|} = 	& - {\tilde f_B(\mu)  \over f^{D'_1}_{ \|} } \,
{ m_B \,m_c \over n \cdot p \, m_b} {\alpha_s C_F \over \pi}\,
\theta(\tau)\theta(\bar\tau)\,\bar \tau
\left[ 1-{f_1\,m_{D_1}\over g_1\,m_c}(\bar \tau-\tau) \right] \,\nonumber\\
&\times\int_{\omega_{s,0}}^{\omega_{s,1}} d \omega'\,
\theta\left(\omega'-{\omega_c\over\bar\tau}\right)
{\rm exp} \left [ {m_{D'_1}^2 - n \cdot p \, \omega' \over n \cdot p \, \omega_M} \right ]
\int_{\omega' - \omega_c / \bar \tau}^\infty d \omega \,
{\phi_B^+ (\omega, \mu) \over \omega }\,.
\end{align}
\subsection{The $B$-meson LCSRs for transverse form factors}
We now concentrate on the transverse correlation functions \eqref{correlationfunction2} and \eqref{correlationfunction4} to extract the transverse form factors $\xi^{R}_{\perp}$ and $\Xi^{R}_{\perp}$. We still have the following decomposition for $\Pi^{R}_{\rho\sigma,\,\perp}$
\begin{align}
   \Pi^{D_1}_{\rho\sigma, \perp}(p, q)  =&n^\mu  \Pi_{\mu\rho\sigma,\perp}(p,q)-{h_2 \,m_{D'_1} \over r_2(n\cdot p)}\,n^\mu n^\nu\,\Pi^*_{\mu\nu\rho\sigma,\perp}(p,q)\,,\\
  \Pi^{D'_1}_{\rho\sigma, \perp}(p, q)  =&n^\mu  \Pi_{\mu\rho\sigma,\perp}(p,q)-{h_1 \,m_{D_1} \over r_1(n\cdot p)}\,n^\mu n^\nu\,\Pi^*_{\mu\nu\rho\sigma,\perp}(p,q)\,.
\end{align}
The two building blocks on the right-hand side are constructed with
$j^3_{\mu\rho}$ and $j^4_{\mu\nu\rho}$, respectively:
\begin{align}
 \Pi_{\mu\rho\sigma,\perp}(p,q)
 ={}&i\int d^4x\,e^{ip\cdot x}
 \langle0|{\rm T}\left\{
 j^3_{\mu\rho}(x),
 (\bar\xi W_c)(0)\gamma_{\sigma\perp}h_v(0)
 \right\}|\bar B_v\rangle\,,
 \\
 \Pi^*_{\mu\nu\rho\sigma,\perp}(p,q)
 ={}&i\int d^4x\,e^{ip\cdot x}
 \langle0|{\rm T}\left\{
 j^4_{\mu\nu\rho}(x),
 (\bar\xi W_c)(0)\gamma_{\sigma\perp}h_v(0)
 \right\}|\bar B_v\rangle\,.
\end{align}
Using the same procedure as for the longitudinal form factors, we
obtain the SCET sum rules for $\xi^R_\perp$
\begin{align}
  \xi^{D_1}_{\perp}= 	& {\tilde f_B(\mu)  \over f^{D_1}_{ \perp} } \,
{ m_B \,m_c \over m_{D_1}\, n \cdot p} \,\int_{\omega_{s,0}}^{\omega_{s,1}} d \omega'\, {\rm exp} \left [ {m_{D_1}^2 - n \cdot p \, \omega' \over n \cdot p \, \omega_M} \right ] \, \nonumber\\
&\times\left [\, \phi_{\|, \, \rm{eff}}^- +  \,\Delta \phi_{\rm eff}^- -{2 h_2\,m_{D'_1}\over r_2 \,m_c}(\phi_{\|, \, \rm{eff}}^{*-} +  \Delta \phi_{\rm eff}^{*-})\right ] (\omega', \mu) \,,\\
\xi^{D'_1}_{\perp}= & {\tilde f_B(\mu)  \over f^{D'_1}_{ \perp} } \,
{ m_B \,m_c \over m_{D'_1} \, n \cdot p} \,\int_{\omega_{s,0}}^{\omega_{s,1}} d \omega'\, {\rm exp} \left [ {m_{D'_1}^2 - n \cdot p \, \omega' \over n \cdot p \, \omega_M} \right ] \, \nonumber\\
&\times\left [\, \phi_{\|, \, \rm{eff}}^- + \, \Delta \phi_{\rm eff}^- -{2 h_1\,m_{D_1}\over r_1 \,m_c}(\phi_{\|, \, \rm{eff}}^{*-} +  \Delta \phi_{\rm eff}^{*-})\right ] (\omega', \mu) \,,
\end{align}
where $\Delta\phi$ functions encode the discrepancy between the transverse spectral representation and the corresponding longitudinal counterparts.
At $\mathcal{O}(\alpha_s)$, the transverse jet functions
$J_{\perp,+}^{C}$ and $J_{\perp,+}^{*C}$ are proportional to $D-4$ and therefore vanish in
four dimensions. Since the corresponding A1-type contribution is
proportional to these jet functions, we obtain
$\xi^R_{\perp,m_c}=0$ at this accuracy. We proceed to write down the resulting sum rules for B1-type transverse form factors $\Xi^{R}_{\perp}$
\begin{align}
    \Xi^{D_1}_{\perp} = 	&  {\tilde f_B(\mu)  \over 2 f^{D_1}_{ \perp} } \,
{ m_B \,m_c\, \, \over  \, m_b m_{D_1}} {\alpha_s C_F \over \pi}\,
\theta(\tau)\theta(\bar\tau)\,\bar \tau
\left[ 1-{h_2 \,m_{D'_1}\over r_2\,m_c}(1-2\tau) \right]
\nonumber\\
 &\times\int_{\omega_{s,0}}^{\omega_{s,1}} d \omega'\,
\theta\left(\omega'-{\omega_c\over\bar\tau}\right)
{\rm exp} \left [ {m_{D_1}^2 - n \cdot p \, \omega' \over n \cdot p \, \omega_M} \right ]
\int_{\omega' - \omega_c / \bar \tau}^\infty d \omega \,
{\phi_B^+ (\omega, \mu) \over \omega } \,,
\\
 \Xi^{D'_1}_{\perp} = 	&  {\tilde f_B(\mu)  \over 2 f^{D'_1}_{ \perp} } \,
{ m_B \,m_c\,\, \over  \, m_b m_{D'_1}} {\alpha_s C_F \over \pi}\,
\theta(\tau)\theta(\bar\tau)\,\bar \tau
\left[ 1-{h_1\,m_{D_1}\over r_1\,m_c}(1-2\tau) \right]
\nonumber\\
 &\times\int_{\omega_{s,0}}^{\omega_{s,1}} d \omega'\,
\theta\left(\omega'-{\omega_c\over\bar\tau}\right)
{\rm exp} \left [ {m_{D'_1}^2 - n \cdot p \, \omega' \over n \cdot p \, \omega_M} \right ]
\int_{\omega' - \omega_c / \bar \tau}^\infty d \omega \,
{\phi_B^+ (\omega, \mu) \over \omega }\,.
\end{align}
\subsection{Renormalization-group improvement}\label{sec:rge}

The hard matching scale is chosen as $\mu_h\sim m_b$, while the
factorization scale $\mu$ is taken at the hard-collinear scale,
$\mu\sim\sqrt{m_b\Lambda_{\rm QCD}}$. For simplicity, we identify the two hard matching scales associated
with the SCET currents and the HQET decay constant,
$\mu_{h1}=\mu_{h2}\equiv\mu_h$. Following the RG treatment of
the $B\to V$ form factors \cite{Gao:2019lta}, we evolve the A0- and
B1-type hard coefficients at NLL and LL accuracy, respectively. The
same framework is employed in the $B\to D^*$ analysis
\cite{cui:2301.12391}. The HQET decay constant is evolved from
$\mu_h$ to $\mu$, while the jet functions are evaluated at $\mu$.
The $B$-meson LCDAs are evolved from the reference scale $\mu_0=1 \,\rm GeV$ to
$\mu$ following \cite{Gao:2112.12674,huang:2212.11624}.

The QCD decay constant $f_B$ is matched onto the HQET decay constant
at the hard scale \cite{Beneke:1110.3228},
\begin{align}
 \widetilde f_B(\mu_h)
 &=
 K^{-1}(\mu_h)f_B,
 &
 K^{-1}(\mu_h)
 &=
 1+\frac{\alpha_s(\mu_h)C_F}{4\pi}
 \left(3\ln\frac{\mu_h}{m_b}+2\right),
 \label{eq:fB-matching}
\end{align}
and subsequently evolved to the hard-collinear scale,
\begin{align}
 \widetilde f_B(\mu)
 =
 U_B(\mu_h,\mu)\widetilde f_B(\mu_h).
 \label{eq:fB-evolution}
\end{align}
For a quantity satisfying
\begin{align}
 \frac{d}{d\ln\mu}F(\mu)
 &=
 \gamma_F(\alpha_s)F(\mu),
 &
 \gamma_F(\alpha_s)
 &=
 \sum_{n=0}^{\infty}
 \gamma_F^{(n)}
 \left(\frac{\alpha_s}{4\pi}\right)^{n+1},
\end{align}
the NLL evolution factor reads
\begin{align}
 U_F(\mu_0,\mu)
 &=
 z^{-\gamma_F^{(0)}/(2\beta_0)}
 \left[
 1+\frac{\alpha_s(\mu_0)}{4\pi}
 \left(
 \frac{\gamma_F^{(1)}}{2\beta_0}
 -\frac{\gamma_F^{(0)}\beta_1}{2\beta_0^2}
 \right)(1-z)
 \right],
 \label{eq:noncusp-evolution}
\end{align}
where $z=\frac{\alpha_s(\mu)}{\alpha_s(\mu_0)}$. The anomalous dimensions
entering $U_B$ are \cite{Broadhurst:9908362,Ji:1991pr}
\begin{align}
 \gamma_B^{(0)}
 &=
 3C_F,
 &
 \gamma_B^{(1)}
 &=
 C_F\left(
 \frac{127}{6}
 +\frac{14\pi^2}{9}
 -\frac{5}{3}n_f
 \right).
\end{align}
When an evolution interval crosses the charm threshold, the evolution
factors obtained with the appropriate values of $n_f$ are multiplied.

The definitions of the interpolating currents imply
\begin{align}
 f_\parallel^R(\mu)
 &=
 U_m(\mu_0,\mu)f_\parallel^R(\mu_0),
 \nonumber\\
 f_\perp^R(\mu)
 &=
 U_m(\mu_0,\mu)U_T(\mu_0,\mu)f_\perp^R(\mu_0),
 \label{eq:fD-evolution}
\end{align}
where $U_m(\mu_0,\mu)$ denotes the $\overline{\rm MS}$ mass-evolution
function. The tensor-current evolution function $U_T$ is obtained from
\eqref{eq:noncusp-evolution} using the two-loop anomalous dimension
\cite{Broadhurst:9410240}
\begin{align}
 \gamma_T^{(0)}
 &=
 -2C_F,
 &
 \gamma_T^{(1)}
 &=
 C_F\left(
 19C_F-\frac{257}{9}C_A+\frac{26}{9}n_f
 \right).
\end{align}

For completeness, we also give the RG equations for the hard matching
coefficients entering the sum rules. The A0-type hard coefficients obey \cite{Hill:0404217,Beneke:0508250}
\begin{align}
 \frac{d}{d\ln\mu}C_i^{({\rm A0})}(n\cdot p,\mu)
 =
 \left[
 \Gamma_{\rm cusp}(\alpha_s)\ln\frac{n\cdot p}{\mu}
 +\gamma(\alpha_s)
 \right]C_i^{({\rm A0})}(n\cdot p,\mu)\,.
 \label{eq:A0-rge}
\end{align}
Their solution can be written as
\begin{align}
 C_i^{({\rm A0})}(n\cdot p,\mu)
 &=
 U_{\rm A0}(n\cdot p,\mu_h,\mu)
 C_i^{({\rm A0})}(n\cdot p,\mu_h),
 \nonumber\\
 U_{\rm A0}(n\cdot p,\mu_h,\mu)
 &=
 \exp\left[
 S(\mu_h,\mu)-a_\gamma(\mu_h,\mu)
 \right]
 \left(\frac{n\cdot p}{\mu_h}\right)^{-a_\Gamma(\mu_h,\mu)},
 \label{eq:A0-evolution}
\end{align}
where
\begin{align}
 S(\nu,\mu)
 &=
 -\int_{\alpha_s(\nu)}^{\alpha_s(\mu)}
 d\alpha\,
 \frac{\Gamma_{\rm cusp}(\alpha)}{\beta(\alpha)}
 \int_{\alpha_s(\nu)}^\alpha
 \frac{d\alpha'}{\beta(\alpha')},
 \nonumber\\
 a_\Gamma(\nu,\mu)
 &=
 -\int_{\alpha_s(\nu)}^{\alpha_s(\mu)}
 d\alpha\,
 \frac{\Gamma_{\rm cusp}(\alpha)}{\beta(\alpha)},
 \nonumber\\
 a_\gamma(\nu,\mu)
 &=
 -\int_{\alpha_s(\nu)}^{\alpha_s(\mu)}
 d\alpha\,
 \frac{\gamma(\alpha)}{\beta(\alpha)}.
 \label{eq:sudakov-functions}
\end{align}
We have written the evolution factor given in
\cite{Beneke:1110.3228} in the compact form above. The anomalous dimensions and the QCD beta function are evaluated at
the orders required for NLL accuracy.

The B1-type coefficients satisfy the nonlocal evolution equation
\begin{align}
 \frac{d}{d\ln\mu}
 C_i^{({\rm B1}),a}(\tau,n\cdot p,\mu)
 &=
 \Gamma_{\rm cusp}(\alpha_s)\ln\frac{n\cdot p}{\mu}\,
 C_i^{({\rm B1}),a}(\tau,n\cdot p,\mu)
 \nonumber\\
 &\quad+
 \int_0^1d\tau'\,
 \gamma_a^{({\rm B1})}(\tau',\tau)
 C_i^{({\rm B1}),a}(\tau',n\cdot p,\mu).
 \qquad (a=\parallel,\perp)
 \label{eq:B1-rge}
\end{align}
Since the tree-level B1 coefficients are independent of $\tau$, their
LL evolution can be written in terms of the approximate solution
given in \cite{Beneke:0508250,Gao:2019lta},
\begin{align}
 C_i^{({\rm B1}),a}(\tau,n\cdot p,\mu)
 &=
 U_{{\rm B1},a}(\tau,n\cdot p,\mu_h,\mu)
 C_i^{({\rm B1}),a}(n\cdot p,\mu_h),
 \nonumber\\
 U_{{\rm B1},a}(\tau,n\cdot p,\mu_h,\mu)
 &=
 \exp\left[S(\mu_h,\mu)\right]
 \left(\frac{n\cdot p}{\mu_h}\right)^{-a_\Gamma(\mu_h,\mu)}
 z^{-\gamma_a^{(0)}(\tau)/(2\beta_0)},
 \label{eq:B1-evolution}
\end{align}
where $z=\frac{\alpha_s(\mu)}{\alpha_s(\mu_h)}$. We note that, compared with the convention used in
\cite{Beneke:0508250}, the $n\cdot p$ dependence has been separated from
the function $S$. The polarization-dependent anomalous dimensions are
\begin{align}
 \gamma_\parallel^{(0)}(\tau)
 &=
 -C_F
 +4\left(C_F-\frac{C_A}{2}\right)
 \frac{\ln\bar\tau}{\tau},
 \nonumber\\
 \gamma_\perp^{(0)}(\tau)
 &=
 -C_F\left(
 1+\frac{4\tau\ln\tau}{\bar\tau}
 \right)
 +4\left(C_F-\frac{C_A}{2}\right)
 \left(
 \frac{1+\tau}{\tau}\ln\bar\tau
 +\frac{\tau\ln\tau}{\bar\tau}
 \right).
 \label{eq:B1-anomalous-dimensions}
\end{align}

For the A1 contribution, the hard matching coefficients
are kept at tree level and are not evolved separately. The RG-improved sum rules used in
the numerical analysis are obtained by incorporating the evolution
functions described above.

\section{Numerical analysis}\label{sec4}
Based on the NLO SCET sum rules derived in the previous sections, we
now evaluate their numerical implications for the semileptonic
$B \to D_1^{(\prime)} \ell \nu_\ell$ transitions. We first collect the
relevant numerical inputs, including the SM parameters, the
$B$-meson LCDA parameters, and the hadronic quantities. The
Bourrely--Caprini--Lellouch (BCL) parameterization
\cite{Bourrely:0504016,Lellouch:9509358,Bourrely:0807.2722}, truncated at
$\mathcal{O}(z)$, is then employed to extrapolate the sum rule results
from the large-recoil region to the full kinematic range. A joint
statistical analysis is performed to properly account for the
correlations among the form factors. Finally, we present predictions for
the differential decay widths, total branching fractions, and the LFU
ratios.

\subsection{Theory inputs}


\begin{table}[htbp!]
\centering
\renewcommand{\arraystretch}{1.3} 
\setlength{\tabcolsep}{20pt} 
\begin{tabular}{|c|c|c|}
\hline
Parameter & Value &  Ref. \\
\hline
\multicolumn{3}{|c|}{\textbf{SM Inputs}} \\
\hline
$\alpha^{\left(5\right)}_s(m_Z)$ & $0.1179 \pm 0.0009$ &  \cite{PDG:2024} \\
$G_F$ & $1.166379 \times 10^{-5} \,\, {\rm GeV}^{-2} $ & \cite{PDG:2024} \\
$m_b^{\rm PS}(2\,\text{GeV})$ & $4.532^{+0.013}_{-0.039}$  GeV & \cite{Beneke:1411.3132}\\
$\overline{m}_b(\overline{m}_b)$ & $4.203\pm 0.011$ GeV & \cite{PDG:2024} \\
$m_c^{\rm pole}$ & $1.67 \pm 0.05$ GeV & \cite{PDG:2024} \\
$\overline{m}_c(\overline{m}_c)$  & $1.273 \pm 0.0028$ GeV & \cite{PDG:2024} \\
$m_\mu$ & $105.658$ MeV & \cite{PDG:2024} \\
$m_\tau$ & $1776.86$ MeV & \cite{PDG:2024} \\
\hline
\multicolumn{3}{|c|}{\textbf{$B$-meson LCDA Parameters (at $\mu_0 = 1.0$ GeV)}} \\
\hline
$\lambda_B$ & $0.460 \pm 0.11$ GeV & \cite{Braun:0309330} \\
$ \{\widehat{\sigma}_{1}(\mu_0), \,
\widehat{\sigma}_{2}(\mu_0)\}$    & $\{0.0, \, \pi^2/6\}$ & \cite{Beneke:2008.12494} \\
  $2 \, \lambda_E^2(\mu_0) + \lambda_H^2(\mu_0)$      & $0.25 \pm 0.15 \,\, {\rm GeV^2}$  & \cite{Beneke:1804.04962} \\
$\lambda_E^2(\mu_0)/\lambda_H^2(\mu_0)$  & $0.50 \pm 0.10 $  & \cite{Beneke:1804.04962} \\
\hline   
\multicolumn{3}{|c|}{\textbf{Hadronic Inputs}} \\
\hline
$f_B$  & $190.0 \pm 1.3$ MeV & \cite{FLAG:2411.04268} \\
$m_B$  & $5279.66$ MeV & \cite{PDG:2024} \\
$\{f_1,f_2\}$  & $\{0.095,\,-0.005\}$ GeV & This Work \\
$\{g_1,g_2\}$  & $\{0.144,\,0.173\}$ GeV & This Work \\
$\{h_1,h_2\}$  &  $\{-0.063\,, 0.094\}$ GeV & This Work \\
$\{r_1,r_2\}$  & $\{-0.095,\,0.005\}$ GeV & This Work \\
$m_{D_1}$ & $2422.1$ MeV &  \cite{PDG:2024} \\
$m_{D'_1}$ & $2412$ MeV &  \cite{PDG:2024} \\
\hline
\multicolumn{3}{|c|}{\textbf{Sum Rule Parameters and Scales}} \\
\hline
$M^2$ & $6.5 \pm 0.5$ $\text{GeV}^2$ & This work\\
$s_1$ & $9 \pm 0.6$ $\text{GeV}^2$ & This work \\
$s_0$ & $5 \pm 0.5$ $\text{GeV}^2$ & This work \\
$\mu$ & $1.5 \pm 0.5 $ GeV & \cite{huang:2212.11624} \\
$\mu_{h}$ & ${m_b}^{+m_b}_{-m_b/2}$ GeV & \cite{huang:2212.11624} \\
\hline 
\end{tabular}
\caption{Summary of the numerical inputs for the $B \to D_1^{(\prime)}$ form factor analysis. }
\label{tab:inputs}
\end{table}
 The numerical values of the necessary theory inputs are compiled in Table~\ref{tab:inputs}.
 The evolution of the strong coupling $\alpha_s(\mu)$ is performed at
five-loop accuracy in the $\overline{\text{MS}}$ scheme using RunDec
\cite{Chetyrkin:0004189,Herren:1703.03751,Schmidt:1201.6149}, with
$\alpha_s^{(5)}(m_Z)=0.1179$ as the boundary condition
\cite{PDG:2024}. To account for heavy-quark decoupling effects, the perturbative matching scales for crossing into the $n_f=4$ and $n_f=3$ active flavor domains are set at the scales $\mu_b = 4.8\,\text{GeV}$ and $\mu_c = 1.3\,\text{GeV}$, respectively \cite{Shen:2009.02723,Beneke:2008.12494}. For the heavy-quark masses, the bottom-quark mass entering the short-distance coefficient functions is defined in the potential-subtracted (PS) scheme \cite{Beneke:9804241} to eliminate the infrared renormalon ambiguity \cite{Bigi:9402360,Beneke:9402364} inherent to the pole mass definition. On the other hand, the pole mass scheme is specified for the charm-quark within the algebraic relations of the hadronic parameters $\{f_i, g_i, h_i, r_i\}$ derived from the leading-order equations of motion (EOMs), in order to capture the corresponding on-shell dynamics. The $\overline{\text{MS}}$ scheme is employed for the charm-quark in the perturbative matching relations, which ensures the convergence of the loop-level radiative corrections at NLO accuracy.
Regarding the $B$-meson LCDAs, we use the three-parameter model \cite{Beneke:1804.04962} (see also \cite{Bell:1308.6114,Wang:1506.00667,Wang:1606.03080,Wang:1803.06667,Shen:2009.02723,Wang:2111.11811,Feldmann:2203.15679}), which inherently respects the constraints from the QCD EOMs and the required infrared asymptotic behaviors at small partonic momenta. 

The eight primary hadronic parameters defined in Sec.~\ref{primary_hadronic_inputs} are evaluated by invoking the QCD EOMs. We consider the following vacuum-to-hadron matrix elements:
\begin{align}
 \langle 0|m_c\, \bar{q} \,\slashed n \gamma_5 c |D_1(p,\epsilon)\rangle &= f_{D_1}\, m^2_{D_1}(n\cdot\epsilon^{D_1})\,,\nonumber\\
 \langle 0|m_c\, \bar{q} \,\slashed n \gamma_5 c |D'_1(p,\epsilon)\rangle &= f_{D'_1}\, m^2_{D'_1}(n\cdot\epsilon^{D'_1})\,,\nonumber\\
 \langle 0| \bar{q} \gamma_5 (i n\cdot\overleftrightarrow{D})  c|D_1(p,\epsilon)\rangle &= g_{D_1}\, m^2_{D_1}(n\cdot\epsilon^{D_1})\,,\nonumber\\
 \langle 0| \bar{q} \gamma_5 (i n\cdot\overleftrightarrow{D})  c|D'_1(p,\epsilon)\rangle &= g_{D'_1}\, m^2_{D'_1}(n\cdot\epsilon^{D'_1})\,,
\end{align}
where the numerical inputs for $\{f_{D_1}, f_{D'_1}, g_{D_1}, g_{D'_1}\}$ are adopted from the QCD sum rule determinations in \cite{Gubernari:2203.08493}. By implementing the identities from the QCD EOMs at tree-level,
\begin{align}
 \langle 0|i \partial^\mu \,(\bar{q} \gamma_\mu \gamma_5 i \overleftrightarrow{D}_\nu\, c ) |R \rangle &= -m_c \langle 0|\bar{q}  \gamma_5 i \overleftrightarrow{D}_\nu\, c |R\rangle  \,, \nonumber\\
 \langle 0| \partial^\mu  (\bar{q} \sigma_{\mu\nu} \gamma_5 \, c ) |R \rangle &=  \langle 0| \bar{q} \gamma_5 i \overleftrightarrow{D}_\nu\, c + m_c \,\bar{q} \gamma_\nu \gamma_5\, c|R\rangle  \,, \nonumber\\
 \langle 0| \partial^\mu  (\bar{q} \sigma_{\mu\rho} \gamma_5 i\overleftrightarrow{D}_\nu\, c ) |R \rangle &=  \langle 0| \bar{q}i \overleftrightarrow{D}_\rho i \overleftrightarrow{D}_\nu\gamma_5 \, c + m_c \,\bar{q} \gamma_\rho \gamma_5 i \overleftrightarrow{D}_\nu\, c|R\rangle  \,, \nonumber\\
 \langle 0| g^{\mu\nu}  (\bar{q} \sigma_{\mu\rho} \gamma_5 \overleftrightarrow{D}_\nu\, c ) |R \rangle &=  \langle 0| m_c\,\bar{q}\gamma_\rho \gamma_5 \, c |R\rangle  \,,
\end{align}
the following matching relations can be established        
\begin{align}
 f_1 &= f_{D_1} \,, & f_2 &= f_{D'_1} \,, \nonumber \\
 r_1 &= -f_{D_1} \,, & r_2 &= -f_{D'_1} \,, \nonumber \\
 g_1 &= -\frac{2 m_c}{m_{D_1}} g_{D_1} + \frac{m_{D_1}}{m_c} f_{D_1} \,, & g_2 &= -\frac{2 m_c}{m_{D'_1}} g_{D'_1} + \frac{m_{D'_1}}{m_c} f_{D'_1} \,, \nonumber \\
 h_1 &= -\frac{m_c}{m_{D_1}} (f_{D_1} + g_{D_1}) \,, & h_2 &= -\frac{m_c}{m_{D'_1}} (f_{D'_1} + g_{D'_1}) \,.
\end{align}
Consequently, the numerical values for the complete set of these eight hadronic inputs are directly determined, as summarized in Table~\ref{tab:inputs}.

The relevant energy scales are varied following the standard conventions in exclusive heavy-hadron decays \cite{Beneke:1804.04962,Beneke:2008.12494,Shen:2009.02723}. Specifically, the matching scale $\mu_h$ is scanned within the interval $[m_b/2, 2\,m_b]$ with a central choice of $m_b$, while the factorization scale $\mu$ is identified as the characteristic hard-collinear scale and varied in the range of $(1.5 \pm 0.5) \, {\rm GeV}$.

\begin{figure}[htbp!]
    \centering
    \begin{tabular}{cc}
    \includegraphics[width=0.48\textwidth]{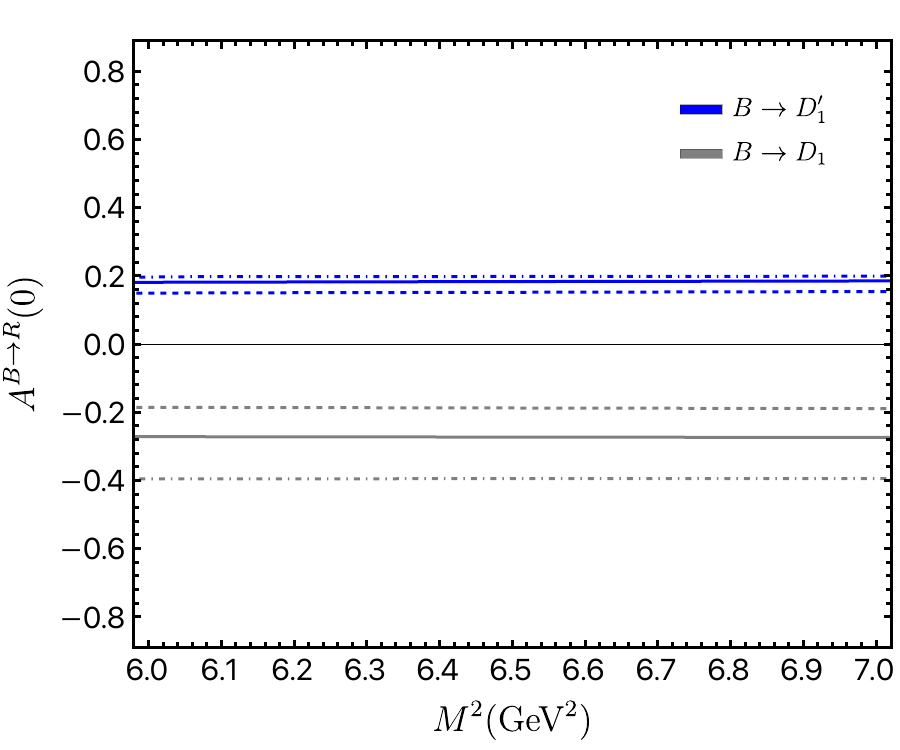}  &
    \includegraphics[width=0.48\textwidth]{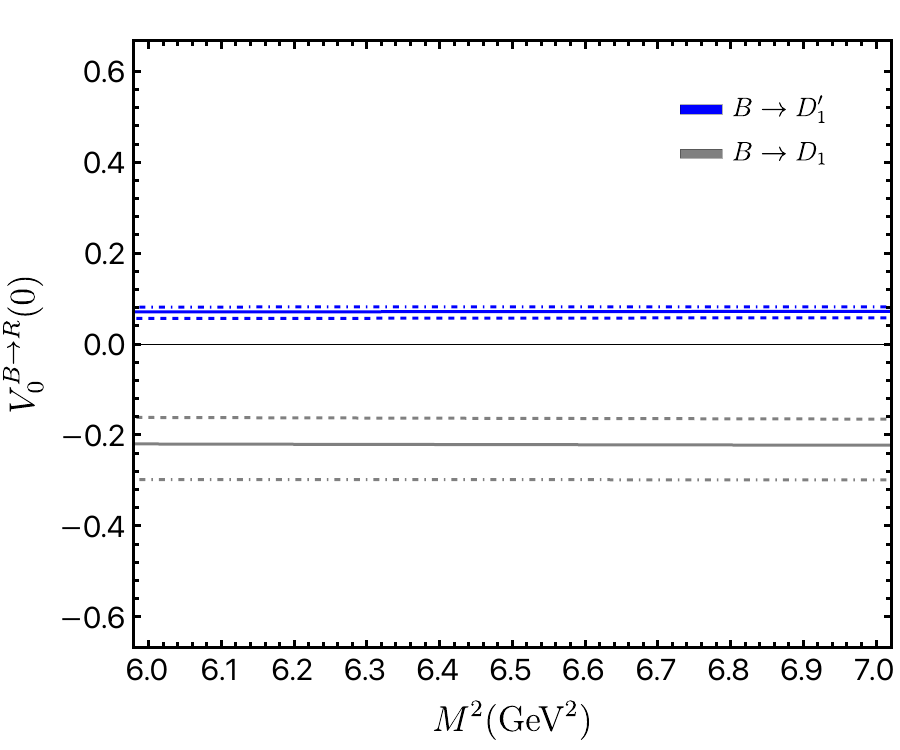}  \\[1.25em]
    \includegraphics[width=0.48\textwidth]{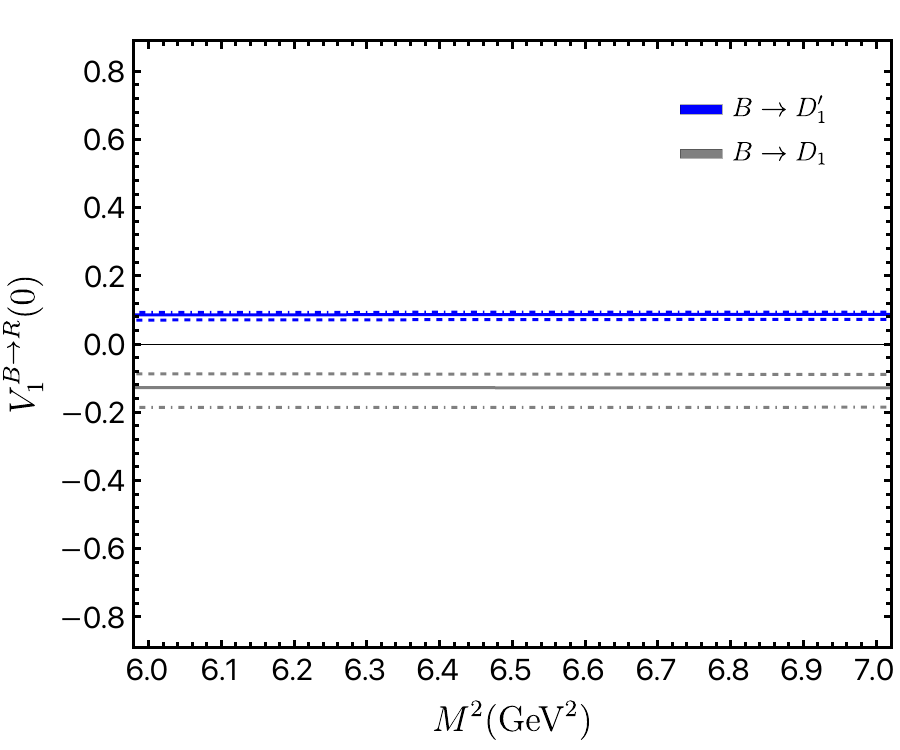}  &
    \includegraphics[width=0.48\textwidth]{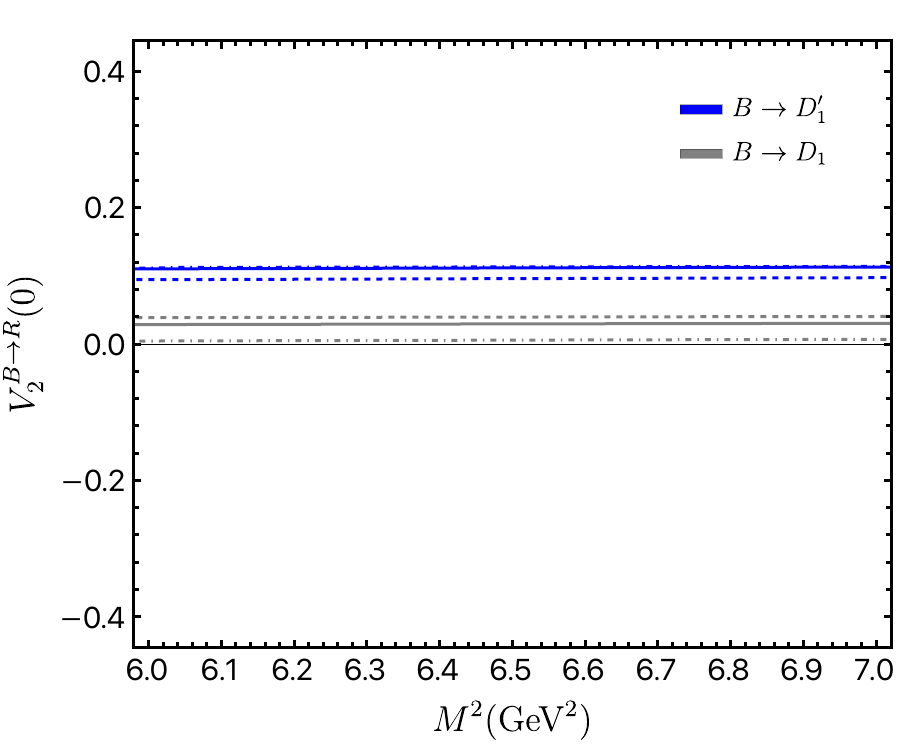}  \\[1.25em]
    \end{tabular}
    \caption{Dependence of the form factors at \(q^2=0\) on the Borel
parameter \(M^2\). The solid, dashed, and dot-dashed curves correspond
to the central, upper, and lower choices of the lower threshold \(s_0\),
respectively.}
    \label{fig:Borel}
\end{figure}

\begin{figure}[htbp!]    
    \centering
    \begin{tabular}{cc}
    \includegraphics[width=0.48\textwidth]{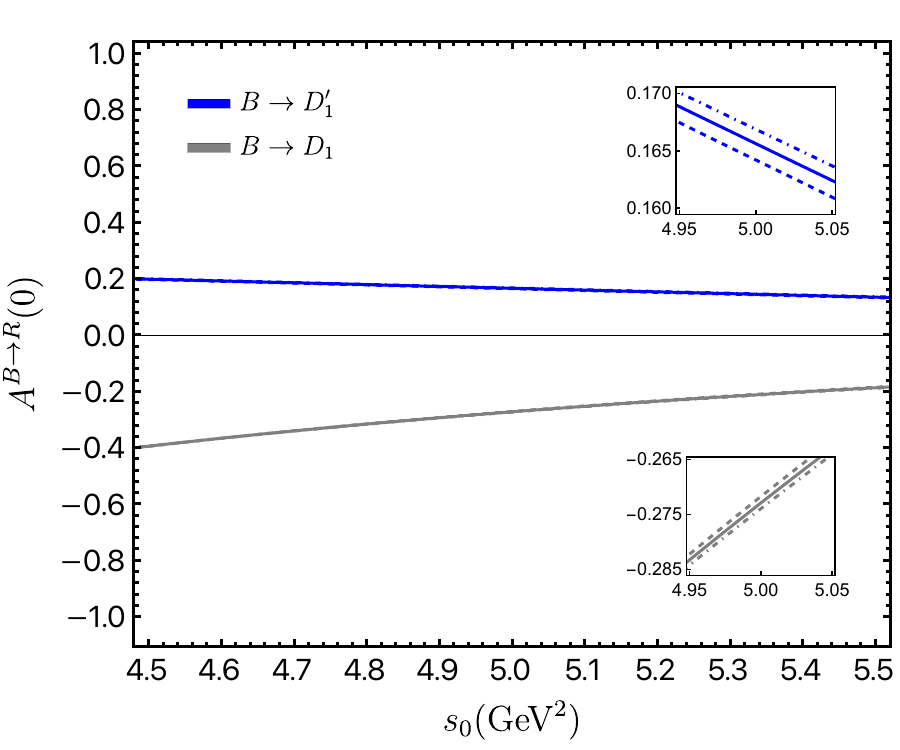}  &
    \includegraphics[width=0.48\textwidth]{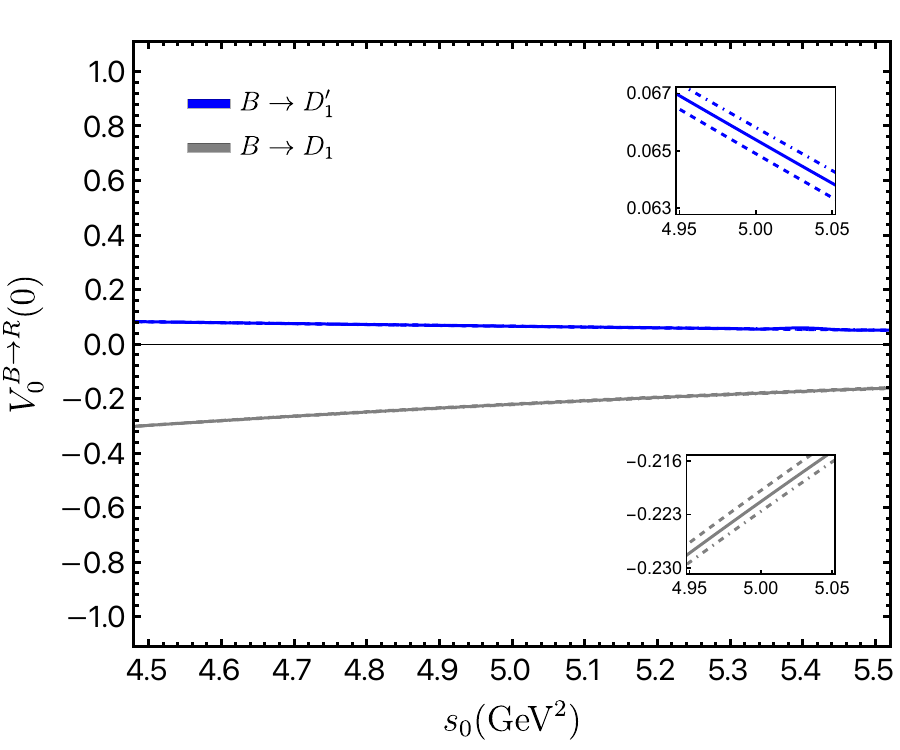}  \\[1.25em]
    \includegraphics[width=0.48\textwidth]{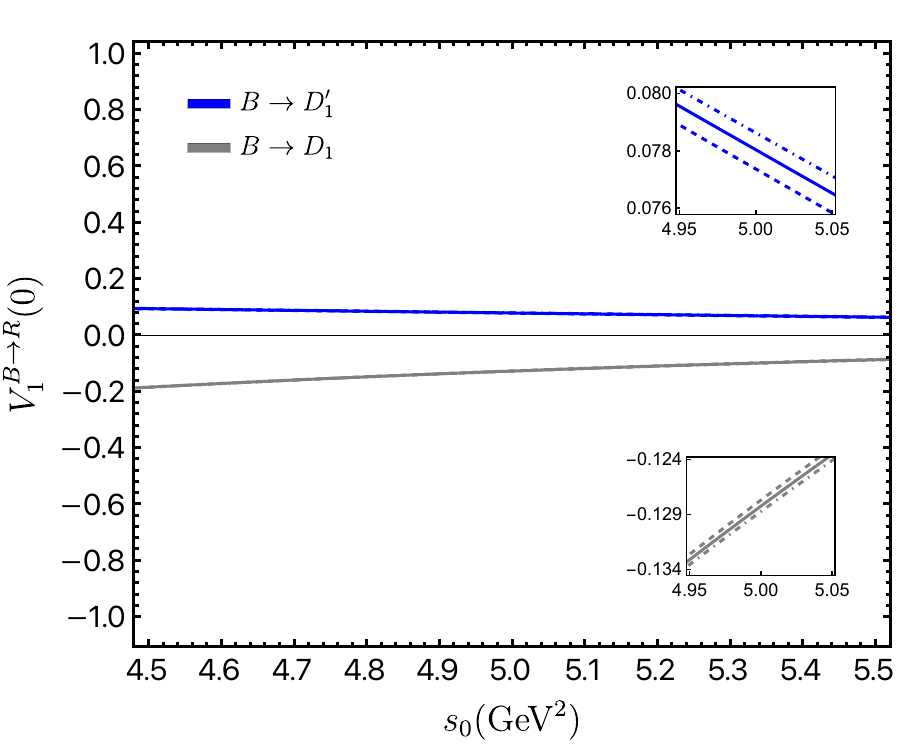}  &
    \includegraphics[width=0.48\textwidth]{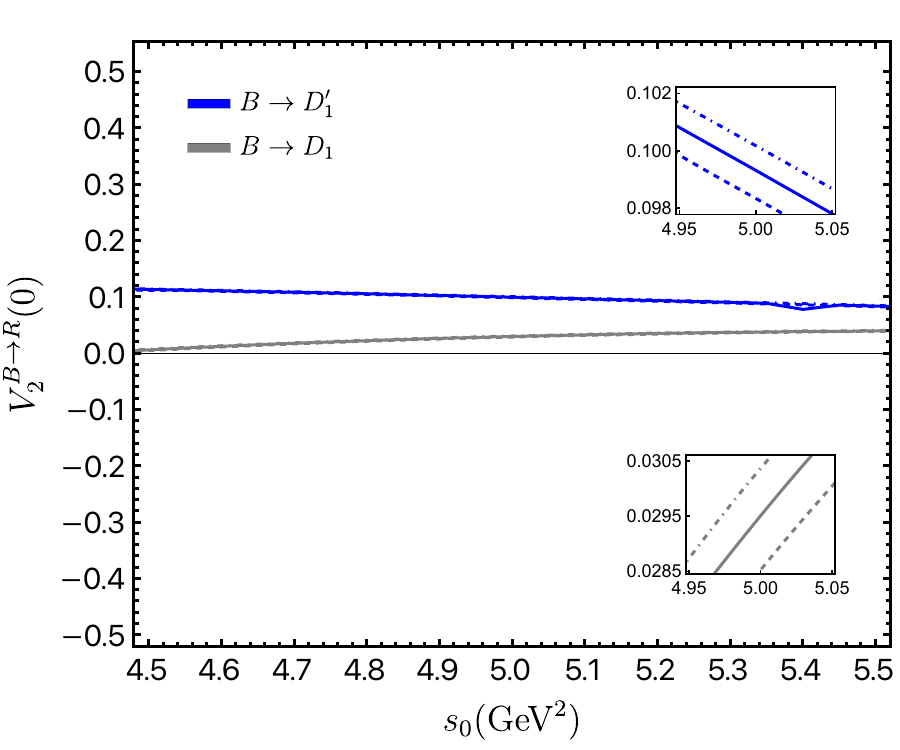}  \\[1.25em]
    \end{tabular}
    \caption{Dependence of the form factors at \(q^2=0\) on the lower
threshold \(s_0\). The solid, dashed, and dot-dashed curves correspond
to the central, upper, and lower choices of the Borel parameter,
respectively. The insets show the region around the central value
\(s_0=5.0\,{\rm GeV}^2\).}
    \label{fig:thre}
\end{figure}

For the numerical evaluation of LCSRs, the Borel parameter $M^2$ must
be carefully chosen to balance the convergence of the OPE series against
the suppression of higher-state contributions. Compared with the
ground-state sum rules \cite{cui:2301.12391}, a moderately larger $M^2$ is adopted here to
relax the exponential suppression on the $P$-wave states of interest,
while still ensuring adequate suppression of the higher excitations. We
vary $M^2$ within the interval
$6\;\text{GeV}^2 \le M^2 \le 7\;\text{GeV}^2$, where the
form factors exhibit only a mild variation, as shown in
Fig.~\ref{fig:Borel}.
The lower threshold $s_0$ determines the upper limit of the
ground-state sum rule and, after subtraction, the lower limit of the
spectral integral for the $P$-wave contribution. We choose $s_0$ within
the mass gap between the ground-state $D^{(*)}$ mesons,
$m_{D^{(*)}}^2\lesssim4.0\,{\rm GeV}^2$, and the $D_1^{(\prime)}$
states, $m_{D_1^{(\prime)}}^2\simeq5.9\,{\rm GeV}^2$, and adopt
$s_0=(5.0\pm0.5)\,{\rm GeV}^2$. As shown in Fig.~\ref{fig:thre}, the form factors show a mild
variation over the adopted range of $s_0$. Their visible dependence on $s_0$ reflects
the fact that this parameter directly sets the lower limit of the
spectral integral for the $P$-wave contribution. This variation is
included in the theoretical uncertainties. The upper threshold is set to $s_1 = 9.0 \pm 0.6\;\text{GeV}^2$,
which lies at the boundary of the predicted $2P$ axial-vector
excitations and provides a sufficiently broad duality interval for the
$D_1^{(\prime)}$ extraction.

\FloatBarrier

\subsection{Form factors and phenomenological predictions}
To extrapolate the form factors from the large-recoil region to the
full kinematic domain, we employ the BCL parameterization. To ensure
rapid convergence of the series expansion, the momentum transfer
squared $q^2$ is transformed into a new expansion parameter $z(q^2)$ via
\begin{align}
  z(q^2) &= \frac{\sqrt{t_+ - q^2} - \sqrt{t_+ - t_0}}
                  {\sqrt{t_+ - q^2} + \sqrt{t_+ - t_0}} \,,
\end{align}
where
\begin{align}
  t_+ = (m_B + m_{D_1^{(\prime)}})^2 \,, \qquad
  t_0  = t_+\!\left(1 - \sqrt{1 - q^2_{\max}/t_+}\right) ,
\end{align}
with $q^2_{\max} = (m_B - m_{D_1^{(\prime)}})^2$.
Each form factor included in the fit is parameterized as
\begin{align}
  F_i(q^2) &= \frac{1}{1 - q^2/m^2_{\mathrm{pole},i}}
  \left[ a_0^{(i)} + a_1^{(i)}
    \bigl(z(q^2) - z(0)\bigr) \right] ,
  \label{eq:BCL}
\end{align}
where $m_{\mathrm{pole},i}$ denotes the mass of the lowest-lying $B_c$
resonance with the quantum numbers matching those of the corresponding
hadronic current. Following \cite{Gubernari:2203.08493}, we adopt the 
lowest-lying $B_c$ masses $m_{0^+} = 6.70$~GeV for $V_0^{BR}$,
$m_{1^-} = 6.33$~GeV for $V_1^{BR}$ and $V_2^{BR}$, as determined
from lattice QCD. Owing to the relation in
\eqref{large-recoil-relations}, $A^{BR}$ is not included as an
independent form factor in the fit, and its $q^2$ dependence is
reconstructed from $V_1^{BR}$.
The $z$-expansion is truncated at $\mathcal{O}(z)$, yielding two
free parameters $\{a_0^{(i)}, a_1^{(i)}\}$ per fitted form factor.

\begin{table}[htbp!]
\centering
\renewcommand{\arraystretch}{1.5} 
\setlength{\tabcolsep}{5pt} 
\resizebox{\textwidth}{!}{
\begin{tabular}{|l|c|cccccccccc|}
\hline
Parameters & Values & $a_0^{V_1}$ & $a_0^{V_1'}$ & $a_0^{V_2}$ & $a_0^{V_2'}$ & $a_1^{V_0}$ & $a_1^{V_0'}$ & $a_1^{V_1}$ & $a_1^{V_1'}$ & $a_1^{V_2}$ & $a_1^{V_2'}$ \\
\hline
$a_0^{V_1}$   & $-0.114_{\,-0.074}^{\,+0.074}$ & 1.000 & 0.363 & 0.854 & 0.409 & -0.148 & 0.013 & 0.598 & 0.223 & 0.488 & 0.154 \\
$a_0^{V_1'}$  & $0.084_{\,-0.037}^{\,+0.038}$  &       & 1.000 & 0.559 & 0.745 & -0.041 & -0.313 & 0.113 & 0.393 & -0.061 & 0.415 \\
$a_0^{V_2}$   & $0.045_{\,-0.119}^{\,+0.120}$  &       &       & 1.000 & 0.568 & -0.109 & -0.137 & 0.393 & 0.171 & 0.149 & 0.122 \\
$a_0^{V_2'}$  & $0.109_{\,-0.040}^{\,+0.040}$ &       &       &       & 1.000 & -0.089 & -0.311 & 0.122 & 0.032 & -0.028 & -0.122 \\
\hline
$a_1^{V_0}$   & $-0.229_{\,-1.058}^{\,+1.049}$  &       &       &       &       & 1.000 & -0.018 & 0.114 & -0.053 & 0.043 & -0.026 \\
$a_1^{V_0'}$  & $-0.818_{\,-0.644}^{\,+0.653}$  &       &       &       &       &       & 1.000 & 0.082 & 0.113 & 0.184 & 0.108 \\
$a_1^{V_1}$   & $-2.859_{\,-1.369}^{\,+1.364}$ &       &       &       &       &       &       & 1.000 & 0.153 & 0.535 & 0.116 \\
$a_1^{V_1'}$  & $-0.092_{\,-0.702}^{\,+0.708}$  &       &       &       &       &       &       &       & 1.000 & 0.170 & 0.637 \\
$a_1^{V_2}$   & $-6.717_{\,-2.519}^{\,+2.549}$ &       &       &       &       &       &       &       &       & 1.000 & 0.123 \\
$a_1^{V_2'}$  & $-0.711_{\,-1.211}^{\,+1.245}$ &       &       &       &       &       &       &       &       &       & 1.000 \\
\hline
\end{tabular}
}
\caption{BCL coefficients and the correlation matrix
for the ten independent fit parameters.}
\label{tab:bcl_para}
\end{table}

The BCL expansion coefficients $a_k^{(i)}$ are determined by a joint
$\chi^2$ fit to the LCSR data points evaluated at
$q^2 = \{-4, -2,0,2,4 \}$~GeV$^2$ within the large-recoil
window where the sum rules are reliable. We note that, in addition to
the relation \eqref{large-recoil-relations}, the kinematic
constraint among $V_1^{BR}(0)$, $V_2^{BR}(0)$, and $V_0^{BR}(0)$ in
\eqref{eq:kinematic-constraints} must also be imposed in the fit.
To avoid artificial singularities in the covariance matrix and ensure
a clean statistical correlation, we eliminate the corresponding
redundant coefficients by imposing these constraints. The remaining ten
independent parameters,
$\{a_0^{V_1}, a_0^{V_1'}, a_0^{V_2}, a_0^{V_2'},
  a_1^{V_0}, a_1^{V_0'},
  a_1^{V_1}, a_1^{V_1'}, a_1^{V_2}, a_1^{V_2'}\}$,
pertaining to the $B \to D_1$ and $B \to D_1'$ channels are then fitted
simultaneously. Through this global fit, the theoretical correlations
induced by shared inputs---such as the Borel parameter,
thresholds, and the LCDA parameters---are properly propagated into
the output covariance matrix. The resulting correlation matrix
is presented in Table~\ref{tab:bcl_para}.

As observed in Table~\ref{tab:bcl_para}, sizeable positive correlations
are present among the normalization parameters within the same decay
channel, reaching $+0.854$ between $a_0^{V_1}$ and $a_0^{V_2}$, and
$+0.745$ between $a_0^{V_1'}$ and $a_0^{V_2'}$. This behavior is
natural, since form factors governing the same final state share common
nonperturbative inputs in their LCSRs, so that their theoretical
uncertainties tend to move coherently. By contrast, although the two
channels share many theoretical input parameters, the inter-channel
correlations do not show a uniformly strong pattern. This is consistent
with the expected separation of the two physical axial-vector states and
indicates that the disentangling procedure works effectively in the
present analysis.

\begin{figure}[htbp!]
    \centering
    \begin{tabular}{cc}
    \includegraphics[width=0.48\textwidth]{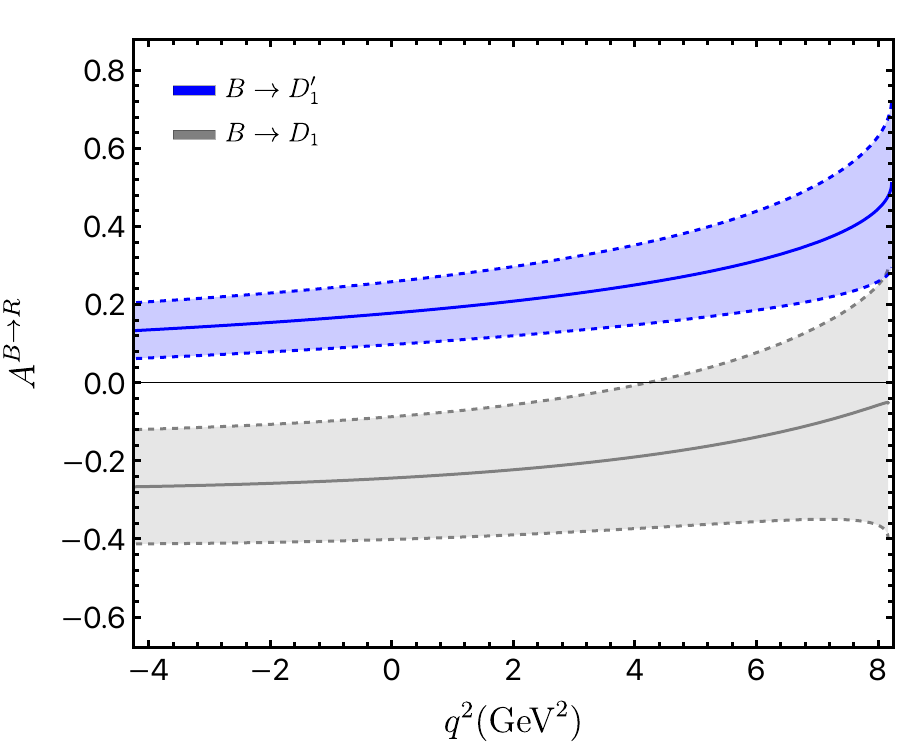}  &
    \includegraphics[width=0.48\textwidth]{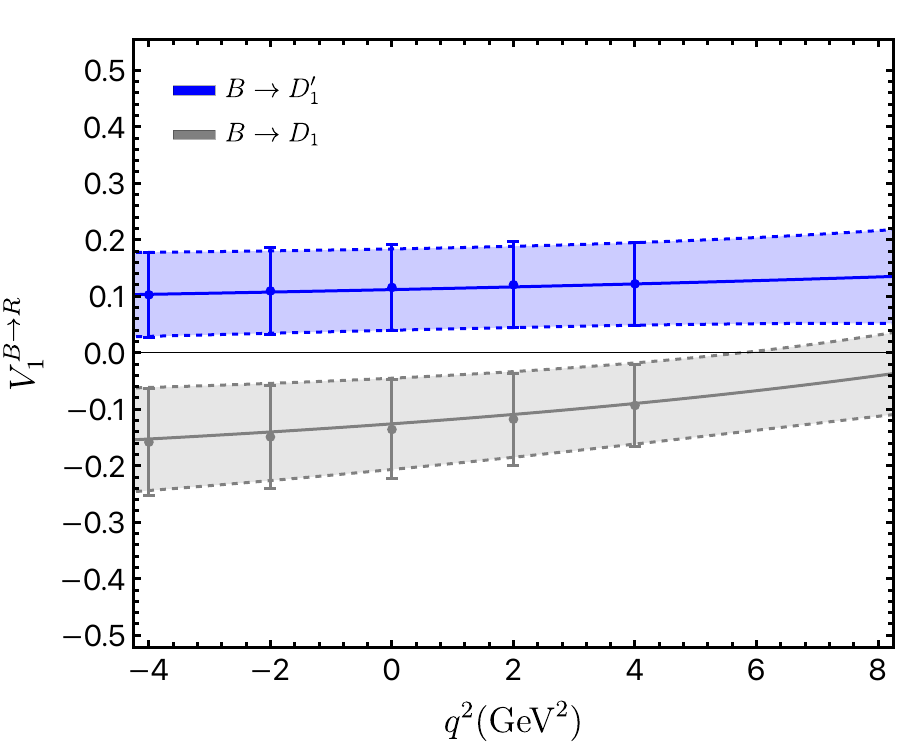}  \\[1.25em]
    \includegraphics[width=0.48\textwidth]{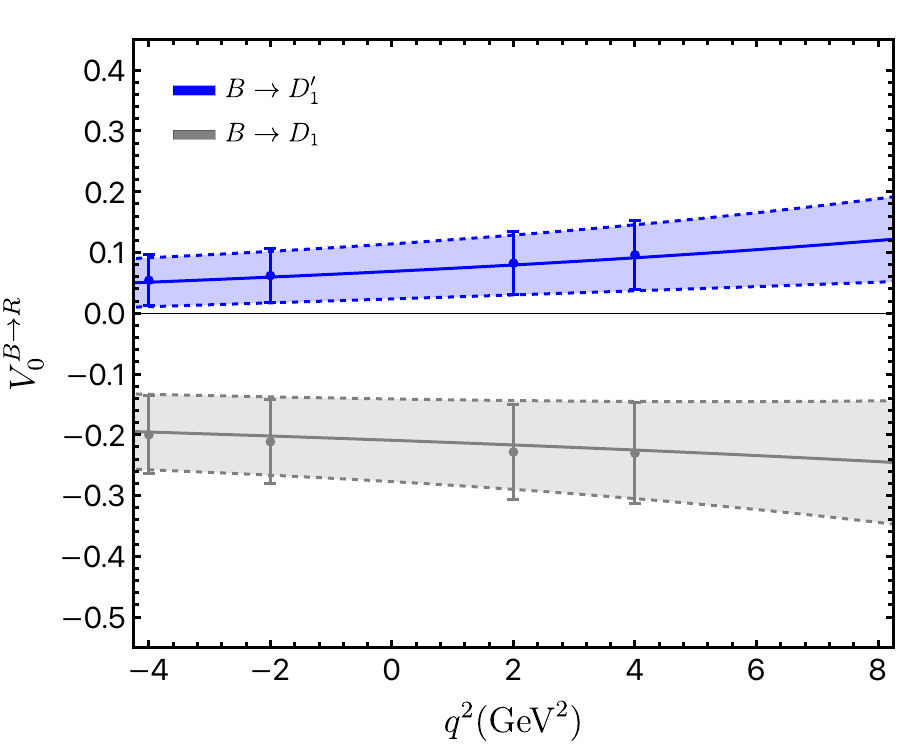}  &
    \includegraphics[width=0.48\textwidth]{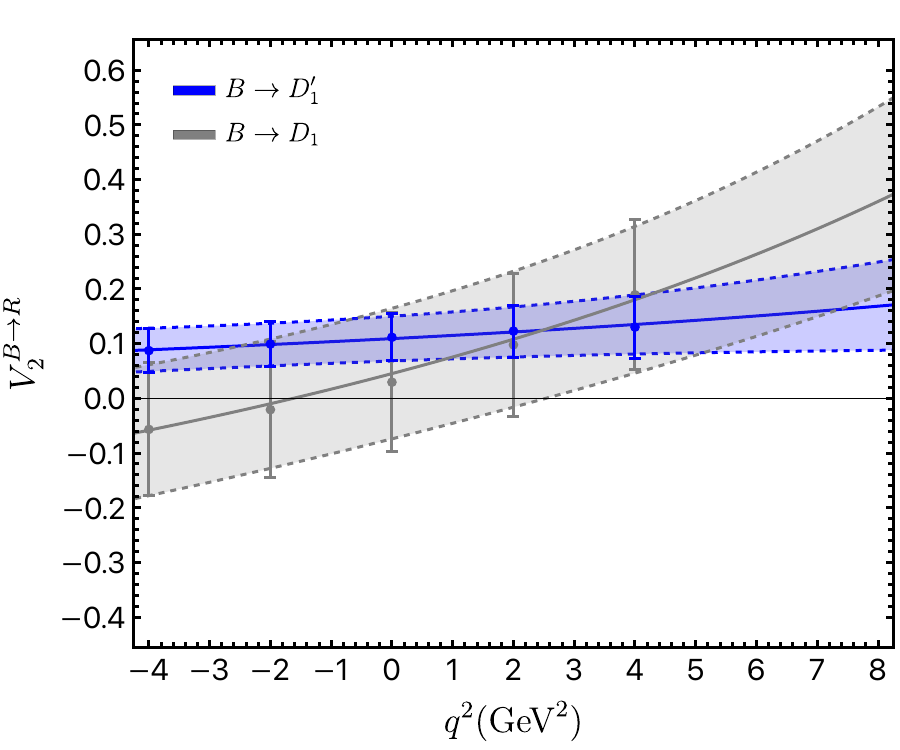}  \\[1.25em]
    \end{tabular}
    \caption{The $q^2$ dependence of $B \to D_1$ and $B \to D'_1$ form factors.}
    \label{fig:ffs}
\end{figure}

\begin{figure}[htbp]
    \centering
    \includegraphics[width=0.49\linewidth]{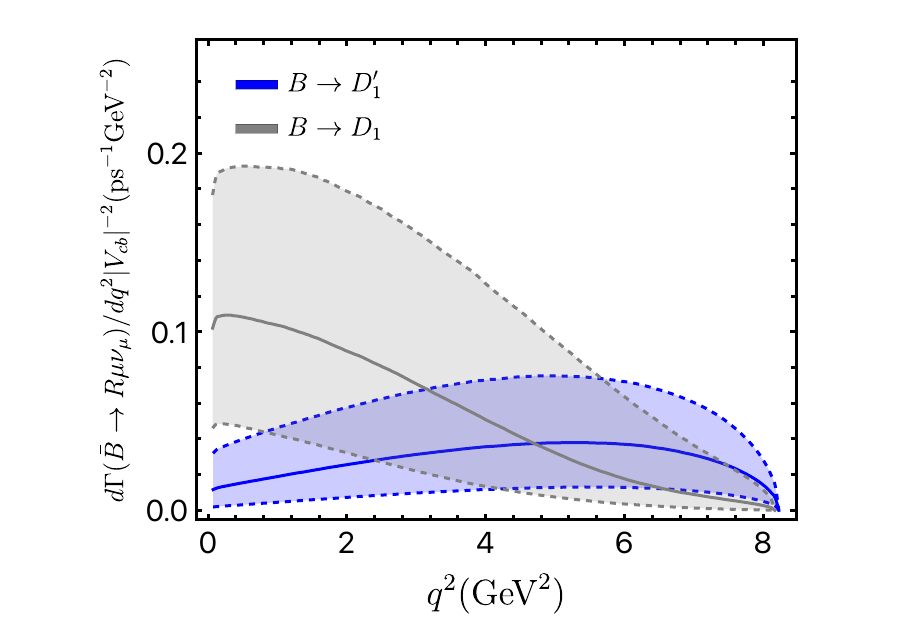}
    \hfill
    \includegraphics[width=0.49\linewidth]{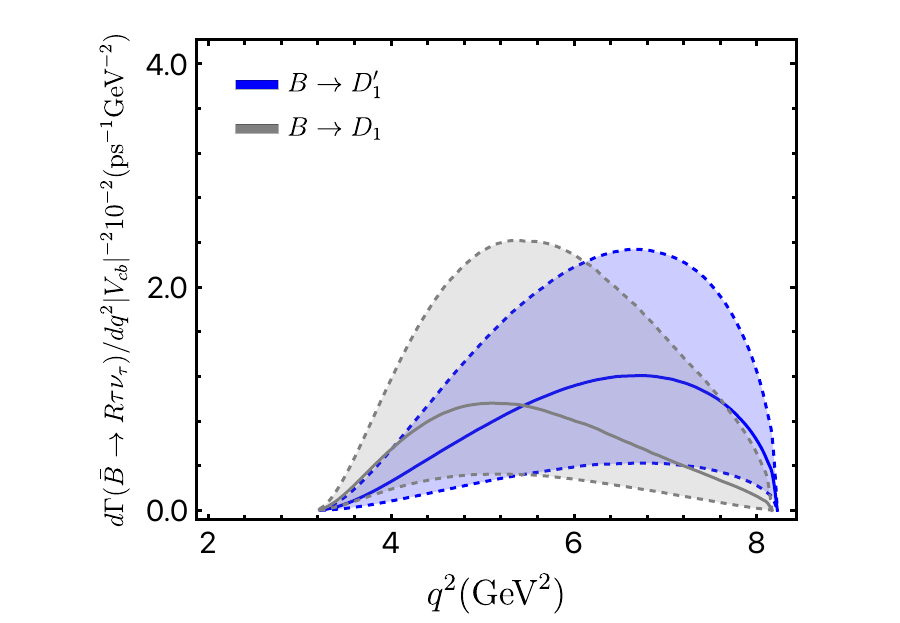}
    \caption{Differential decay widths normalized by \(|V_{cb}|^2\) for the
light-lepton and tauonic modes.}
    \label{fig:dw}
\end{figure}

With the BCL parameters fully determined, we display the resulting
$q^2$ dependence of the form factors over the physical semileptonic
region in Fig.~\ref{fig:ffs}. The uncertainty bands are obtained from
the covariance matrix of the fitted BCL coefficients in
Table~\ref{tab:bcl_para}. Having obtained the complete physical form factors, we proceed to
evaluate the phenomenological observables for the semileptonic decays
$B \to R \ell \bar{\nu}_\ell$. The differential decay width can be written in terms of the
helicity amplitudes as
\begin{align}
  \frac{d\Gamma(B \to R \ell \bar{\nu}_\ell)}{dq^2} =&
  \frac{G_F^2 |V_{cb}|^2 q^2 \lambda_R^{1/2}}{192 \pi^3 m_B^3}
  \left( 1 - \frac{m_\ell^2}{q^2} \right)^2
 \nonumber \\  &\times \bigg[  \left( 1 + \frac{m_\ell^2}{2q^2} \right)
  \left( |H_+|^2 + |H_-|^2 + |H_0|^2 \right) 
  + \frac{3m_\ell^2}{2q^2} |H_t|^2 \bigg] \,,
  \label{eq:dGamma}
\end{align}
where $G_F$ is the Fermi coupling constant, $V_{cb}$ is the CKM matrix
element, $m_\ell$ is the lepton mass ($\ell = e, \mu, \tau$), and
$\lambda_R = \lambda(m_B^2,m_R^2,q^2)$ is the standard K\"all\'en
function,
\begin{align}
  \lambda(a,b,c)=a^2+b^2+c^2-2(ab+ac+bc)\, .
\end{align}
With the form-factor convention in \eqref{qcdff}, the helicity
amplitudes are given by
\begin{align}
  H_\pm(q^2)
  =&\ i\left[
  (m_B+m_R)V_1^{BR}(q^2)
  \mp {\lambda_R^{1/2}\over m_B+m_R}A^{BR}(q^2)
  \right],
  \nonumber\\
  H_0(q^2)
  =&\ {i(m_B+m_R)\over 2m_R\sqrt{q^2}}
  \left[
  (m_R^2+q^2-m_B^2)V_1^{BR}(q^2)
  +{\lambda_R\over (m_B+m_R)^2}V_2^{BR}(q^2)
  \right],
  \nonumber\\
  H_t(q^2)
  =&\ -i{\lambda_R^{1/2}\over\sqrt{q^2}}V_0^{BR}(q^2)\, .
  \label{eq:helicity-amplitudes}
\end{align}
The resulting differential distributions for the light-lepton and
tauonic modes are shown in Fig.~\ref{fig:dw}.

The total decay widths are obtained by integrating \eqref{eq:dGamma}
over the physical region $m_\ell^2 \le q^2 \le (m_B-m_R)^2$. The
corresponding branching fractions are then given by
\begin{align}
  \mathcal{B}(B \to R \ell \bar{\nu}_\ell)
  = \tau_B \int_{m_\ell^2}^{(m_B-m_R)^2} dq^2\,
  \frac{d\Gamma(B \to R \ell \bar{\nu}_\ell)}{dq^2} \, .
  \label{eq:branching-fraction}
\end{align}
For the branching fractions, $m_B$ and $\tau_B$ are taken to be the
charged-$B$ mass and lifetime, with
$\tau_{B^\pm}=1.638\,{\rm ps}$.
For the light-lepton modes, we obtain
\begin{align}
  \mathcal{B}(B \to D_1 \ell \bar{\nu}_\ell)
  &= \left(1.13^{+1.31}_{-0.69}\right)\times 10^{-3}\,,
  \nonumber\\
  \mathcal{B}(B \to D'_1 \ell \bar{\nu}_\ell)
  &= \left(0.62^{+0.65}_{-0.42}\right)\times 10^{-3}\,.
  \qquad (\ell=e,\mu)
  \label{eq:light-lepton-branching-fractions}
\end{align}
The numerical results are of order \(10^{-3}\). Despite
having central values lower than the corresponding PDG averages
extracted from \(B\to D^*\pi\ell\bar\nu_\ell\), the predictions remain
compatible with these PDG averages within the theoreticalx uncertainties
\cite{PDG:2024}. The sizeable uncertainties mainly reflect the present limited knowledge
of the nonperturbative inputs. Among them, the decay constants of the
interpolating currents play a dominant role, since they strongly affect
the overall size of the form factors extracted from the sum rules. More
precise determinations of these decay constants would therefore directly
reduce the uncertainties of the branching fractions.

We further consider the LFU ratios
\begin{align}
  R(X) =
  \frac{\mathcal{B}(B \to X \tau \bar{\nu}_\tau)}
       {\mathcal{B}(B \to X \ell \bar{\nu}_\ell)}
  =
  \frac{\int_{m_\tau^2}^{(m_B-m_X)^2} dq^2\,
  d\Gamma(B \to X \tau \bar{\nu}_\tau)/dq^2}
       {\int_{m_\ell^2}^{(m_B-m_X)^2} dq^2\,
  d\Gamma(B \to X \ell \bar{\nu}_\ell)/dq^2}\,,
  \quad (X=D_1,D'_1)
  \label{eq:lfu-ratio}
\end{align}
where $\ell=e,\mu$. The dependence on $|V_{cb}|$ cancels in these ratios, which are also
less affected by the nonperturbative uncertainties from the form factors
than the branching fractions. The numerical
results read
\begin{align}
  R(D_1) &= 0.070^{+0.028}_{-0.018}\,,
  \nonumber\\
  R(D'_1) &= 0.159^{+0.032}_{-0.025}\, .
  \label{eq:lfu-results}
\end{align}
At present there is no direct experimental determination of these LFU
ratios. Compared with the tree-level LCSR estimates in \cite{Gubernari:2203.08493},
$R(D_1)=0.10\pm0.02$ and $R(D'_1)=0.10\pm0.03$, our results are still
compatible within uncertainties, with a lower central value for $D_1$
and a higher one for $D'_1$. Both ratios are much smaller than unity, as expected from the
phase-space suppression of the tauonic modes. Since parametric
uncertainties partly cancel in the ratios, the resulting uncertainties
are relatively small compared with those of the branching
fractions.

\section{Conclusion}
\label{sec5}

In this work, we have calculated the $B\to D_1(2420)$ and
$B\to D'_1(2430)$ form factors at $\mathcal{O}(\alpha_s)$ using
$B$-meson LCSRs within SCET. We first matched the QCD transition currents
onto the A0-type and B1-type operators in $\mathrm{SCET}_{\mathrm{I}}$.
For a finite charm-quark mass, the $\mathrm{SCET}_{\mathrm{I}}$ current
basis contains the additional A1-type operator
$O_{\parallel}^{(\mathrm{A1},m_c)}$, which is absent in heavy-to-light
transitions with a massless quark in the final state. The corresponding
vacuum-to-$B$-meson correlation functions were then factorized into
jet functions and $B$-meson LCDAs in
$\mathrm{SCET}_{\mathrm{II}}$. Matching their spectral representations
to the hadronic dispersion relations gave the sum rules for the
longitudinal and transverse effective form factors.

Since each of the original interpolating currents couples to both
axial-vector states, we constructed separate linear combinations for
$D_1$ and $D'_1$. The coefficients of these combinations were fixed by
requiring the current associated with either state to have a vanishing
matrix element with the other state. The ground-state contributions were
then removed by subtracting the corresponding $B\to D^{(*)}$ sum rules.
The resulting form factors remain stable under the allowed variations
of the Borel parameter and the duality thresholds. We also found that no
uniformly strong pattern exists in the correlations between the two
channels, which supports the effectiveness of the state-separation
procedure in the present calculation.

The A0-type and B1-type contributions reproduce the leading-power
large-energy relations among the QCD form factors. The additional
A1-type operator generates the longitudinal form factor
$\xi^R_{\parallel,m_c}$, whose sum rule depends only on the $B$-meson
LCDA $\phi_B^+$. At the same order, the corresponding $\phi_B^+$
contribution vanishes in the transverse A0-type correlation function,
and hence no transverse counterpart of $\xi^R_{\parallel,m_c}$ is
generated. The resulting physical form factors were then extrapolated to the full
semileptonic region using the BCL parameterization.

The predicted branching fractions for the light-lepton modes are of
order $10^{-3}$, as shown in
\eqref{eq:light-lepton-branching-fractions}. Their central values are below the corresponding PDG averages extracted
from $B\to D^*\pi\ell\bar{\nu}_\ell$, but the predictions remain
compatible with these averages within the present large uncertainties
\cite{PDG:2024}. At the present precision, this comparison mainly tests
the overall size of the decay rates. Future measurements of the differential $q^2$ distributions would
provide further information on the shapes of form factors and help separate
the $D_1$ and $D'_1$ contributions in the
$B\to D^*\pi\ell\bar{\nu}_\ell$ channel. The LFU ratios obtained in \eqref{eq:lfu-results} are compatible
within uncertainties with the previous LCSR estimates
\cite{Gubernari:2203.08493}. No direct experimental measurements of
these ratios are currently available. The dependence on $|V_{cb}|$
cancels in these ratios, while the nonperturbative uncertainties from
the form factors are partially reduced. Measurements at Belle~II and
LHCb would directly test LFU in semileptonic decays to excited charm
mesons and improve the treatment of the $D^{**}$ contributions in
measurements of $R(D^{(*)})$.

The largest numerical uncertainties currently arise from the decay
constants entering the combinations of interpolating currents. More
precise nonperturbative determinations of these quantities would
therefore improve the phenomenological predictions. A further
improvement concerns the treatment of the ground-state contribution.
In this work, the $P$-wave form factors were obtained by subtracting the
ground-state sum rules from sum rules containing both the ground and
excited states. If lattice QCD determinations of the ground-state decay
constants associated with the tensor interpolating currents with derivatives used here became available, the known $D^{(*)}$ pole contributions could
instead be subtracted explicitly on the hadronic side. The $D_1^{(\prime)}$ contributions could then be extracted from a
single sum rule. This would eliminate the additional duality
approximation introduced by the separate ground-state sum rule and
reduce the corresponding uncertainty. Other
improvements include the calculation of subleading-power and
higher-twist contributions and the treatment of finite-width effects
for the broad $D'_1$ state.

\acknowledgments

We are grateful to Yu-Ming Wang for helpful discussions. This
work is supported by the National Natural Science
Foundation of China with Grant No. 12475097 and No.
12535006, and by the Natural Science Foundation of
Tianjin with Grant No. 25JCZDJC01190, and by
the Fundamental Research Funds for the Central Universities
with Grant No. 63261180.

\appendix
\section{Spectral representations of the effective LCDAs}
\label{secA1}

For completeness, we collect the spectral representations of the
effective LCDAs entering the SCET sum rules. For compactness, we use
\begin{align}
  \theta_1=\theta(\omega'-\omega-\omega_c)\,,
  \qquad
  \theta_2=\theta(\omega+\omega_c-\omega')\,.
\end{align}
In the main text, we use the compact notation
\begin{align}
  \phi^+_{\parallel,\rm eff}(\omega',\mu)
  \equiv m_c\,\phi^+_{m,\rm eff}(\omega',\mu)\, .
\end{align}
With the conventions adopted here, the non-starred effective LCDAs
coincide with those obtained in
\cite{cui:2301.12391}.

The effective LCDAs entering \eqref{specj1} are
\begin{align}
  \phi^-_{\parallel,\rm eff}(\omega',\mu)
  ={}&
  \bigg\{\phi_B^-(\omega'-\omega_c,\mu)
  +{\alpha_s C_F\over 4\pi}
  \int_0^\infty d\omega\,
  \bigg[
  \left(\rho_1(\omega,\omega')+\rho_2(\omega,\omega')\right)
  {d\over d\omega}\phi_B^-(\omega,\mu)
  \nonumber\\
  &
  +\rho_3(\omega,\omega')\,\phi_B^-(\omega,\mu)
  \bigg]\bigg\}\theta(\omega'-\omega_c)
  \nonumber\\
  &-{\alpha_s C_F\over 4\pi}\,
  \omega_c\left(6\ln{\mu^2\over m_c^2}+8\right)
  {d\over d\omega'}
  \left[
    \theta(\omega'-\omega_c)
    \phi_B^-(\omega'-\omega_c,\mu)
  \right] ,
  \label{app:phi-par-minus}
  \\
  \phi^+_{m,\rm eff}(\omega',\mu)
  =&-\theta(\omega'-\omega_c){\alpha_s C_F\over 4\pi}
  \int_0^\infty d\omega\,\bigg\{
  \theta_2\left[
  \left({\omega'-\omega_c\over \omega'}\right)^2
  \ln{(\omega+\omega_c-\omega')\omega_c
  \over(\omega'-\omega_c)^2}
  -\ln{\omega_c\over \omega'}\right]
  \nonumber\\
  &
  +\theta_1\left[
  \left({\omega'-\omega_c\over \omega'}\right)^2
  \ln{\omega'-\omega-\omega_c\over \omega'-\omega_c}
  -\ln{\omega'-\omega\over \omega'}
  -{\omega\,\omega_c\over \omega'(\omega'-\omega)}
  \right]\bigg\}
  {d\over d\omega}{\phi_B^+(\omega,\mu)\over \omega}\, .
  \label{app:phi-m-eff}
\end{align}
The kernels in \eqref{app:phi-par-minus} are
\begin{align}
  \rho_1(\omega,\omega')
  =&\ \theta_1
  \ln{\omega'-\omega-\omega_c\over \omega'-\omega_c}
  \left[
  \left(1-{\omega_c\over\omega'}\right)^2
  -4\ln{\omega_c\over\omega'-\omega-\omega_c}
  -2\ln{\mu^2\over n\cdot p\,\omega'}-2
  \right],
  \nonumber\\
  \rho_2(\omega,\omega')
  =&\ \theta_2\bigg[
  \ln{\omega+\omega_c-\omega'\over \omega'-\omega_c}
  \left(
  \left(1-{\omega_c\over\omega'}\right)^2
  +2\ln{\mu^2\omega'\over n\cdot p\,(\omega'-\omega_c)^2}
  +2
  \right)
  \nonumber\\
  &\hspace{1.8cm}
  +2{\rm Li}_2\left({\omega_c\over\omega_c-\omega'}\right)
  +\ln{\omega_c\over\omega'-\omega_c}
  \left(
  \left(1-{\omega_c\over\omega'}\right)^2
  -2\ln{\omega_c\over\omega'}-2
  \right)
  \nonumber\\
  &\hspace{1.8cm}
  -\ln^2{\mu^2\over n\cdot p\,(\omega'-\omega_c)}
  -{5\pi^2\over 6}-{\omega_c\over\omega'}\bigg],
  \nonumber\\
  \rho_3(\omega,\omega')
  =&\ {\theta_1\over\omega+\omega_c-\omega'}
  \left[
  4\ln{\omega'-\omega\over\omega_c}
  -\left({\omega'-\omega-\omega_c\over\omega'-\omega}\right)^2
  \right]
  +{2\theta_2\over\omega}.
  \label{app:rho123}
\end{align}

The additional effective LCDAs entering \eqref{specj2} are
\begin{align}
  \phi^{*-}_{\parallel,\rm eff}(\omega',\mu)
  ={}&{1\over 2}\phi^-_{\parallel,\rm eff}(\omega',\mu)
  +{1\over 2}{\alpha_s C_F\over 4\pi}
  \bigg\{
  \bigg[
  -{52\over9}-{10\omega_c\over3\omega'}
  +{4\omega_c^2\over3\omega'^2}
  \nonumber\\
  &+{4\omega_c^3-12\omega_c^2\omega'\over3\omega'^3}
  \ln\left|{\omega_c-\omega'\over\omega_c}\right|
  -{8\over3}
  \ln\left|{\mu^2\over n\cdot p\,(\omega_c-\omega')}\right|
  \bigg]\phi_B^-(\omega'-\omega_c,\mu)
  \nonumber\\
  &  +\int_0^\infty d\omega\,
  \theta(\omega'-\omega_c)\,
  {4\omega_c^3-12\omega_c^2\omega'+8\omega'^3
  \over 3(\omega+\omega_c-\omega')\omega'^3}
  \phi_B^-(\omega,\mu)
  \bigg\},
  \label{app:phi-star-minus}
  \\
  \phi^{*+}_{\parallel,\rm eff}(\omega',\mu)
  =&\ m_c\,{\alpha_s C_F\over 4\pi}
  \bigg\{
  \theta(\omega'-\omega_c)
  {\omega'(-3\omega'+4\omega_c)
  -(\omega'^2-5\omega'\omega_c+4\omega_c^2)
  \ln\left|{\omega_c\over\omega'-\omega_c}\right|
  \over 6\omega'^3}
  \nonumber\\
  &
  \times\phi_B^+(\omega'-\omega_c,\mu)
  \nonumber\\
  &+\int_0^\infty d\omega\,
  \bigg[
  {(\omega'-4\omega_c)(\omega'-\omega_c)^2
  \over 6\omega\,\omega'^3(\omega'-\omega-\omega_c)}
  \theta(\omega'-\omega_c)
  \nonumber\\
  &
  -{\omega'^2-2\omega'\omega+\omega^2
  -5\omega'\omega_c+5\omega\omega_c+4\omega_c^2
  \over 6\omega(\omega'-\omega)^3}
  \theta(\omega'-\omega-\omega_c)
  \bigg]\phi_B^+(\omega,\mu)
  \bigg\}\,.
  \label{app:phi-star-plus}
\end{align}
For the A1-type contribution, the effective LCDA enters with
the prefactor $-m_c/(n\cdot p)$.

The difference between the transverse and longitudinal A0-type spectral
representations is encoded in
\begin{align}
  \Delta\phi^-_{\rm eff}(\omega',\mu)
  ={}& \theta(\omega'-\omega_c){\alpha_s C_F\over 4\pi}
  \int_0^\infty d\omega\,\bigg\{
  \theta_1\,{\omega'^2-\omega_c^2\over\omega'^2}
  \ln{\omega'-\omega_c-\omega\over\omega'-\omega_c}
  \nonumber\\
  &
  +\theta_2\bigg[
  \ln{\mu^2\over m_c^2}
  +{\omega'^2-\omega_c^2\over\omega'^2}
  \ln{\omega_c(\omega+\omega_c-\omega')
  \over(\omega'-\omega_c)^2}
  +{\omega_c\over\omega'}
  \bigg]\bigg\}
  {d\over d\omega}\phi_B^-(\omega,\mu)\,,
  \label{app:delta-phi-minus}
  \\
  \Delta\phi^{*-}_{\rm eff}(\omega',\mu)
  ={}& {1\over 2}{\alpha_s C_F\over 4\pi}
  \bigg\{
  \left[
  -{1\over9}+{\omega_c\over6\omega'}
  -{2\omega_c^2\over3\omega'^2}
  -{1\over6}\ln{\mu^2\over n\cdot p\,\omega'}
  \right.
  \nonumber\\
  &\left.
  -{\omega_c^2(-3\omega'+4\omega_c)\over6\omega'^3}
  \ln\left|{\omega_c-\omega'\over\omega_c}\right|
  +{1\over6}\ln\left|{\omega'-\omega_c\over\omega'}\right|
  \right]\phi_B^-(\omega'-\omega_c,\mu)
  \nonumber\\
  &+\int_0^\infty d\omega\,
  {1\over \omega'-\omega-\omega_c}
  \left[
  {1\over6}\theta(\omega')
  -{\omega_c^2(-3\omega'+4\omega_c)\over6\omega'^3}
  \theta(\omega'-\omega_c)
  -{1\over6}\theta(\omega_c-\omega')
  \right]
  \nonumber\\
  &\times\phi_B^-(\omega,\mu)
  \bigg\}\, .
  \label{app:delta-phi-star-minus}
\end{align}
\bibliographystyle{JHEP}
\bibliography{ref}

@article{Beneke:9804241,
    author = "Beneke, M.",
    title = "{A Quark mass definition adequate for threshold problems}",
    eprint = "hep-ph/9804241",
    archivePrefix = "arXiv",
    reportNumber = "CERN-TH-98-120",
    doi = "10.1016/S0370-2693(98)00741-2",
    journal = "Phys. Lett. B",
    volume = "434",
    pages = "115--125",
    year = "1998"
}

@article{Hill:0404217,
    author = "Hill, R. J. and Becher, T. and Lee, Seung J. and Neubert, M.",
    title = "{Sudakov resummation for subleading SCET currents and heavy-to-light form-factors}",
    eprint = "hep-ph/0404217",
    archivePrefix = "arXiv",
    reportNumber = "SLAC-PUB-10412, CLNS-04-1865",
    doi = "10.1088/1126-6708/2004/07/081",
    journal = "JHEP",
    volume = "07",
    pages = "081",
    year = "2004"
}

@article{Broadhurst:9410240,
    author = "Broadhurst, David J. and Grozin, A. G.",
    title = "{Matching QCD and HQET heavy-light currents at two loops and beyond}",
    eprint = "hep-ph/9410240",
    archivePrefix = "arXiv",
    
    reportNumber = "OUT-4102-52",
    doi = "10.1103/PhysRevD.52.4082",
    journal = "Phys. Rev. D",
    volume = "52",
    pages = "4082--4098",
    year = "1995"
}

@article{Broadhurst:9908362,
    author = "Broadhurst, David J. and Grozin, A. G.",
    title = "{Two loop renormalization of the effective field theory of a static quark}",
    eprint = "hep-ph/9908362",
    archivePrefix = "arXiv",
    reportNumber = "OUT-4102-30",
    doi = "10.1016/0370-2693(91)90532-U",
    journal = "Phys. Lett. B",
    volume = "267",
    pages = "105--110",
    year = "1991"
}

@article{Ji:1991pr,
    author = "Ji, Xiang-Dong and Musolf, M. J.",
    title = "{Subleading logarithmic mass dependence in heavy meson form-factors}",
    doi = "10.1016/0370-2693(91)91916-J",
    journal = "Phys. Lett. B",
    volume = "257",
    pages = "409--413",
    year = "1991"
}

@article{Beneke:1110.3228,
    author = "Beneke, M. and Rohrwild, J.",
    title = "{B meson distribution amplitude from $B\to \gamma \ell \nu$}",
    eprint = "1110.3228",
    archivePrefix = "arXiv",
    primaryClass = "hep-ph",
    reportNumber = "TTK-11-51, SFB-CPP-11-56",
    doi = "10.1140/epjc/s10052-011-1818-8",
    journal = "Eur. Phys. J. C",
    volume = "71",
    pages = "1818",
    year = "2011"
}

@article{Leibovich:0303099,
    author = "Leibovich, Adam K. and Ligeti, Zoltan and Wise, Mark B.",
    title = "{Comment on Quark Masses in SCET}",
    eprint = "hep-ph/0303099",
    archivePrefix = "arXiv",
    reportNumber = "FERMILAB-PUB-03-035-T, LBNL-52196, CALT-68-2432",
    doi = "10.1016/S0370-2693(03)00565-3",
    journal = "Phys. Lett. B",
    volume = "564",
    pages = "231--234",
    year = "2003"
}

@article{Feldmann:2203.15679,
    author = {Feldmann, Thorsten and L{\"u}ghausen, Philip and van Dyk, Danny},
    title = "{Systematic parametrization of the leading $B$-meson light-cone distribution amplitude}",
    eprint = "2203.15679",
    archivePrefix = "arXiv",
    primaryClass = "hep-ph",
    reportNumber = "SI-HEP-2022-05, P3H-22-029, TUM-HEP-1388/22",
    doi = "10.1007/JHEP10(2022)162",
    journal = "JHEP",
    volume = "10",
    pages = "162",
    year = "2022"
}

@article{Briceno:1706.06223,
    author = "Briceno, Raul A. and Dudek, Jozef J. and Young, Ross D.",
    title = "{Scattering processes and resonances from lattice QCD}",
    eprint = "1706.06223",
    archivePrefix = "arXiv",
    primaryClass = "hep-lat",
    reportNumber = "JLAB-THY-17-2495, ADP-17-28-T1034",
    doi = "10.1103/RevModPhys.90.025001",
    journal = "Rev. Mod. Phys.",
    volume = "90",
    number = "2",
    pages = "025001",
    year = "2018"
}

@article{Wang:2111.11811,
    author = "Wang, Chao and Wang, Yu-Ming and Wei, Yan-Bing",
    title = "{QCD factorization for the four-body leptonic $B$-meson decays}",
    eprint = "2111.11811",
    archivePrefix = "arXiv",
    primaryClass = "hep-ph",
    reportNumber = "TUM-HEP-1332/21",
    doi = "10.1007/JHEP02(2022)141",
    journal = "JHEP",
    volume = "02",
    pages = "141",
    year = "2022"
}

@article{Wang:1803.06667,
    author = "Wang, Yu-Ming and Shen, Yue-Long",
    title = "{Subleading-power corrections to the radiative leptonic $B \to \gamma \ell \nu$ decay in QCD}",
    eprint = "1803.06667",
    archivePrefix = "arXiv",
    primaryClass = "hep-ph",
    doi = "10.1007/JHEP05(2018)184",
    journal = "JHEP",
    volume = "05",
    pages = "184",
    year = "2018"
}

@article{Wang:1606.03080,
    author = "Wang, Yu-Ming",
    title = "{Factorization and dispersion relations for radiative leptonic $B$ decay}",
    eprint = "1606.03080",
    archivePrefix = "arXiv",
    primaryClass = "hep-ph",
    reportNumber = "UWTHPH-2016-10",
    doi = "10.1007/JHEP09(2016)159",
    journal = "JHEP",
    volume = "09",
    pages = "159",
    year = "2016"
}

@article{Wang:1506.00667,
    author = "Wang, Yu-Ming and Shen, Yue-Long",
    title = "{QCD corrections to $B\to \pi$ form factors from light-cone sum rules}",
    eprint = "1506.00667",
    archivePrefix = "arXiv",
    primaryClass = "hep-ph",
    reportNumber = "TUM-HEP-995-15",
    doi = "10.1016/j.nuclphysb.2015.07.016",
    journal = "Nucl. Phys. B",
    volume = "898",
    pages = "563--604",
    year = "2015"
}

@article{Bell:1308.6114,
    author = "Bell, Guido and Feldmann, Thorsten and Wang, Yu-Ming and Yip, Matthew W Y",
    title = "{Light-Cone Distribution Amplitudes for Heavy-Quark Hadrons}",
    eprint = "1308.6114",
    archivePrefix = "arXiv",
    primaryClass = "hep-ph",
    reportNumber = "OUTP-13-15P, SI-HEP-2013-05, QFET-2013-05, TTK-13-19, SFB-CPP-13-60, IPPP-13-55, DCPT-13-110, TUM-HEP-899-13",
    doi = "10.1007/JHEP11(2013)191",
    journal = "JHEP",
    volume = "11",
    pages = "191",
    year = "2013"
}

@article{Beneke:9402364,
    author = "Beneke, M. and Braun, Vladimir M.",
    title = "{Renormalons, pole masses and residual masses in HQET}",
    eprint = "hep-ph/9402364",
    archivePrefix = "arXiv",
    reportNumber = "MPI-PHT-94-9, UM-TH-94-4",
    doi = "10.1016/0550-3213(94)90314-X",
    journal = "Nucl. Phys. B",
    volume = "426",
    pages = "301--343",
    year = "1994"
}

@article{Bigi:9402360,
    author = "Bigi, Ikaros I. Y. and Shifman, Mikhail A. and Uraltsev, N. G. and Vainshtein, A. I.",
    title = "{The Pole mass of the heavy quark. Perturbation theory and beyond}",
    eprint = "hep-ph/9402360",
    archivePrefix = "arXiv",
    reportNumber = "TPI-MINN-94-4-T, UMN-TH-1239-94, CERN-TH-7171-94, UND-HEP-94-BIG03",
    doi = "10.1103/PhysRevD.50.2234",
    journal = "Phys. Rev. D",
    volume = "50",
    pages = "2234--2246",
    year = "1994"
}

@article{Shen:2009.02723,
    author = "Shen, Yue-Long and Wang, Yu-Ming and Wei, Yan-Bing",
    title = "{Double radiative bottom-meson decays in SCET}",
    eprint = "2009.02723",
    archivePrefix = "arXiv",
    primaryClass = "hep-ph",
    doi = "10.1007/JHEP12(2020)169",
    journal = "JHEP",
    volume = "12",
    pages = "169",
    year = "2020"
}

@article{MILC:1503.07237,
    author = "Bailey, Jon A. and others",
    collaboration = "Fermilab Lattice, MILC",
    title = "{$B\to D\ell\nu$ form factors at nonzero recoil and $|V_{cb}|$ from $2+1$-flavor lattice QCD}",
    eprint = "1503.07237",
    archivePrefix = "arXiv",
    primaryClass = "hep-lat",
    reportNumber = "FERMILAB-PUB-15-107-T",
    doi = "10.1103/PhysRevD.92.034506",
    journal = "Phys. Rev. D",
    volume = "92",
    number = "3",
    pages = "034506",
    year = "2015"
}

@article{MILC:2105.14019,
    author = "Bazavov, A. and others",
    collaboration = "Fermilab Lattice, MILC",
    title = "{Semileptonic form factors for $B\to D^*\ell\nu$ at nonzero recoil from $2+1$-flavor lattice QCD}",
    eprint = "2105.14019",
    archivePrefix = "arXiv",
    primaryClass = "hep-lat",
    reportNumber = "FERMILAB-PUB-21-261-T~, FERMILAB-PUB-21/261-T",
    doi = "10.1140/epjc/s10052-022-10984-9",
    journal = "Eur. Phys. J. C",
    volume = "82",
    number = "12",
    pages = "1141",
    year = "2022",
    note = "[Erratum: Eur.Phys.J.C 83, 21 (2023)]"
}

@article{HPQCD:2304.03137,
    author = "Harrison, Judd and Davies, Christine T. H.",
    collaboration = "HPQCD",
    title = "{$B\to D^*$ and $B_s\to D_s^*$ form factors from lattice QCD}",
    eprint = "2304.03137",
    archivePrefix = "arXiv",
    primaryClass = "hep-lat",
    doi = "10.1103/PhysRevD.109.094515",
    journal = "Phys. Rev. D",
    volume = "109",
    number = "9",
    pages = "094515",
    year = "2024"
}

@article{Gao:2112.12674,
    author = "Gao, Jing and Huber, Tobias and Ji, Yao and Wang, Chao and Wang, Yu-Ming and Wei, Yan-Bing",
    title = "{$B\to D\ell\nu_\ell$ form factors beyond leading power and extraction of $|V_{cb}|$ and $R(D)$}",
    eprint = "2112.12674",
    archivePrefix = "arXiv",
    primaryClass = "hep-ph",
    reportNumber = "TUM-HEP-1331/21, SI-HEP-2021-37, P3H-21-103",
    doi = "10.1007/JHEP05(2022)024",
    journal = "JHEP",
    volume = "05",
    pages = "024",
    year = "2022"
}

@article{Wang:1701.06810,
    author = {Wang, Yu-Ming and Wei, Yan-Bing and Shen, Yue-Long and L{\"u}, Cai-Dian},
    title = "{Perturbative corrections to $B\to D$ form factors in QCD}",
    eprint = "1701.06810",
    archivePrefix = "arXiv",
    primaryClass = "hep-ph",
    reportNumber = "UWTHPH-2016-29",
    doi = "10.1007/JHEP06(2017)062",
    journal = "JHEP",
    volume = "06",
    pages = "062",
    year = "2017"
}

@article{Faller:0809.0222,
    author = "Faller, S. and Khodjamirian, A. and Klein, Ch. and Mannel, Th.",
    title = "{$B\to D^{(*)}$ form factors from QCD light-cone sum rules}",
    eprint = "0809.0222",
    archivePrefix = "arXiv",
    primaryClass = "hep-ph",
    reportNumber = "SI-HEP-2008-13",
    doi = "10.1140/epjc/s10052-009-0968-4",
    journal = "Eur. Phys. J. C",
    volume = "60",
    pages = "603--615",
    year = "2009"
}

@article{FLAG:2411.04268,
    author = "Aoki, Y. and others",
    collaboration = "Flavour Lattice Averaging Group (FLAG)",
    title = "{FLAG review 2024}",
    eprint = "2411.04268",
    archivePrefix = "arXiv",
    primaryClass = "hep-lat",
    reportNumber = "CERN-TH-2024-192, FERMILAB-PUB-24-0785-T",
    doi = "10.1103/nfzp-p5dn",
    journal = "Phys. Rev. D",
    volume = "113",
    number = "1",
    pages = "014508",
    year = "2026"
}

@article{Gambino:2020jvv,
    author = "Gambino, P. and others",
    title = "{Challenges in semileptonic $B$ decays}",
    eprint = "2006.07287",
    archivePrefix = "arXiv",
    primaryClass = "hep-ph",
    reportNumber = "FERMILAB-PUB-20-235-T",
    doi = "10.1140/epjc/s10052-020-08490-x",
    journal = "Eur. Phys. J. C",
    volume = "80",
    number = "10",
    pages = "966",
    year = "2020"
}

@article{Gao:2019lta,
    author = {Gao, Jing and L\"u, Cai-Dian and Shen, Yue-Long and Wang, Yu-Ming and Wei, Yan-Bing},
    title = "{Precision calculations of $B \to V$ form factors from soft-collinear effective theory sum rules on the light-cone}",
    eprint = "1907.11092",
    archivePrefix = "arXiv",
    primaryClass = "hep-ph",
    doi = "10.1103/PhysRevD.101.074035",
    journal = "Phys. Rev. D",
    volume = "101",
    number = "7",
    pages = "074035",
    year = "2020"
}

@article{Beneke:1808.04742,
    author = "Beneke, Martin and Garny, Mathias and Szafron, Robert and Wang, Jian",
    title = "{Anomalous dimension of subleading-power $N$-jet operators. Part II}",
    eprint = "1808.04742",
    archivePrefix = "arXiv",
    primaryClass = "hep-ph",
    reportNumber = "TUM-HEP-1155/18",
    doi = "10.1007/JHEP11(2018)112",
    journal = "JHEP",
    volume = "11",
    pages = "112",
    year = "2018"
}

@article{Isgur:1991wq,
    author = "Isgur, Nathan and Wise, Mark B.",
    title = "{Spectroscopy with heavy quark symmetry}",
    reportNumber = "CEBAF-TH-91-01, WM-91-101, CALT-68-1704",
    doi = "10.1103/PhysRevLett.66.1130",
    journal = "Phys. Rev. Lett.",
    volume = "66",
    pages = "1130--1133",
    year = "1991"
}

@article{Bauer:0109045,
    author = "Bauer, Christian W. and Pirjol, Dan and Stewart, Iain W.",
    title = "{Soft collinear factorization in effective field theory}",
    eprint = "hep-ph/0109045",
    archivePrefix = "arXiv",
    reportNumber = "UCSD-PTH-01-15",
    doi = "10.1103/PhysRevD.65.054022",
    journal = "Phys. Rev. D",
    volume = "65",
    pages = "054022",
    year = "2002"
}

@article{Beneke:1411.3132,
    author = "Beneke, M. and Maier, A. and Piclum, J. and Rauh, T.",
    title = "{The bottom-quark mass from non-relativistic sum rules at NNNLO}",
    eprint = "1411.3132",
    archivePrefix = "arXiv",
    primaryClass = "hep-ph",
    reportNumber = "TUM-HEP-966-14, SFB-CPP-14-88",
    doi = "10.1016/j.nuclphysb.2014.12.001",
    journal = "Nucl. Phys. B",
    volume = "891",
    pages = "42--72",
    year = "2015"
}

@article{Beneke:1804.04962,
    author = "Beneke, M. and Braun, V. M. and Ji, Yao and Wei, Yan-Bing",
    title = "{Radiative leptonic decay $B\to \gamma \ell \nu_\ell$ with subleading power corrections}",
    eprint = "1804.04962",
    archivePrefix = "arXiv",
    primaryClass = "hep-ph",
    reportNumber = "TUM-HEP-1135-18",
    doi = "10.1007/JHEP07(2018)154",
    journal = "JHEP",
    volume = "07",
    pages = "154",
    year = "2018"
}

@article{Beneke:0311335,
    author = "Beneke, M. and Feldmann, T.",
    title = "{Factorization of heavy-to-light form factors in soft-collinear effective theory}",
    eprint = "hep-ph/0311335",
    archivePrefix = "arXiv",
    reportNumber = "PITHA-03-11, CERN-TH-2003-286",
    doi = "10.1016/j.nuclphysb.2004.02.033",
    journal = "Nucl. Phys. B",
    volume = "685",
    pages = "249--296",
    year = "2004"
}

@article{Barca:2508.09006,
    author = "Barca, Lorenzo",
    title = "{Current-enhanced excited states in lattice QCD three-point functions}",
    eprint = "2508.09006",
    archivePrefix = "arXiv",
    primaryClass = "hep-lat",
    reportNumber = "DESY-25-115",
    doi = "10.1103/69yc-d74z",
    journal = "Phys. Rev. D",
    volume = "112",
    number = "9",
    pages = "L091503",
    year = "2025"
}

@article{Blossier:0902.1265,
    author = "Blossier, Benoit and Della Morte, Michele and von Hippel, Georg and Mendes, Tereza and Sommer, Rainer",
    title = "{Generalized eigenvalue method for energies and matrix elements}",
    eprint = "0902.1265",
    archivePrefix = "arXiv",
    primaryClass = "hep-lat",
    reportNumber = "DESY-09-014, SFB-CPP-09-10, MKPH-T-09-01, LPT-ORSAY-09-05",
    doi = "10.1088/1126-6708/2009/04/094",
    journal = "JHEP",
    volume = "04",
    pages = "094",
    year = "2009"
}

@article{Luscher:1990ck,
    author = "Luscher, Martin and Wolff, Ulli",
    title = "{Elastic scattering matrix from finite-volume simulations}",
    reportNumber = "DESY-90-010",
    doi = "10.1016/0550-3213(90)90540-T",
    journal = "Nucl. Phys. B",
    volume = "339",
    pages = "222--252",
    year = "1990"
}

@article{Michael:1985ne,
    author = "Michael, Christopher",
    title = "{Adjoint Sources in Lattice Gauge Theory}",
    reportNumber = "LTH 125",
    doi = "10.1016/0550-3213(85)90297-4",
    journal = "Nucl. Phys. B",
    volume = "259",
    pages = "58--76",
    year = "1985"
}

@article{Belle:1904.08794,
    author = "Abdesselam, A. and others",
    collaboration = "Belle",
    title = "{Measurement of $\mathcal{R}(D)$ and $\mathcal{R}(D^{\ast})$ with a semileptonic tagging method}",
    eprint = "1904.08794",
    archivePrefix = "arXiv",
    primaryClass = "hep-ex",
    month = "4",
    year = "2019"
}

@article{BaBar:1205.5442,
    author = "Lees, J. P. and others",
    collaboration = "BaBar",
    title = "{Evidence for an excess of $\bar{B} \to D^{(*)} \tau^-\bar{\nu}_\tau$ decays}",
    eprint = "1205.5442",
    archivePrefix = "arXiv",
    primaryClass = "hep-ex",
    reportNumber = "BABAR-PUB-12-012, SLAC-PUB-15028",
    doi = "10.1103/PhysRevLett.109.101802",
    journal = "Phys. Rev. Lett.",
    volume = "109",
    pages = "101802",
    year = "2012"
}

@article{Cheng:1110.2249,
    author = "Cheng, Hai-Yang",
    title = "{Revisiting Axial-Vector Meson Mixing}",
    eprint = "1110.2249",
    archivePrefix = "arXiv",
    primaryClass = "hep-ph",
    doi = "10.1016/j.physletb.2011.12.013",
    journal = "Phys. Lett. B",
    volume = "707",
    pages = "116--120",
    year = "2012"
}

@article{Suzuki:1993yc,
    author = "Suzuki, M.",
    title = "{Strange axial - vector mesons}",
    doi = "10.1103/PhysRevD.47.1252",
    journal = "Phys. Rev. D",
    volume = "47",
    pages = "1252--1255",
    year = "1993"
}

@article{Neubert:1993mb,
    author = "Neubert, Matthias",
    title = "{Heavy quark symmetry}",
    eprint = "hep-ph/9306320",
    archivePrefix = "arXiv",
    reportNumber = "SLAC-PUB-6263",
    doi = "10.1016/0370-1573(94)90091-4",
    journal = "Phys. Rept.",
    volume = "245",
    pages = "259--396",
    year = "1994"
}

@article{Lu:1991px,
    author = "Lu, Ming and Wise, Mark B. and Isgur, Nathan",
    title = "{Heavy quark symmetry and $D_1(2420)\to D^*\pi$ decay}",
    reportNumber = "CEBAF-TH-91-16, CALT-68-1741",
    doi = "10.1103/PhysRevD.45.1553",
    journal = "Phys. Rev. D",
    volume = "45",
    pages = "1553--1556",
    year = "1992"
}

@article{huang:2212.11624,
    author = "Cui, Bo-Yan and Huang, Yong-Kang and Shen, Yue-Long and Wang, Chao and Wang, Yu-Ming",
    title = "{Precision calculations of $B_{d,s}\to \pi,K$ decay form factors in soft-collinear effective theory}",
    eprint = "2212.11624",
    archivePrefix = "arXiv",
    primaryClass = "hep-ph",
    doi = "10.1007/JHEP03(2023)140",
    journal = "JHEP",
    volume = "03",
    pages = "140",
    year = "2023"
}

@article{PDG:2024,
    author = "Navas, S. and others",
    collaboration = "Particle Data Group",
    title = "{Review of particle physics}",
    doi = "10.1103/PhysRevD.110.030001",
    journal = "Phys. Rev. D",
    volume = "110",
    number = "3",
    pages = "030001",
    year = "2024"
}

@misc{cui:2301.12391,
      title="{Shedding new light on ${\cal R}(D_{(s)}^{(\ast)})$ and $|V_{cb}|$ from semileptonic $\bar B_{(s)} \to D_{(s)}^{(\ast)} \ell \bar\nu_{\ell}$ decays}", 
      author={Bo-Yan Cui and Yong-Kang Huang and Yu-Ming Wang and Xue-Chen Zhao},
      year={2023},
      eprint={2301.12391},
      archivePrefix={arXiv},
      primaryClass={hep-ph},
      url={https://arxiv.org/abs/2301.12391}, 
}

@article{Beneke:0008255,
   title="{Symmetry-breaking corrections to heavy-to-light $B$-meson form factors at large recoil}",
   volume={592},
   ISSN={0550-3213},
   url={http://dx.doi.org/10.1016/S0550-3213(00)00585-X},
   DOI={10.1016/s0550-3213(00)00585-x},
   number={1–2},
   journal={Nuclear Physics B},
   publisher={Elsevier BV},
   author={Beneke, M. and Feldmann, Th.},
   year={2001},
   month=jan, pages={3–34} }

@article{Becher:0408344,
   title="{Loop corrections to heavy-to-light form factors and evanescent operators in SCET}",
   volume={2004},
   ISSN={1029-8479},
   url={http://dx.doi.org/10.1088/1126-6708/2004/10/055},
   DOI={10.1088/1126-6708/2004/10/055},
   number={10},
   journal={Journal of High Energy Physics},
   publisher={Springer Science and Business Media LLC},
   author={Becher, Thomas and Hill, Richard J},
   year={2004},
   month=oct, pages={055–055} }

@article{Bourrely:0504016,
    author = "Bourrely, Claude and Caprini, Irinel",
    title = "{Bounds on scalar $K\pi$ form factors at zero momentum transfer}",
    eprint = "hep-ph/0504016",
    archivePrefix = "arXiv",
    reportNumber = "CPT-2005-P-017",
    doi = "10.1016/j.nuclphysb.2005.06.013",
    journal = "Nucl. Phys. B",
    volume = "722",
    pages = "149--165",
    year = "2005"
}

@article{Bourrely:0807.2722,
    author = "Bourrely, Claude and Caprini, Irinel and Lellouch, Laurent",
    title = "{Model-independent description of $B\to \pi\ell\nu$ decays and a determination of $|V_{ub}|$}",
    eprint = "0807.2722",
    archivePrefix = "arXiv",
    primaryClass = "hep-ph",
    reportNumber = "CPT-P36-2007",
    doi = "10.1103/PhysRevD.82.099902",
    journal = "Phys. Rev. D",
    volume = "79",
    pages = "013008",
    year = "2009",
    note = "[Erratum: Phys.Rev.D 82, 099902 (2010)]"
}

@article{Lellouch:9509358,
    author = "Lellouch, Laurent",
    title = "{Lattice constrained unitarity bounds for $\bar B^0\to \pi^+\ell^-\bar\nu_\ell$ decays}",
    eprint = "hep-ph/9509358",
    archivePrefix = "arXiv",
    reportNumber = "CPT-95-P-3236",
    doi = "10.1016/0550-3213(96)00443-9",
    journal = "Nucl. Phys. B",
    volume = "479",
    pages = "353--391",
    year = "1996"
}

@article{Beneke:0508250,
    author = "Beneke, M. and Yang, D.",
    title = "{Heavy-to-light $B$ meson form-factors at large recoil energy: Spectator-scattering corrections}",
    eprint = "hep-ph/0508250",
    archivePrefix = "arXiv",
    reportNumber = "PITHA-05-10",
    doi = "10.1016/j.nuclphysb.2005.11.027",
    journal = "Nucl. Phys. B",
    volume = "736",
    pages = "34--81",
    year = "2006"
}

@article{Schmidt:1201.6149,
    author = "Schmidt, Barbara and Steinhauser, Matthias",
    title = "{CRunDec: running and decoupling of the strong coupling and quark masses}",
    eprint = "1201.6149",
    archivePrefix = "arXiv",
    primaryClass = "hep-ph",
    reportNumber = "SFB-CPP-12-03, TTP12-02",
    doi = "10.1016/j.cpc.2012.03.023",
    journal = "Comput. Phys. Commun.",
    volume = "183",
    pages = "1845--1848",
    year = "2012"
}

@article{Herren:1703.03751,
    author = "Herren, Florian and Steinhauser, Matthias",
    title = "{Version 3 of RunDec and CRunDec}",
    eprint = "1703.03751",
    archivePrefix = "arXiv",
    primaryClass = "hep-ph",
    reportNumber = "TTP17-011",
    doi = "10.1016/j.cpc.2017.11.014",
    journal = "Comput. Phys. Commun.",
    volume = "224",
    pages = "333--345",
    year = "2018"
}

@article{Chetyrkin:0004189,
    author = "Chetyrkin, K. G. and Kuhn, Johann H. and Steinhauser, M.",
    title = "{RunDec: running and decoupling of the strong coupling and quark masses}",
    eprint = "hep-ph/0004189",
    archivePrefix = "arXiv",
    reportNumber = "DESY-00-034, TTP-00-05",
    doi = "10.1016/S0010-4655(00)00155-7",
    journal = "Comput. Phys. Commun.",
    volume = "133",
    pages = "43--65",
    year = "2000"
}

@article{Burdman:2000ku,
    author = "Burdman, Gustavo and Hiller, Gudrun",
    title = "{Semileptonic form factors from $B\to K^*\gamma$ decays in the large energy limit}",
    eprint = "hep-ph/0011266",
    archivePrefix = "arXiv",
    reportNumber = "SLAC-PUB-8708, BUHEP-00-20, FERMILAB-PUB-00-299-T",
    doi = "10.1103/PhysRevD.63.113008",
    journal = "Phys. Rev. D",
    volume = "63",
    pages = "113008",
    year = "2001"
}

@article{Beneke:2008.12494,
    author = "Beneke, Martin and Bobeth, Christoph and Wang, Yu-Ming",
    title = "{$B_{d,s}\to\gamma\ell\bar{\ell}$ decay with an energetic photon}",
    eprint = "2008.12494",
    archivePrefix = "arXiv",
    primaryClass = "hep-ph",
    reportNumber = "TUM-HEP-1280/20",
    doi = "10.1007/JHEP12(2020)148",
    journal = "JHEP",
    volume = "12",
    pages = "148",
    year = "2020"
}

@article{Braun:0309330,
    author = "Braun, V. M. and Ivanov, D. Yu. and Korchemsky, G. P.",
    title = "{The $B$ meson distribution amplitude in QCD}",
    eprint = "hep-ph/0309330",
    archivePrefix = "arXiv",
    reportNumber = "LPT-ORSAY-03-63",
    doi = "10.1103/PhysRevD.69.034014",
    journal = "Phys. Rev. D",
    volume = "69",
    pages = "034014",
    year = "2004"
}

@article{Gubernari:2203.08493,
    author = "Gubernari, Nico and Khodjamirian, Alexander and Mandal, Rusa and Mannel, Thomas",
    title = "{$B \to D_1 (2420)$ and $B \to D_1^\prime (2430)$ form factors from QCD light-cone sum rules}",
    eprint = "2203.08493",
    archivePrefix = "arXiv",
    primaryClass = "hep-ph",
    reportNumber = "SI-HEP-2022-04, P3H-22-027",
    doi = "10.1007/JHEP05(2022)029",
    journal = "JHEP",
    volume = "05",
    pages = "029",
    year = "2022"
}

@article{Beneke:0211358,
   title="{Multipole-expanded soft-collinear effective theory with non-Abelian gauge symmetry}",
   volume={553},
   ISSN={0370-2693},
   url={http://dx.doi.org/10.1016/S0370-2693(02)03204-5},
   DOI={10.1016/s0370-2693(02)03204-5},
   number={3–4},
   journal={Physics Letters B},
   publisher={Elsevier BV},
   author={Beneke, M. and Feldmann, Th.},
   year={2003},
   month=feb, pages={267–276} }

@article{Boos:0504005,
    author = "Boos, H. and Feldmann, Th. and Mannel, T. and Pecjak, B. D.",
    title = "{Shape functions from $\bar B\to X_c\ell\bar\nu_\ell$}",
    eprint = "hep-ph/0504005",
    archivePrefix = "arXiv",
    reportNumber = "SI-HEP-2005-02",
    doi = "10.1103/PhysRevD.73.036003",
    journal = "Phys. Rev. D",
    volume = "73",
    pages = "036003",
    year = "2006"
}

@article{Grozin:9607366,
   title={Asymptotics of heavy-meson form factors},
   volume={55},
   ISSN={1089-4918},
   url={http://dx.doi.org/10.1103/PhysRevD.55.272},
   DOI={10.1103/physrevd.55.272},
   number={1},
   journal={Physical Review D},
   publisher={American Physical Society (APS)},
   author={Grozin, A. G. and Neubert, M.},
   year={1997},
   month=jan, pages={272–290} }
\end{document}